\documentclass[reprint,amsmath,amssymb,aps,prb]{revtex4-1}
\usepackage{comment}
\usepackage{graphicx}% Include figure files
\usepackage{dcolumn}% Align table columns on decimal point
\usepackage{bm}% bold math
\usepackage{hyperref}% add hypertext capabilities
\usepackage{amsmath,amsfonts}
\usepackage{color}
\newcolumntype{P}[1]{>{\centering\arraybackslash}p{#1}}

\begin{document}

%\preprint{APS/123-QED}

\title{
%Predictable control of 
Alternative stable states in a model of 
microbial community
%ecosystems
\\ 
limited by multiple essential nutrients}
%communities by multiple essential resources}
% Force line breaks with \\

\author{Veronika Dubinkina}
\thanks{These three authors contributed equally}
\affiliation{Department of Bioengineering and Carl R. Woese Institute for Genomic Biology,
University of Illinois at Urbana-Champaign, Urbana, IL 61801, USA}
\author{Yulia Fridman}
\thanks{These three authors contributed equally}
\affiliation{National Rserach Center "Kurchatov Institute", 
Akademika Kurchatova pl., Moscow, 123182, Russia}
\author{Parth Pratim Pandey}
\thanks{These three authors contributed equally}
\affiliation{Carl R. Woese Institute for Genomic Biology and 
National Center for Supercomputing Applications, 
University of Illinois at Urbana-Champaign, Urbana, IL 61801, USA}
\author{Sergei Maslov}
\thanks{maslov@illinois.edu}
\affiliation{
Department of Bioengineering and Carl R. Woese Institute for Genomic Biology,
University of Illinois at Urbana-Champaign, Urbana, IL 61801, USA.}

\date{\today}

\begin{abstract}
% Microbial communities play an important role in human health, 
% agriculture, global, climate and industrial biotechnology. 
Microbial communities routinely have several alternative stable states observed for the same environmental parameters. Sudden and irreversible  transitions between these states make external manipulation of these systems more complicated. To better understand the mechanisms and origins of multistability in microbial communities, we introduce and study a model of a microbial ecosystem colonized by multiple specialist species selected from a fixed pool. Growth of each species can be limited by essential nutrients of two types, e.g. carbon and nitrogen, each represented in the environment by multiple metabolites. We demonstrate that our model has an exponentially large number of potential stable states realized for different environmental parameters. Using game theoretical methods adapted from the stable marriage problem we predict all of these states based only on ranked lists of competitive abilities of species for each of the nutrients. We show that for every set of nutrient influxes, several mutually uninvadable stable states are generally feasible and we distinguish them based upon their dynamic stability. We further explore an intricate network of discontinuous transitions (regime shifts) between these alternative states both in the course of community assembly, or upon changes of nutrient influxes. 

\end{abstract}

%\pacs{Valid PACS appear here}
%Use showkeys class option if keyword
\maketitle
%\tableofcontents
\section*{Introduction}
%\textcolor{red}{Logic: bacteria grow on essential nutrients $->$ in reality they have several non-substitutable types $->$ concept of limitation $->$ lack of models incorporating this fact, only Tilman, but Tilman operates with only two types, while it is not true (organic/inorganic C N sources and diff bacteria speciation) $->$ our model}\\
Microbial communities play an important role in medicine (human microbiome \cite{Qin2010,HMP2012}),
agriculture (soil \cite{chaparro2012manipulating}, 
plant root \cite{pii2015microbial}, and animal \cite{hobson2012rumen}
microbiomes), climate, (via carbon cycle feedbacks \cite{bardgett2008microbial}), 
and technology (industrial bioreactors \cite{Briones2003}, 
wastewater digesters \cite{wagner2002microbial}, etc.)
They are often characterized by more than one 
stable state %realized 
observed for the same set of 
environmental parameters 
\cite{zhou2007differences,lahti2014tipping,Lozupone2012,Zhou2013}.
% the next sentence is too specific for the intro...
Such alternative stable states 
\cite{sutherland1974multiple,holling1973resilience,may1977thresholds,Fukami2011,bush2017oxicanoxic} 
have several hallmark properties discussed in Ref. \cite{Schroder2005} 
including ``discontinuity in the response 
to an environmental driving parameter'' 
(referred to as regime shifts in the ecosystems literature), 
lack of recovery after a perturbation (hysteresis), and 
``divergence due to different initial conditions'' or 
due to the order in which species were introduced during 
the initial colonization process \cite{goyal2018multiple}. 

To be able to predict the behavior of a 
microbial community, one
needs to understand the mechanisms 
that favor one such state 
over the other and the factors 
triggering transitions between them. In many practical 
situations we would also like to be able to manipulate and control 
a microbial ecosystem in a predictable manner, 
and the existence of more than one stable state 
greatly complicates this task 
\cite{gonze2017multi}.
%\textcolor{red}{
% To reliably manipulate and control these systems to a desired 
% state one needs mechanistic understanding of their dynamics.
% \begin{comment}
% In many practical situations we would like to manipulate and control 
% microbial communities in a predictable and robust manner, but the 
% existence of multiple states complicates this task. 
% Thus we need a mechanistic understanding of the factors that 
% determine the properties of all the states (and hence enumerate 
% all possible states) and the factors responsible for triggering 
% transitions between them.
% \end{comment}

% A major factor determining 
% the state of a microbial ecosystem (defined as its species composition) 
% are nutrients present in the environment. 

%state of a microbial ecosystem  
%defined as the 
Growth rates and, ultimately, abundances 
of microbial species are affected by multiple factors, with 
availability of nutrients being among 
the most important ones. 
Thus environmental concentrations and 
influx rates of externally supplied nutrients
play a crucial role in determining the 
state (or multiple states) of a microbial ecosystem defined 
by its species composition.  Changes in nutrient concentrations
can also trigger transitions (regime 
shifts \cite{bush2017oxicanoxic}) between these states  
(see Ref. \cite{Shade2012} for a recent literature survey on microbial communities' response to disturbances). Nutrients required for growth of a microbial (or any other) species 
exist in the form of multiple metabolites of several 
essential types (i.e. sources of 
C, N, P, Fe, etc.). The growth of each species is usually 
limited by the most scarce type of nutrient (for an exception to this rule see 
Ref. \cite{browning2017nutrient} demonstrating 
that oceanic phytoplankton can be co-limited by more than one essential nutrient).
%%% add sentence about specialists vs generalists to justify our 
%assumption about microbes using only particular carbon or nitrogen source
%\textcolor{red}{Moreover, different microbial species can have different 
%metabolic capabilities to utilize the nutrients of a specific type  \cite{}.}

% I'm not sure that we need this paragraph, it might be better directly proceed with the models considering resources in our intro

Here we introduce and study a new mathematical model of a microbial community 
limited by multiple essential nutrients. To put our model in context of previously 
studied ones, we briefly review common approaches to  modelling of microbial 
communities.

One of the simplest and thereby most popular approaches in ecological modelling \cite{may1972will,  mounier2008microbial,allesina2012stability,faust2012microbial,stein2013ecological,marino2014mathematical,fisher2014identifying,berry2014deciphering,coyte2015ecology,gibson2016origins,friedman2017community,bunin2017ecological,fried2017alternative,xiao2017mapping} is based on variants of generalized Lotka-Volterra (gLV) model  \cite{lotka1926elements,volterra1926variations}. The gLV model does 
not explicitly consider nutrients, replacing them with the effective 
direct inter-species interactions %\textcolor{red}{(see Ref. \cite{Momeni2017} for limitations of this approach: this is repeated below!!!)}.
%%%% might want to remove:
%In these models it is assumed that a species’ growth can be written as the sum of its basal growth and the %effect of its pairwise interactions with all the species in the community.
While gLV models have provided valuable insights due to their simplicity, 
they have also been criticized when applied to multispecies communities \cite{Momeni2017}. 
%\textcolor{red}{how to cite the 2 Gore papers?}
%\textcolor{red}{Veronika, cite all the criticism papers. 
% Also include all the higher order critics.
%should we mention this, reviewers may give us a hard time on this. 
%Safe not to say this in this way. But gLV fails to predict some things
%\cite{Momeni2017, abreu2018mortality}
%}

%Another popular type is consumer-resource models was introduced by MacArthur \cite{macarthur1964competition}, where 

Another popular approach is based on variants of the classic MacArthur consumer-resource model \cite{macarthur1964competition, macarthur1970species} in which the growth rate 
of each species is given by a linear combination of concentrations of several 
fully substitutable resources \cite{huisman2001biological, tikhonov2017collective, 
posfai2017metabolic, goldford2018emergent, goyal2018multiple, butler2018stability}. This 
corresponds to a logical OR-gate operating on all nutrient inputs of a given species.  
A more general case in which growth rate can be arbitrary non-linear function of concentrations of {\it just two resources} has been considered in the 
foundational work by Tilman \cite{tilman1982resource}.
%, miller2005critical}. 
The modeling framework and the geometric interpretation of resource dynamics 
developed in Ref. \cite{tilman1982resource} proved to be useful 
for interpreting experimental data describing ecology and plankton communities 
\cite{burson2018competition} and remains an active field of research \cite{menge2012nitrogen, brauer2012nutrient, koffel2018facilitation}. While bistability between a pair of 
species  
%alternative stable states with a single species in each state 
has already been mentioned in Ref. \cite{tilman1982resource}, more complex 
scenarios with multiple species and/or more than two alternative stable states, to the best 
of our knowledge has not been
described (see \cite{fried2017alternative} for the analysis of alternative stable states 
in gLV model). 
%systematically investigated. 

%However up to date most of them consider only 2-3 resources which allows for a small 
%number of alternative stable states.
%  The important property of a \textcolor{red}{Tilman-like} 
%  models is that both nutrients are necessary for growth (logical {\it AND-gate} on inputs). 
%That is rather different from another popular type of consumer-resource models introduced by MacArthur \cite{macarthur1964competition}, where all resources are completely substitutable (logical {\it OR-gate} on inputs). 
%%% add sentence about advantages of tilman model..

% \begin{comment}
% For feasibility cite \cite{grilli2017feasibility}
% Stability: \cite{Ives2007}
% \end{comment}

Our study fills this gap by 
generalizing the model of Ref. \cite{tilman1982resource} to more than two metabolites.
The population dynamics in our model is shaped by species competing for {\it multiple metabolites} 
of two {\it essential types} to which we refer to as 
sources of carbon and nitrogen, while any other pair of essential nutrients is equally possible. 
In our model multiple different metabolites (e.g. different sugars) 
can serve as carbon sources, and another set of metabolites - 
as nitrogen sources (for simplicity we ignore the possibility 
of the same metabolite providing both carbon and nitrogen). 
The ecosystem in our model is colonized by highly specialized species, 
with each species capable of utilizing just one specific pair of metabolites as its 
carbon and one nitrogen sources. Using specialist species greatly 
simplifies our calculations but we will also propose variants of our model incorporating generalist species.
%Below we discuss how such specialization could effectively emerge due to diauxic shifts \textcolor{blue}{(PARTH:Are we really discussing below how such specialization emerge???)} as well as propose variants of our model incorporating generalist species.

%Thus we have competition for each pair of essential resources, which allows us to observe a lot more %alternative states than in simple two resource Tilman case \cite{tilman1982resource}. 

We show that our model is characterized by exponentially large number of steady states, 
each realized for different sets of environmental parameters.
Using game theoretical methods adapted from the well-known 
stable marriage problem \cite{gale1962college,gusfield1989stable}, 
we show that all of these states can be identified based only on 
ranked lists of competitive abilities of species for each of the resources.
For any set of nutrient influxes a few {\it mutually uninvadable stable states} 
are generally feasible. They may or may not be dynamically stable, and our 
methods allow us to infer dynamic stability of each of them 
for any set of species' C:N stoichiometries. As in Ref. 
\cite{macarthur1970species,tilman1982resource}, 
multistability (alternative stable states) is only possible when stoichiometric 
ratios of different species are not identical.
Our model allows us to explore the intricate network of discontinuous 
transitions (regime shifts) between these alternative states in the 
course of community assembly and changing nutrient influxes. The aim 
of our study is to provide an intuitive understanding of the basic rules 
governing the existence of alternative communities in microbial 
ecosystems growing on multiple essential resources 
and of transitions between these states.

While we formulate our model for microbial ecosystems, nothing in its 
rules prevent it from describing macroscopic ecosystems, e.g. that dominated by plants.
In fact, the model of Ref. \cite{tilman1982resource}, which our model generalizes, has been 
successfully applied to a broad variety of natural and artificial ecosystems.

\section*{Model and Results}
Our model describes an ecosystem colonized by 
microbes selected from a pool of $S$ species.
Growth of species in our community is limited by two types 
of essential nutrients, which we will refer to as 
``carbon'' and ``nitrogen'' sources. 
In principle, these could be any two types of nutrients essential for life: 
C, N, P, Fe, etc. A straightforward generalization 
of our model involves three or more types of essential nutrients. 
Carbon and nitrogen sources exist 
in the environment in the form of $K$ distinct metabolites 
containing carbon, and $M$ other metabolites containing nitrogen.
To allow for a mathematical understanding of steady states in our model 
%we assume that each of our $S$ species is an extreme specialist, capable of utilizing just one pair of carbon and nitrogen metabolites.
we assume that each of our $S$ species is an extreme specialist, capable of utilizing a single pair resources, i.e., one carbon and one nitrogen metabolites.
%Although we made this assumption mostly for simplicity it is not completely unnatural, since ... cite diff resource utilization capabilities
We further assume 
that the growth rate $g_{\alpha}$ of a species $\alpha$ is determined by 
the concentration of the rate-limiting resource via Liebig's 
law of the minimum \cite{de1994liebig}: 
\begin{equation}
g_{\alpha}(c_i,n_j)=\min(\lambda_{\alpha}^{(c)}c_i,\lambda_{\alpha}^{(n)}n_j) \quad ,
\label{growth_law}
\end{equation}
where $c_i$ and $n_j$, are the environmental concentrations of the 
carbon resource $i$ and the nitrogen resource $j$ consumed by this species $\alpha$, 
while $\lambda_{\alpha}^{(c)}$ and $\lambda_{\alpha}^{(n)}$ 
are, respectively, its competitive abilities for these resources.
%%%%%%%% probably should be in the discussion
% As discussed in the Supplementary Information, our results can be easily 
% extended to a more general functional form of microbial growth 
% $g_{\alpha}(c_i,n_j)$ including Monod equation for two essential nutrients. 
% For example, Ref. \cite{tkachenko2017onset} one of us 
% explored the properties of a model in which $g_{\alpha}(c_i,n_j)=
% \lambda_{ij} c_i \cdot n_j$ (in a different context of polymer growth)
% and arrived to results similar to this study. 
The dynamics of microbial populations $B_{\alpha}$ is defined by: %(\textcolor{red}{should we say B is the population density rather than population})
\begin{equation}
\frac{dB_{\alpha}}{dt}=
B_{\alpha}\left[g_{\alpha}(c_i,n_j)-\delta \right] 
\quad . 
\label{eqn:dbdt}
\end{equation}
Here we assumed that microbes have no 
maintenance costs
%however a general case of maintenance cost can be handled within the same framework without changing our results (See SI). 
and that 
both microbes and their resources 
are in a chemostat-like environment 
subject to a constant dilution rate $\delta$. 
However, all our results remain unchanged in a more general 
case of non-zero (and microbe-specific) 
microbial maintenance cost that could be 
different from the dilution rate of resources.
The resources are externally supplied to our system 
at fixed influxes $\phi_i^{(c)}$ and $\phi_j^{(n)}$ 
and their concentrations follow the equations:
\begin{eqnarray}
\frac{dc_i}{dt}&=&\phi_i^{(c)} -\delta \cdot c_i- 
%\nonumber \\
%&-& 
\sum_{\text{all }\alpha \text{ using }c_i}
B_{\alpha}\frac{g_{\alpha}(c_i,n_j)}{Y_{\alpha}^{(c)}}\quad, \nonumber\\
\frac{dn_j}{dt}&=&\phi_j^{(n)} -\delta \cdot n_j- 
%\nonumber \\
%&-& 
\sum_{\text{all }\alpha \text{ using }n_j}
B_{\alpha}\frac{g_{\alpha}(c_i,n_j)}{Y_{\alpha}^{(n)}} \quad .
\label{eqn:dcdt}
\end{eqnarray}
Here $Y_{\alpha}^{(c)}$ and $Y_{\alpha}^{(n)}$ are the carbon and nitrogen growth
yields of the species $\alpha$ quantifying the number 
%(or rather a small  fraction) 
of microbial cells generated per unit of concentration 
of each of the two consumed resources.  
It is easy to show that our system satisfies mass conservation laws:
\begin{eqnarray}
c_i+\sum_{\text{all }\alpha \text{ using }c_i}\frac{B_{\alpha}}{Y_{\alpha}^{(c)}}=\frac{\phi_i^{(c)}} {\delta} \nonumber \quad,\\
n_j+\sum_{\text{all }\alpha \text{ using }n_j}\frac{B_{\alpha}}{Y_{\alpha}^{(n)}}=\frac{\phi_i^{(n)}} {\delta} \quad.
\label{eqn:conservation_laws}
\end{eqnarray}

%%%% we have it in results and discussion, to avoid repetition and distraction of the readers from the model let's comment it for now

%In this study we explore the properties of the steady states of this microbial ecosystem. 
%Our results shown below strongly suggest the existence of a 
%(generally non-convex) Lyapunov function that would preclude other types of 
%stationary solutions to these equations such as stable cycles or chaos.
%%%%%%%%%%%%%
The concentrations of all surviving bacteria and all resources 
in a steady state are determined by setting the left hand sides of Eqs. 
\ref{eqn:dbdt},\ref{eqn:dcdt} to zero and solving them for the 
steady state concentrations $B_{\alpha}$, $c_i$, and $n_j$. 
% As shown in SI 
% (see the section "Constraints on steady states from 
% microbial and nutrient dynamics"), 
As we will show the steady state equations 
impose a number of constraints on competitive abilities of 
surviving microbes, their yields, and nutrient fluxes, where 
a given steady state is feasible.
One type of constraints comes from the competitive exclusion principle 
\cite{gause1932experimental,gause1934book} 
contained in the equations \ref{eqn:dbdt}. In models with substitutable 
resources of one type (say, multiple sources of carbon), the specialist species 
with the largest competitive ability $\lambda^{(c)}$ 
generally wins the battle
for each carbon source (see e.g. Ref.  \cite{goyal2018diversity}, where it 
was employed to describe the colonization dynamics of an ecosystem 
with cross-feeding). 
In the case of non-substitutable essential resources of two
(or more) types, that is the subject of this study, 
this simple rule is replaced with the following 
{\it two competitive exclusion rules}:
\begin{itemize}
\item Exclusion Rule 1: Each nutrient (either carbon or nitrogen source) 
can limit the growth of no more than one species $\alpha$.
From this it follows that (barring special circumstances) 
the number of surviving species in any given steady state cannot be 
larger than $K+M$, the total number of nutrients. 
\item Exclusion Rule 2: Each nutrient (say, a specific carbon source) 
can be used by any number of species in a 
non-rate-limiting fashion (that is to say, where it does not influence 
species' growth rate by setting the value of the minimum in Eq. \ref{growth_law}). 
However, any such species $\beta$ has to have $\lambda_{\beta}^{(c)} >\lambda_{\alpha}^{(c)}$, 
where $\lambda_{\alpha}^{(c)}$ is the competitive ability of 
the species whose growth is limited by this nutrient. 
In case of a non-rate-limiting use of a nitrogen source, the constraint becomes 
$\lambda_{\beta}^{(n)} >\lambda_{\alpha}^{(n)}$. 
\end{itemize}

%%% we are not using abbreviations in the following text anyways, and Nat Comm guidelines says we should avoid uncommon abbreviations
As we will show below, any set of microbes 
with the rate-limiting nutrient specified for each microbe, 
that satisfy the above two constraints imposed by the exclusion rules is 
a steady state in our system for some set of nutrients influxes (but this does not characterize the dynamic stability of the states). 
We will refer to them as states allowed by the exclusion rules, 
or simply ``allowed states''. %(AS).

\begin{figure}
%\centerline{\includegraphics[width=1.0\textwidth]{fig1n.eps}}
\centerline{\includegraphics[width=\linewidth]{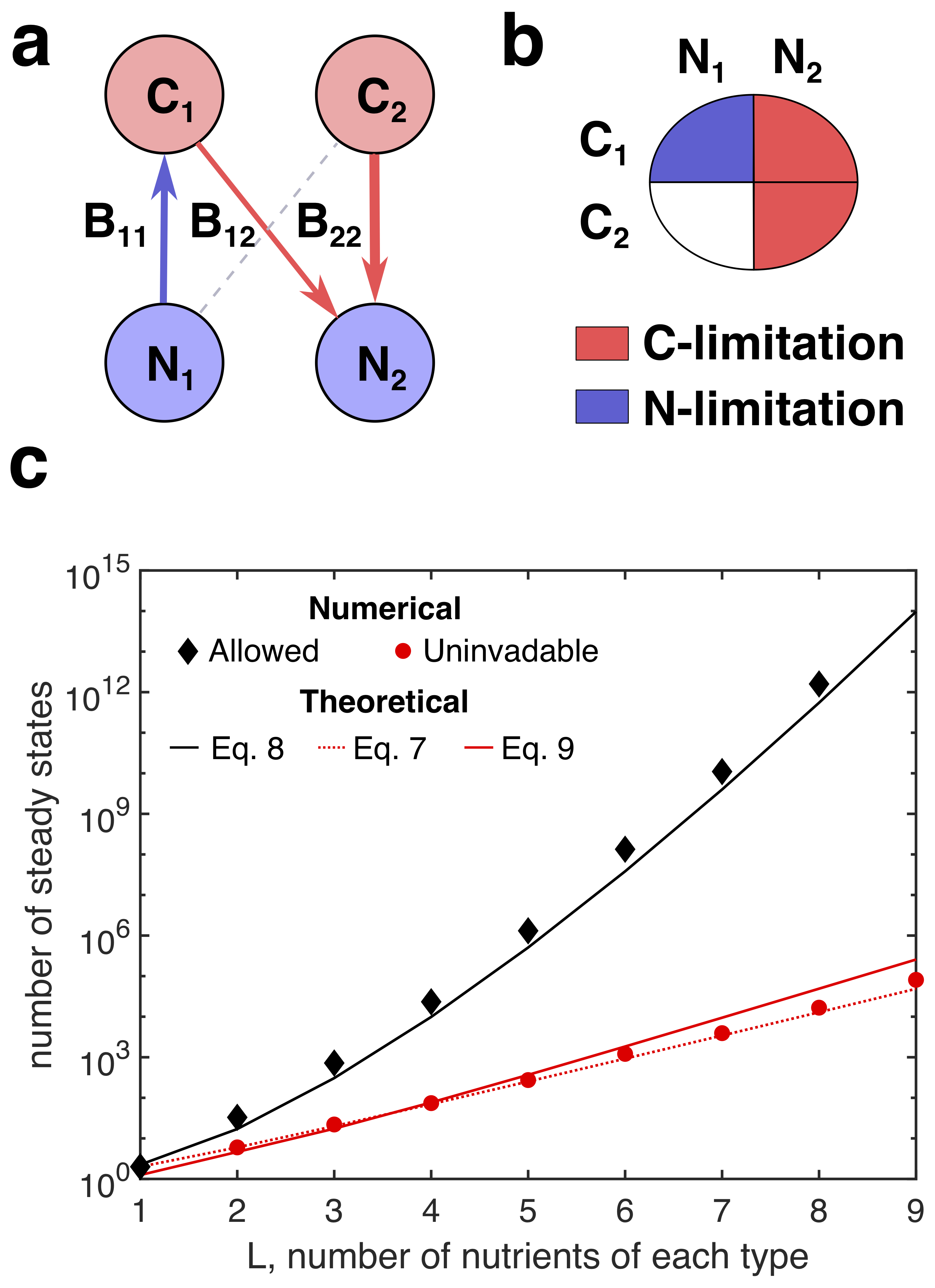}}
\caption{{\bf Number of allowed and uninvadable
states.} 
(\textbf{a-b}) Two equivalent schematic representations of a simple community guided by exclusion rules in our model. (\textbf{a}) Bipartite network representation of an allowed state. Nodes of two types represent different nutrients, arrows represent different bacteria species, direction and color of each arrow corresponds to the nutrient limitation of each species (arrow goes from the limiting nutrient to non-limiting one), 
(\textbf{b}) %\textcolor{red}{Veronika, I think we should have (i,j) written inside each cell. Then panel b will have all info that panel a has. Also you have mentioned the species as B$_{ij}$. But that is the biomass of species (i,j). So we can write (i,j) instead of B$_{ij}$ in panel a and b} 
Matrix representation of an allowed state. The allowed state shown in panel A can be represented as a matrix with $K=2$ rows (representing the 2 carbon nutrients) and $M=2$ columns (representing the 2 nitrogen nutrients). Cell $(i,j)$ of the matrix represents the species consuming $i^{th}$ carbon source and $j^{th}$ nitrogen source. A colored cell implies the presence of the species and an empty cell signifies its absence. If the species is present, it can be limited by either the carbon it is utilizing (in which case the cell is colored red) or the nitrogen it is consuming (colored blue).
%To convey the limiting nutrient for the $(i,j)^{th}$ species we color the $(i,j)^{th}$ element of the state matrix by red if it is carbon limited or blue if the species is nitrogen limited
%Each cell corresponds to bacteria capable of utilizing a specific pair of resources. The color of the cell identifies the presence of a particular specie in a community (white means bacteria is absent) and it's limitation in this case (red - carbon limitation, blue - nitrogen limitation).
%Algorithm of enumeration of allowed and uninvadable states. On step 1 we are choosing set of outgoing links for one type of nutrient and on step 2 we are filling all possible combinations (allowed by our exclusion rules) of outgoing links for another type.
(\textbf{c})
The number of allowed states
% (AS) 
including both invadable and uninvadable steady states, 
red squares, and the subset of uninvadable states
% (UIS)
, black circles, obtained by exhaustive testing of 
all possible states against the exclusion rules 1 and 2. 
The x-axis is the number of nutrients of each type ($L$ carbon sources and 
$L$ nitrogen sources). The pool has $L^2$ species - one for each pair of nutrients. Red and black %dashed 
lines are the theoretical estimates for each of the numbers %(long-dashed 
for continuous approximation. 
%and dashed for the discrete 
%approximation as described in the SI).
Red dashed line is the lower bound to the number of uninvadable states based on the stable marriage model and given by the Eq. \ref{NUIS lower bound for L}. Note the logarithmic scale on the y-axis. %\textcolor{red}{Should we try to have this figure in page 3. We don't have any figure for a long time in the text}
}
\label{fig1n}
\end{figure}

Each allowed state could be conveniently 
visualized in terms of a bipartite directed network
with vertices corresponding to individual resources 
and edges connecting carbon and nitrogen sources - to surviving species (see Fig. \ref{fig1n}A). 
We choose the direction of each edge to go from the rate-limiting 
resource for this species to the non-rate-limiting one. Our 
exclusion rules can be reformulated in the network language as follows: 
each vertex can have at most one outgoing link and any 
number of incoming links (Rule 1). 
All incoming links have to have larger values of $\lambda$ than the 
outgoing link (if any) (Rule 2). Hence, the task of discovering and enumerating 
all possible steady states realized for 
different nutrient influxes is equivalent to finding the set of 
all directed graphs satisfying the above constraints. An allowed state can also be conveniently represented as a matrix with K rows (representing the K carbon nutrients) and M columns (representing the M nitrogen nutrients) where each element $(i,j)$ represents the specialist species consuming $i^{th}$ carbon source and $j^{th}$ nitrogen source. To convey the limiting nutrient for the $(i,j)^{th}$ species we color the $(i,j)^{th}$ cell of the state matrix by red if it is carbon limited or blue if the species is nitrogen limited. A cell is left empty/uncolored if the species allowed to consume that pair of resources is absent from the community. In this formulation the constraints imposed by the exclusion rules translate to: Rule 1 - each row of the state matrix can have at most one red species (i.e., limited by carbon) and each column can have at most one blue species (i.e., limited by nitrogen) and Rule 2 - In each row $\lambda^{(c)}$ of all blue species should be larger than the $\lambda^{(c)}$ of the red species (if any) and similarly in every column all red species should have a larger $\lambda^{(n)}$ than the $\lambda^{(n)}$ of the blue species (if any).
\ref{fig1n}B) shows the corresponding matrix form of the state described in Fig \ref{fig1n}A. We will use this matrix representation of states in the following figures.
%Our exclusion rules could be transferred to the matrix representation in a following sense: in one row or column there could not be more than one bacteria limited by the corresponding type of nutrient, i.e. if $i^{th}$ row corresponds to all bacteria capable of utilizing $C_i$ only one of them could be limited by it.

It is useful to single out a subset of 
``uninvadable states'' %(UIS) 
among all of the 
steady states allowed by the exclusion rules. These are defined by the 
condition that not a single microbe (among $S$ species in our pool)
that is missing from the current steady state can successfully 
grow in it, thereby invading the ecosystem and modifying the 
steady state. To be able to grow, both $\lambda^{(c)}$ and 
$\lambda^{(n)}$ of the invading microbe has to be larger 
than those constants for each of the resident microbes (if it exists) 
currently limited by these resources. 
% Since  allowed states treat the same microbe
% limited by C and N differently, it is natural to extend 
% our condition for uninvadability by testing that no 
% species already present in the community can change its 
% limiting resource and grow in the opposite 
% (represented in our network representation
% by inverting this species' arrow). 

%

\subsection*{Total number of allowed and uninvadable states}

A natural question to ask is how many allowed states %(AS) 
and, separately, 
how many uninvadable states %(UIS) 
are in principle possible 
for a given pool of species (each state will be realized for different environmental parameters). 
Our exclusion rules allow one to identify all of them
{\it based only on the set of  ranked tables of competitive 
abilities of all microbes for each of the nutrients}. 
%\textcolor{red}{VERONIKA: Ok, I see what we probably wanted to emphasize here - yields (another set of parameters in our model) has nothing to do with determining allowed/non-allowed states. And the above measurements remark belongs to the discussion.}
For large $K$,$M$, and $S$ this is computationally expensive. Indeed, in a 
brute force method one has to check all of the $3^S$ 
candidate states (each of $S$ species 
could be either absent, or, if present - limited by either 
carbon or nitrogen) for compliance with the exclusion 
rules 1 and 2. Each of the allowed states then need to 
be checked for invasion against up to $S-1$ missing species to 
verify their uninvadability. %\textcolor{red}{A smarter application of the exclusion rule 1 can reduce this number somewhat, but it still remains exponentially large. Do we need this sentence? At least rephrase}

To help the process of search for allowed and uninvadable states
we mapped the problem of finding them to that of finding 
all stable matchings in the celebrated stable 
marriage and college admissions 
problem in game theory and economics \cite{gale1962college}, 
which is algorithmically well-studied \cite{gusfield1989stable}.
% In particular, we discovered a one-to-one correspondence 
% between uninvadable steady states in our model and the set of 
% stable matchings 
This connection is described in detail in \ref{SI-Note3}. In a nutshell, we found a one-to-one correspondence 
between the set of uninvadable steady states in our model and 
the set of stable matchings in Gale-Shapley college 
admission problem. One first allocates the number of ``partners''
(in-degrees in our network representation of steady states) 
for every resource of a given nutrient type (say N). 
As described in Methods Section, one can then use the mathematical machinery of the stable 
marriage problem to discover all stable matchings in which all 
C sources have out-degree 1 and each of N sources has in-degree
prescribed by our selected allocation.
%%%%%% VERONIKA: We have this part in methods, let's just refer to it
%Each stable matching describes a set of C-limited species (edges going from C to N). When supplemented with a prescribed set of species limited by N (edges going in the opposite direction, from N to C), it describes all allowed and exactly one uninvadable state with this set of in-degrees on sources of N. The process has to be repeated for all possible allocations of numbers of ``partners'' on N sources (in-degrees) and then for all partner allocations on another resource type (in-degrees on C in our example). 

Since the number of ways of selecting the number of partners 
%distributing $K$ carbon resources among 
%$M$ nitrogen resources and vice versa 
is exponentially large (the combinatoric factor is shown below), and that for 
each such distribution the Gale-Shapley theorem guarantees at least 
one stable matching, the overall number of uninvadable states is also 
exponentially large and is bounded from below by: 
\begin{equation}  
  N_{UIS}(K,M) \ge {{K+M}\choose{K}} \quad .
  %\simeq \frac{2^{K+M}}{\sqrt{\pi L}} \qquad .
  \label{NUIS lower bound}
\end{equation}
For equal (and large) number of carbon and nitrogen sources $K=M=L$
this estimate can be further simplified to give:
\begin{equation}  
  N_{UIS}(L) \ge \frac{4^{L}}{\sqrt{\pi L}} \quad
  %\simeq \frac{2^{K+M}}{\sqrt{\pi L}} \qquad .
  \label{NUIS lower bound for L}
\end{equation}
(see \ref{SI-Note3} for details). Note that the connection between uninvadable states in our system and 
the stable marriage problem is rather different than that
in Ref. \cite{goyal2018multiple}. Indeed, while in Ref. 
\cite{goyal2018multiple} ``stable marriages'' are established between 
microbes and sequentially-used (diauxic) substitutable resources,
in the present study the ``marriages'' are between 
different sources of carbon and nitrogen, while microbes 
play the role of ``matchmakers'' (connectors).

% Using methods different from the stable marriage problem we 
% were able to estimate the number of allowed and uninvadable states 
% in a continuous approximation: $S/M, S/K \gg 1$.
% The answer is given by a 2-dimensional integral 
% (or more generally $N_e$-dimensional
% integral for $N_e$ types of essential resources) 
% derived and evaluated in the saddle point approximation 
% in the SI (chapter "Number of steady states: 
% Continuous approximation").

To verify these mathematical results 
we carried out numerical simulations
of the model with equal number - $L$ - of carbon and nitrogen sources 
and a pool of $L^2$ species, with exactly one microbe using 
each pair of nutrients. %(see Methods Section for the algorithm generating all the allowed states). 
The number of states revealed by our numerical simulations is 
indeed very large. For example, for only 9 carbon resources, 
9 nitrogen resources and a pool of 81 species,
the microbial community 
is  capable of 81,004 distinct uninvadable states and roughly
$10^{14}$ allowed states. 
Fig.\ref{fig1n} shows the numerical results, which 
are in agreement with our theoretical predictions.
The number of allowed states increases 
faster than exponential. 
%(it can be shown 
%to be $(0.569L)^{2L}$).
In the continuous approximation 
(solid black curve in Fig. \ref{fig1n}C) it 
is asymptotically described by 
\begin{equation}
 N_{AS}(L) \simeq 0.827 (0.569 L + 0.901)^{2L}
\end{equation}
(see %Eq. \ref{SM_continuous_NAS} in
\ref{SI-Note4} for details).
The number of uninvadable steady states also 
rapidly increases with $L$. While for $L \leq 9$ 
it rather closely follows the lower bound
given by Eq. \ref{NUIS lower bound for L}
(dashed line in Fig. \ref{fig1n}C), 
for larger values of $L$ we saw a crossover 
to a faster-than-exponential regime 
where it grows as 
\begin{equation}
 N_{UIS}(L)=\frac{11.8769}{L}\left(0.1411\!L +0.2419\right)^L \qquad.
\end{equation}
This asymptotic formula derived in \ref{SI-Note4} is accurate for 
values of $L$ much larger than those shown in Fig. \ref{fig1n}C.
However, the numerical integration of the expression derived in 
the continuous approximation (solid red line in Fig. \ref{fig1n}C) 
is close to the exact number of uninvadable states (see \ref{SI-Note4}). The discrepancy is likely due to the fact that the 
continuous approximation assumes that the distribution of the
number of species per each pair of resources is Poisson with mean 
equal to 1 (instead of exactly one species in our numerical 
simulations).
% , approximately as 
% ${{2L}\choose{L}} \simeq 
% 2^{2L}/\sqrt{\pi L}$ (black dashed line in 
% Fig. \ref{fig1n}), which is the lower bound given 
% by the stable marriage model.
Note that this growth is much faster than sub-exponential expression 
recently calculated for Lotka-Volterra model 
with strong interactions \cite{fried2016communities}. 
However, unlike Ref. \cite{fried2016communities}, 
we calculate the total number of possible stable uninvadable states 
feasible for different values of environmental parameters.

\subsection*{Feasible regions of nutrient influxes for each of the steady states}
Steady states allowed by Eqs. \ref{eqn:dbdt}, \ref{eqn:dcdt} (satisfying the constraints imposed by our two exclusion rules) are further constrained 
by Eqs. \ref{eqn:conservation_laws}. That is to say,  for a given set of the $K+M$ nutrient influxes
only a very small subset of exponentially large number of 
allowed states would be feasible. 
From another angle, according to Eqs. \ref{eqn:conservation_laws} each allowed state has a finite region of nutrient influxes 
where it is feasible.
Similar to Ref. \cite{grilli2017feasibility} 
(for Generalized Lotka-Volterra model) and Ref. \cite{butler2018stability} 
(for consumer-resource MacArthur model)
testing if a given Allowed State %(AS)
is feasible at a specific nutrient influx
requires inverting the matrix in Eqs. \ref{eqn:conservation_laws} (see Eqs. \ref{eqn:Matrix-form} and \ref{eqn:Matrix-form2} in Methods Section %\nameref{Meth:FeasibleVolumeAlgo}
) to get the unique set of bacterial 
populations $B_{\alpha}$ for all species present in a given 
allowed state.  
%%%%%% VERONIKA: we have these details in Methods
%When the number of surviving species ($S_{surv}$) is smaller than $K+M$ (the number of nutrients), the process of inversion provides the population of the $S_{surv}$ species and the $K+M-S_{surv}$ resources not limited by any the of $S_{surv}$ species. The allowed state is feasible if and only if all the $S_{surv}$ microbial populations and non-limiting resource concentrations are positive. In Methods Section we describe this algorithm in more detail. 

In principle one can also find the set of all nutrient influxes for which a given allowed state is feasible in the opposite way that does not involve matrix inversion. One just 
needs to span (or sample by a Monte-Carlo simulation) the $K+M$-dimensional 
space of positive microbial populations and non-limiting nutrient 
concentrations in this allowed state. For each point in this set 
Eqs. \ref{eqn:conservation_laws} trivially define all nutrient 
influxes required to realize these populations/concentrations 
in the steady state defined by a given allowed state (see Methods Section for details).
%\textcolor{red}{VERONIKA: Can we write this paragraph in a more concise way?}
%and find the set of all nutrient influxes for which a given allowed state is feasible. This is an even simpler task since that does not involve matrix inversion. One just needs to span (or sample by a Monte-Carlo simulation) the $K+M$-dimensional space of positive microbial populations and non-limiting nutrient concentrations in this allowed state. For each point in this set Eqs. \ref{eqn:dbdt} trivially define all nutrient influxes required to realize these populations/concentrations in the steady state defined by a given allowed state. 

The volume of the region of feasible 
influxes quantifies the structural stability \cite{rohr2014structural}
of the steady state. States with larger volumes are expected to be more 
robust in case of fluctuating nutrient influxes.
In order for this volume to remain finite, 
we impose an upper bound on the influx of each nutrient: 
$\phi^{(c)}_i, \phi^{(n)}_j \leq \phi_0$.
Another way to quantify the structural stability of 
each allowed state is to calculate the volume of all influxes 
as bacterial abundances and non-limiting nutrient  
concentrations vary within a given positive range.
Structural stability defined this way is proportional 
to $\det{\hat{Y}^{-1}}$, where $\hat{Y}^{-1}$ is the matrix of 
inverse yields by which the vector of bacterial populations is 
multiplied in the Eq. \ref{eqn:conservation_laws}.

Each region of feasible influxes is generally 
bounded by multiple hyperplanes in a $K+M$-dimensional 
space and thus is difficult to visualize.
\begin{figure*}
%\centerline{\includegraphics[width=\linewidth]{fig2n.eps}}
\centerline{\includegraphics[width=1.0\textwidth]{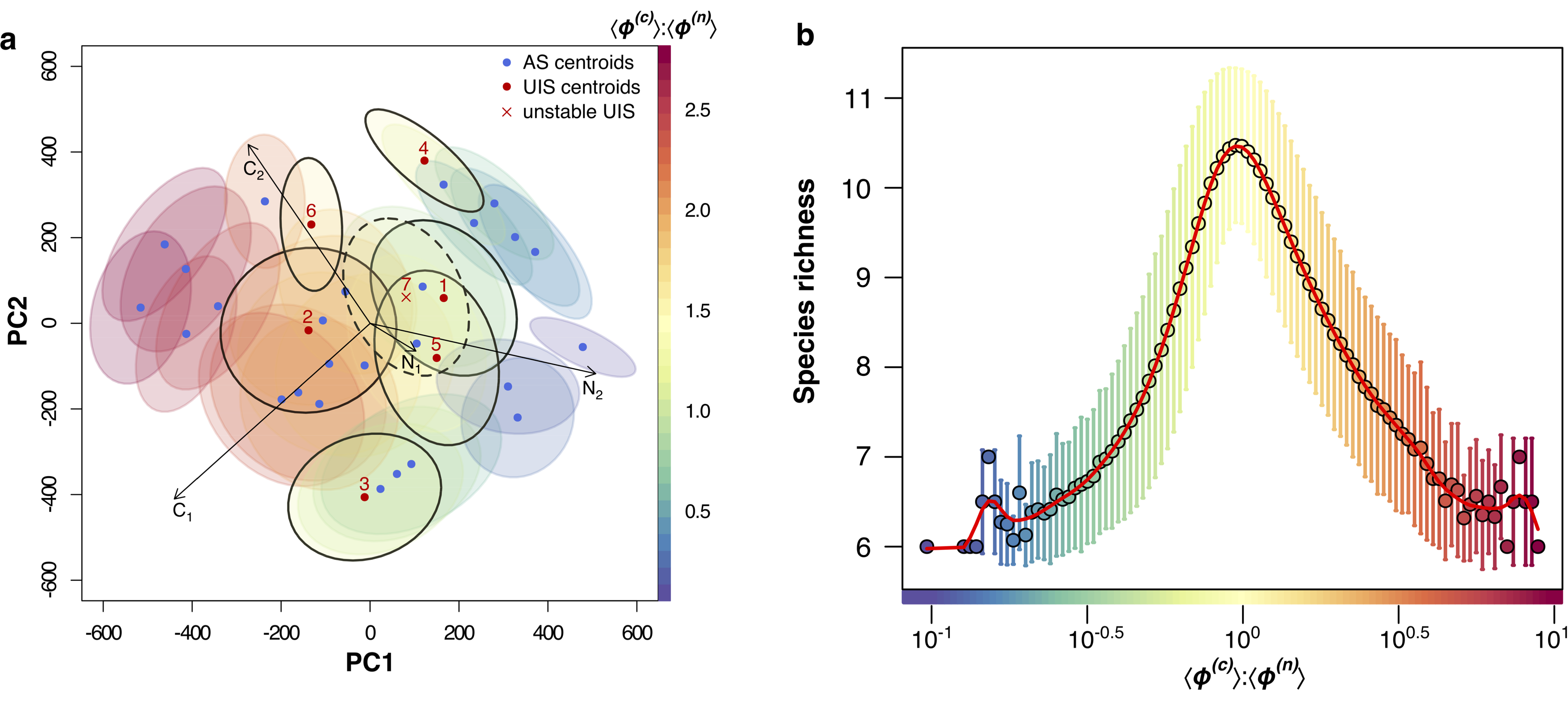}}
%\centerline{\includegraphics[width=\linewidth]{C6_N6_overlap_AllStates_network_curved.pdf}}
\caption{{\bf Feasible volume of states 
%in a 2Cx2Nx4S model 
%and networks of overlaps 
% and boundaries  
%between them.
and average species richness.} (\textbf{a}) Principle Component 
Analysis (PCA) of the 4-dimensional vectors of average nutrient fluxes 
feasible for the 33 allowed states in our 2Cx2Nx4S example (see $\lambda$ and yields in Supplementary Tables \ref{tab:Lambdas_L2}, \ref{tab:Yields_L2}). 
%in a microbial community grown on $K=2$ 
%carbon and $M=2$ nitrogen sources and $S=4$ species. 
% Each dot represents an allowed state with red symbols corresponding to 7 uninvadable states, while blue symbols to the remaining 26 invadable allowed states. The cross marks the uninvadable state 7 that is dynamically unstable - hence generally not observed in community dynamics. 
The ellipse around each dot approximates 
the boundary encompassing 25\% of feasible nutrient influxes for each of the 33 allowed state. 
The arrows in the middle correspond to the direction of changes of
%changes in each of the 
4 nutrient fluxes in PCA 
coordinates. 
(\textbf{b}) Dependence of \textit{species richness} on the averaged %C:N 
$\phi^{(c)}$: $\phi^{(n)}$ flux ratios for our 6Cx6Nx36S example (See Supplementary Tables \ref{tab:Lambdas_c_L6}, \ref{tab:Lambdas_n_L6}, \ref{tab:Yields_c_L6}, \ref{tab:Yields_n_L6} for the values of $\lambda$ and yields used for this example). 
%For this example we were able to classify 1211 uninvadable states.
%\textcolor{red}{SERGEI: WHY DO WE NEED TO DESCRIBE IT HERE? BY NOW PEOPLE SHOULD KNOW HOW MANY STABLE STATES WE HAVE IN 6X6 SYSTEM. IF THIS IS NOT NEEDED YOU CAN COPY-PASTE TO WHERE IT NEEDS TO BE DESCRIBED. 1211 uninvadable states out of which 1058 were dynamically stable and 137 dynamically unstable and 16 had volumes that are too small to distinguish between these two options.} \textcolor{blue}{PARTH: I thought when we are talking about species richness, it is important to mention that we use only the dynamically stable UIS. Also the fact that we can ignore the no. of unstable UIS and compute the species richness with all the 1211 UIS is also contingent upon the fact that we classify the stable and unstable ones.}
Each point represents the average number of surviving species in all the stable uninvadable states which are feasible in an interval of nutrient influxes with average %C:N 
$\phi^{(c)}$: $\phi^{(n)}$ ratio, partitioned into 100 bins.
%Each interval of nutrient flux is chosen such that it partitions the flux space into bins  characterized by their average C:N ratios average richness values for every C:N ratio interval (100 intervals in total). 
Error bars show standard deviation of the species richness around each interval and the solid red curve is a trend line. Colors on both plots corresponds to similar %C:N 
$\phi^{(c)}$: $\phi^{(n)}$ ratio.
% (B) Feasible flux regions of 
% six uninvadable and dynamically-stable states S1-S6
% (blue, ...) visualized in 3D PCA coordinates. Note the 
% boundaries and overlaps between states as well as the range 
% of structural stabilities (volumes) of different states ranging 
% from the largest three states S?? to the smaller volumes of 
% S??.
%
%(\textbf{b}) The network of overlaps (bistability)between feasible influx regions of 6 stable uninvadable states. Numbers near the links show the overlap volume normalized to the overall volume of a particular state.
%
%(C) The network of all boundaries between feasible regions of the 6 stable uninvadable states. Numbers near the links show the boundary size normalized to the total boundary size of a particular state.
%
%(\textbf{c}) The network of all overlaps in $L=6$ case normalized by the smaller volume of two overlapping states). The size of a node reflects its degree (the number of other states it overlaps with). %
% Stable states are green and unstable states are red.
}
\label{fig2n} 
\end{figure*}
In Fig. \ref{fig2n}A we show feasible volumes of different 
allowed states in a model with $K=M=2$ 
nutrients (i.e., two carbon and two nitrogen sources) 
and $S=4$ species with exactly one species for each 
pair of these nutrients. Hereafter we denote this example as the 2Cx2Nx4S and use similar nomenclature for other examples.
For a particular choice of $\lambda$-values 
used in our numerical simulations of the 2Cx2Nx4S system
(see Supplementary Tables \ref{tab:Lambdas_L2}, \ref{tab:Yields_L2} for values of $\lambda$ and yields) 
we get 33 allowed states (plus 1 empty state without microbes) 
that do not violate the two rules of competitive 
exclusion. It is well below $3^S=3^4=81$ candidate states 
possible before competitive exclusion rules were imposed. 
%%%%% repeat of page 3 in model description
%Here $3$ reflects the fact that each microbe could be absent from 
%a state or, when present, it could be either limited by its carbon
%or its nitrogen source. 
We labeled these allowed states in such a way that the 
first seven of them (S1-S7) are also uninvadable. 
Fig. \ref{fig2n}A visualizes the feasible region of 
each state as an ellipse, with its
center positioned at the center of mass of feasible influxes, 
and its area selected to cover 25\% of feasible influxes. 
To better separate the feasibility regions of these states we performed principle component analysis using center of mass for each state.
%Fig. S\ref{fig2n_supp} 
The website given in Supplementary Information 
attempts a more realistic visualization 
of the 6 uninvadable states, which are also dynamically stable
(see section on dynamical stability below) as linearly confined 
regions in 
%All fluxes wheres each of these states is feasible are visualized 
%as different colors and are projected onto 
the 3-dimensional (out of the total of 4 dimensions)
PCA space. As one can see from this figure, structural stabilities 
(feasible volumes) of different states vary over a broad range 
with  S1, S2, and S3 being the most structurally stable with large feasible volumes, 
while S4, S5, and S6 are considerably smaller (they are the narrow stripes, 
sandwiched between the three largest states). 

%\textcolor{blue}{
We further explored a more complex model with $K=M=6$ nutrients and %6x6=
36 %specialist
species, i.e., 6Cx6Nx36S (see Supplementary Tables \ref{tab:Lambdas_c_L6}, \ref{tab:Lambdas_n_L6}, \ref{tab:Yields_c_L6}, \ref{tab:Yields_n_L6} for the values of $\lambda$ and yields). For this choice of $\lambda$ we obtained a total of 134,129,346 allowed states out of which 1211 were uninvadable. %In the following section we discuss the process of classification of the states based upon their dynamic stability.
Using Monte-Carlo simulations over a region of nutrient influx space (see Methods Section for details) we obtained the feasible volumes of each of the uninvadable states. Utilizing this data we explored how environmental parameters (in our case nutrient influxes) affect the number of surviving species. In Fig. \ref{fig2n}B we show that species richness peaks when the fluxes of the two available nutrients are most balanced and it falls with increasing disproportionate between the two.

\subsection*{Dynamical stability of steady states}
\begin{figure*}
\centerline{\includegraphics[width=0.9\textwidth]{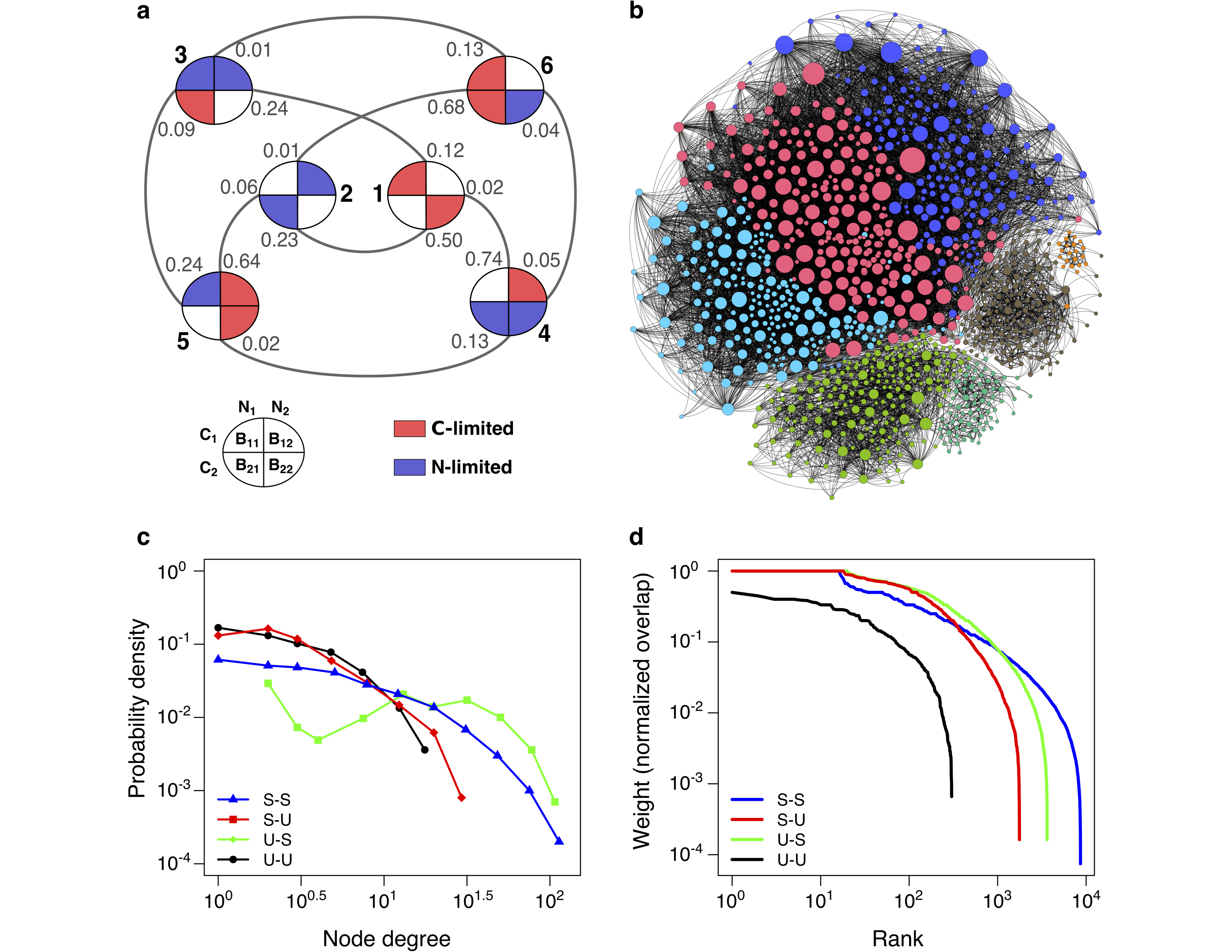}}
\caption{{\bf Networks of overlaps 
between feasible volumes of uninvadable states} (\textbf{a}) The network of overlaps among 6 stable uninvadable states in the 2Cx2Nx4S example (same as in Fig. \ref{fig2n}A). Each state is represented in its matrix configuration (see Fig. \ref{fig2n}A). A link between 2 states represents an overlap of their feasible volumes. The fraction of the volume of each 
state that overlaps with another state is the weight of a directed edge indicated on the link connecting these states. Hence, the sum of all these fractions denoted on the links near a state indicates the total fraction of its volume that overlaps with all the other uninvadable state.
(\textbf{b}) The network of overlaps of the 1195 uninvadable states (both stable and unstable) 
in our 6Cx6Nx36S example (same as in Fig. \ref{fig1n}B). Nodes and links 
%represent the uninvadable states and overlapping volumes. 
%Strength of a link denotes the overlapping volume normalized by the volume of the smaller of the 2 states. 
are the same as in panel a).
Size of a node reflects its degree (i.e., the total number of states it overlaps with). The color of each node corresponds to the modularity cluster/class (8 in total) it belongs to (see Methods Section for details).
%\textcolor{red}{We identified densely interconnected clusters in the network using the algorithm implemented as a part of Gephi 0.9.2 software package used to visualize this network [CITE ALGORITHM. LOOK IN GEPHI FOR CITATION]. VERONIKA: THESE DETAILS BELONG TO METHODS}
%Each state is colored to indicate the cluster (out of 8) it belongs to.
(\textbf{c}) Degree distribution of the network in panel (b) with different colors representing 
different types of degrees: the number of stable states neighboring each of the stable nodes (blue triangles), the number of unstable states neighboring each of the stable nodes (red squares), the number of stable states neighboring each of the unstable nodes (green diamonds), the number of unstable states neighboring each of the unstable nodes (black circles)
%\textcolor{red}{Too complicated to describe} 
%Different curves represent overlaps between two different types of nodes - stable (S) and unstable (U).  
(\textbf{d}) 
%\textcolor{red}{Too complicated to describe} 
Rank ordered distribution of weights of the network from panel (b). The weights normalized as in (a) represent normalized overlaps of states. Different lines represent overlaps between pairs of two different types of nodes - stable (S) and unstable (U) with the same colors/labels as in panel (c). 
}
\label{fig3n} 
\end{figure*}
So far we avoided an important question of dynamical stability 
of steady states in our model.
We tested the stability of all allowed states in our 2Cx2Nx4S 
example by performing computer simulations in which each of the 33 allowed states was subjected
to small perturbations of all microbial populations present in a given state (see Methods Section for details). 
Naturally, an invadable state will be dynamically unstable against introducing small populations of successful invaders, which does 
not count as its dynamical instability.  In our example only 
one of the states (S7) was 
found to be dynamically unstable. Interestingly, it was unstable for all combinations of nutrient influxes we tested, while the remaining 32 allowed states were always dynamically stable.
% . This unstable (and also uninvadable) state
% shares a  part of its feasible influx range  with each of the remaining 
% six uninvadable states.  Furthermore, S7 was dynamically unstable while 
% the rest of the states were stable for all values of nutrient influxes.
This property of our model is different from, e.g. MacArthur model, where stability of a state generally depends on nutrient influxes and concentrations in the environment \cite{posfai2017metabolic}. For another variant of the MacArthur model all steady states were found to be stable \cite{butler2018stability}.

%
% Trying ecosystems with more carbon and nitrogen sources
% and more microbes we did not do a complete cartography of
% those but were able to check the dynamic stability of individual 
% uninvadable and allowed states. 

%This sentence should be later? It seems that it is inferred data, not real simulations as described for 2x2.

%out of $1211$ uninvadable states. 
%... are stable and ... are unstable) %\textcolor{blue}{
%The final count is that the number of stable states is $1022+36=1058$ and the number of unstable ones is $173-36=137$. Feasible volumes of 
%
% (\textcolor{red}{but don't we need to emphasize that we find feasible UIS which are dynamically unstable. This distinguishes us from James O'Dwyer paper and this leads to finding up to 9 coexisting UIS for L=6 case.}) 
% In fact, there are ecosystems in
% which all steady states, both allowed and uninvadable, are 
% dynamically stable. This numerical result is in agreement with 
% our mathematical estimates. 
%(see SI, "Tilman plane in higher dimensions"). 

%To carry out this classification we did not attempt to \textcolor{red}{quantify? (YF)} the response of each state to small perturbations for each of the combination of influxes. This would have been computationally prohibitively expensive. Instead, we computationally inferred dynamic stability using the following arguments:

%\textcolor{blue}{
For our 6Cx6Nx36S example the number of allowed states is too large, hence we wish to classify only the 1211 uninvadable states based upon its dynamical stability. Further, to carry out this classification we use a computationally inexpensive algorithm (as against the direct test of stability used for our 2Cx2Nx4S example) as described below.
%}
Since for each set of influxes there must be {\it at least one dynamically stable uninvadable state} providing 
the endpoint of model's dynamics, an unstable steady state 
can never be alone in the influx space: wherever feasible, 
it is bound to decay into one of the stable uninvadable 
states feasible for these environmental parameters.  
% Thus, in our 
% multidimensional map of uninvadable steady states, an unstable 
% states cannot be found alone.  
That provides an intriguing way to use influx maps to infer stabilities of states. Indeed, if for a given state one could find a flux region in which no other uninvadable states are feasible - then this state has to be automatically dynamically stable. In our numerical simulations we found that at each specific influx point $V$ stable uninvadable states are always accompanied by $V-1$ dynamically unstable ones. 
%\textcolor{blue}{
Application of this indirect algorithm on our 6Cx6Nx36S example revealed $137$ dynamically unstable and $1058$ dynamically stable states out of the 1211 uninvadable states. This method could not infer the stability of 16 uninvadable states with very small feasible volumes.
%}
%\textcolor{red}{ 
Using Monte-Carlo simulations we computed the feasible volume of each uninvadable state for our 6Cx6Nx36S example. We find that the distribution of volumes of all these states is log-normal (for the distribution of log of volume parameters are: $\mu=-8.87 \pm 0.06$, $\sigma=2.08 \pm 0.04$). We also find that the there is no significant difference between distributions of volumes for the stable and unstable uninvadable states (two-sample Kolmogorov-Smirnov test: $p-value = 0.94$).
%}
%Describe the distribution of state volumes, lognormal:\\
%Distribution:    Normal\\
%Log likelihood:  -2571.07\\
%Domain:          -Inf < y < Inf\\
%Mean:            -8.86555\\
%Variance:        4.33192\\
%Parameter  Estimate  Std. Err.\\
%mu         -8.86555  0.0602083\\
%sigma       2.08133  0.0426004\\
%for Stable and unstable there is no significant difference in distributions 
%\textcolor{red}{Move boundaries discussion to the next subsection?}
%\textcolor{red}{VERONIKA: Should we say smth about justification of out V, V-1 rule by lower-dimensional simulations? (L=4?)}
%%%%%%%%% move to discussion? or SI? we don't need it there...
% This pattern suggests the existence of a special type of non-convex 
% %one-dimensional version of the 
% Lyapunov function bounded from above 
% that always has $V$ maxima (corresponding to stable states) 
% and $V-1$ minima or other unstable critical points other than 
% maxima (corresponding to unstable ones). 
% In 1D any smooth function bounded from 
% above and reaching $-\infty$ at $x=\pm \infty$ always 
% has $V$ maxima and $V-1$ minima. In higher dimensions
% ($K+M$ in our case) this property imposes additional constraints 
% on indices on critical points other than maxima 
% which are dictated by the Morse theory \cite{milnor1963morse}.
% We leave the search for this Lyapunov function for future studies.
%%%%%%%%%%%%%%%%%%%%%
%MAYBE DRAGONGFLY FIGURE HERE.
\subsection*{Multistability of microbial ecosystem, alternative stable states for the same environmental parameters}
\begin{figure*}
%\centerline{\includegraphics[width=\linewidth]{fig4n.eps}}
\centerline{\includegraphics[width=1.0\textwidth]{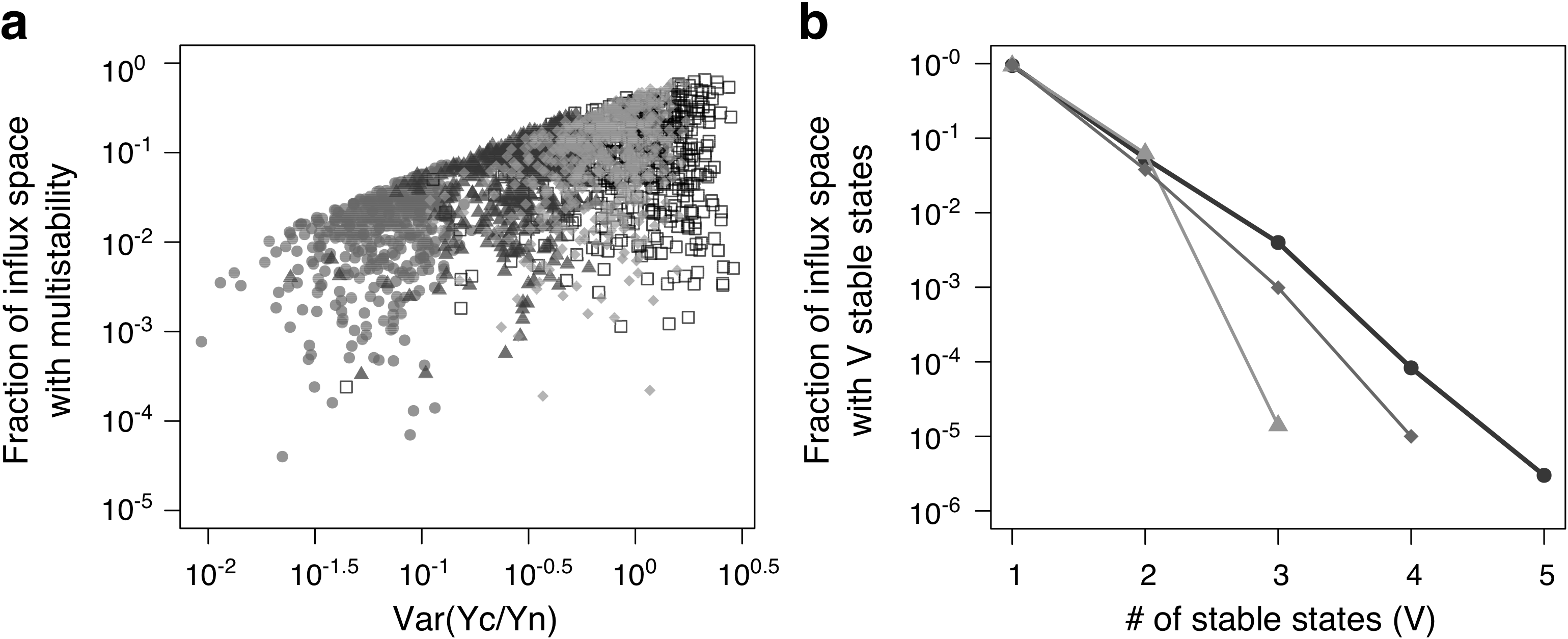}}
\caption{{\bf Statistics of multistability for different combinations of yields.
%and feasible volumes of uninvadable states for larger number of resources %($L=6$ carbon and $L=6$ nitrogen sources). %(\textcolor{red}{Panel A is for L=2 case !!})
} 
(\textbf{a}) Fraction of the nutrient influx space with multistability for different set of yields in the 2Cx2Nx4S example ($\lambda$ are the same as in Fig. \ref{fig2n}A). Y-coordinate of each point represents the fraction of the nutrient influx space (out of $10^5$ influx points we sampled) where multistability was observed for a given set of yields (while keeping competitive abilities of all species fixed). Its x-coordinate represents the variance between the ratios of yields ($Y^{(c)}/Y^{(n)}$) across all species. 
In total, 4000 numerical simulations were done, each with a different set of yields. The 4 different types of points indicate the 4 different (uniform) distributions (with different variance) from which the yields of different species were chosen (see Methods Section for details). %1000 sets of yields were chosen per distribution. 
%Out of the 4000 data points we plot only 2069 where a non-trivial fraction of multistability was observed. In the remaining roughly half of yield combinations, the system had a unique and stable uninvadable state for any set of influxes.
%(\textcolor{red}{I am confused about the x and y axis - VD: yep, it's a bit hard to digest, but I think that Variance is the simplest explanation that we can have (it allows us to drop this scary log(sd(log(X)))). I'm not sure about y axis - now I'm using an absolute number, but it might be better to have a fraction - can you tell me the total number of points in these simulations? Is it 1M?})
(\textbf{b}) For 3 different sets of yields we plot the fraction of nutrient influx space with V stable uninvadable states for the 6Cx6Nx36S model ($\lambda$ are the same as in Fig. \ref{fig2n}B). For each set of yields we explored the fraction of the nutrient influx space (out of $10^6$ influx points) that contains V=1,2,3,4,5,.. stable uninvadable states. The figure shows that a difference in the yields of the species results in difference in multistability, although the values of $\lambda$ (and hence the set of uninvadable states) are the same in all the three cases. Bold black curve corresponds to the example shown in Fig. \ref{fig3n}B. %\textcolor{red}{Change label of y-axis to "Fraction of nutrient influx space with multistability"}
%(\textbf{c}) Species richness depending on C:N flux ratio. Points represent average richness values for every C:N ratio interval (100 intervals in total). Error bars show standard deviation for the richness value around a particular value of C:N ratio. Red curve shows trend line. 
%(\textbf{d}) Prevalence distribution of species (U-shaped curve) for our numerical simulation and experimental data.
%Dragonfly plot (?) or diversity vs stoichiometry (?)...
% (B) Distribution of sizes of state boundaries for uninvadable states for case of $L=9$ carbon and $L=9$ nitrogen sources. 
% (C) Percentage of overall volume occupied by multi-stable cases depending on $L$.
% (D) Distribution of volumes occupied by $V$-stable states plotted for different 
% values of stoichiometric spread.
}
\label{fig4n}
\end{figure*}
Our model is generally capable of bistability or even multistability when two or more uninvadable states are feasible for the same environmental conditions given by nutrient influxes. This happens when the feasibility regions of multiple stable uninvadable states overlap with each other. For influxes in the intersection area,  all of the overlapping states are feasible and, since each of them is uninvadable, they cannot transition to each other through addition of other species from the pool. 
Fig. \ref{fig3n}A shows the network of such overlaps between the 6 stable uninvadable states in our 2Cx2Nx4S example. The fractional number shown on each edge represents the fraction of the feasible volume of each state, over which it overlaps with its neighboring state. 
%\textcolor{red}{Unlike edge weights in \textcolor{red}{SFig. \ref{fig2n}C}, these numbers are not normalized to 1.}
%%
% since for each state we have some flux region where only this state is feasible.
%%

The network of overlaps of the 1030 uninvadable states (both stable and unstable) that have at least one neighbor 
in our 6Cx6Nx36S example (same as in Fig. \ref{fig2n}B). Nodes and links are the same as in panel a).  
Size of a node reflects its degree (i.e., the total number of states it overlaps with).

%\textcolor{blue}{
In Fig. \ref{fig3n}B we plot the network of overlaps between feasible volumes of the 1195 uninvadable states in our 6Cx6Nx36S example. 
%(we leave out the 16 states whose dynamic stability remained ambiguous). 
%shows how complicated this network could be even in a slightly higher-dimensional case of 6Cx6Nx36S. 
We performed the standard modularity analysis on this network (see Methods Section for details) and obtained 8 clusters indicating that the states are not randomly distributed but clustered in the flux space.
%}
%\textcolor{blue}{\textbf{For Sergei: Mention what we learn from Fig. \ref{fig3n}C,D} !!!!!}.
\textcolor{black}{We distinguished two types of nodes in this network - dynamically stable (S) and unstable (U) ones. 
%Then, the edges connecting stable nodes represent all multistable cases Fig. \ref{fig3n}C. 
%What else we can say about S-U, U-S links? The existence of U-U links indicates the presence of multistability characterized by $V>2$ alternative stable states.
%(as we mentioned above in V, V-1 rule) 
%Then, in general it is possible to look for some patterns in this network like S-U-S, S-U-S-U-S, etc. 
Thus in Fig. \ref{fig3n}C 
we distinguish between four different types of degrees: the number of stable states neighboring each of the stable nodes (blue triangles), the number of unstable states neighboring each of the stable nodes (red squares), the number of stable states neighboring each of the unstable nodes (green diamonds), the number of unstable states neighboring each of the unstable nodes (black circles). One can see that all four type of degrees vary over a broad range with the largest degrees
of the four types listed above equal to 164, 41, 115, and 21 correspondingly. Thus the biggest hub among stable states is connecting to around 15\% of other stable states.
%\textcolor{red}{Too complicated to describe} 
%Different curves represent overlaps between two different types of nodes - stable (S) and unstable (U).  
%\textcolor{red}{Too complicated to describe} 
Rank-ordered distribution (Zipf plot) of edge weights 
of the network from Fig. \ref{fig3n}B are 
are shown in Fig \ref{fig3n}D.  
Different colors and symbols 
represent overlaps between different types of 
nodes (S-S, S-U, U-S, and U-U). 
Here the x-axis shows the rank of the weight of a certain type 
(1 being the largest) and the y-axis shows the value of this weight.
%Edge types (except for the U-U type) have some weights equal to 1 corresponding 
%to a states fully contained within another state. For S-states this is an artifact.
Figure \ref{fig3n}D shows that different types of edges 
have different probability distributions of weights (all of them broad).
%In this case rank ordered distribution of them shows that there exists 36(?) unstable states which has their feasibility volumes completely covered by some stable state. 
%Also, one can see that the curve representing links between stable states has an unexpected step(?), which indicates that 1M points for simulation was not enough and we missed some flux points where stable state was alone.
}

%\textcolor{blue}{
We can define \textit{coexisting} states as a set of states which are \textit{simultaneously} feasible in a finite region in the nutrient influx space. This determines the multistability of the system.
%} 
The number of such coexisting stable uninvadable states never goes above 2 in our 2Cx2Nx4S example with $\lambda$ and yields as in Fig. \ref{fig3n}A. However, for a larger number of resources we do observe multistability of more than 2 stable uninvadable states. 
%\textcolor{red}{SERGEI: DON'T DISCUSS THREE CHOICES OF YIELDS HERE.} 
E.g., for our 6Cx6Nx36S example (with $\lambda$ and yields same as chosen for Fig \ref{fig3n}B) we notice up to 5 coexisting stable uninvadable states (see bold black line in Fig \ref{fig4n}B). It is still a far cry from an exponentially large number of all uninvadable states possible for different combinations of nutrient influxes. Also notice that the volume of the influx space occupied by $V$-stable states falls of with $V$ faster than exponentially.

We further explored the factors that determine this multistability. Like in its simpler special limit studied in Ref. \cite{tilman1982resource}, the multistability in our model is only possible if individual microbial species have different C:N stoichiometry, quantified by the ratio of their carbon and nitrogen yields. Our numerical simulations (see Methods Section for details) strongly support that when all species have the same stoichiometry $S^{(C:N)}_{\alpha}=Y^{(c)}_{\alpha}/Y^{(n)}_{\alpha}$, for every set of nutrient fluxes there is a unique uninvadable state. The same is true in the MacArthur model provided that the ratio of yields of different nutrients is the same for all microbes \cite{maslov2018unpublished}. %\textcolor{red}{Sergei unpublished - cite this}.
%in which microbes co-utilize perfectly substitutable nutrients, once again  
Simulations on the 2Cx2Nx4S example (see Fig. \ref{fig4n}A) supports our claim that the more narrow is the spread of $Y^{(c)}_{\alpha}/Y^{(n)}_{\alpha}$
oin our pool of species, the smaller is the average volume of multistable states realized among all possible combinations of nutrient fluxes. Interestingly, in our experiments with varying stoichiometry roughly a half of yield combinations resulted in multistability (i.e., out of 4000 numerical experiments we performed, only for 2069 a non-zero fraction of multistability was observed).

We also simulated our 6Cx6Nx36S system (with the same values of $\lambda$) for many different sets of yields. Fig \ref{fig3n}B shows the fraction of nutrient influx space that permits the coexistence of $V$-stable states for 3 different sets of yields, one of which (as mentioned above) permits multistability up to 5. The other two sets of yields show a smaller multistability of 3 and 4 uninvadable stable states.
%The three curves correspond to simulations performed with 3 different choices of set of yields for the species.
%It should be noted that the uninvadable states are identical for all the three set of simulations because the set of $\lambda$ is common. It is the set of yields that distinguishes the three cases. (\textcolor{red}{refer Suppl Note for values of lambda and yields}). 
%Notice the two other curves in Fig 4B. with maximum multistability of stable uninvadable states up to 3 and 4. These simulations were performed for the same set of $\lambda$ but different yields.
 
%\textcolor{red}{Our simulation strongly support this hypothesis, showing that for for any flux point there is a unique state.}
%representing the unique endpoint of any 
%colonization sequence. 

%When stoichiometry is the same, by measuring each nutrient concentration in units of its yield, without loss of generality all yields can be set to $1$. 

\subsection*{Colonization dynamics}
\begin{figure*}
%\centerline{\includegraphics[width=\linewidth]{fig3n.eps}}
\centerline{\includegraphics[width=\textwidth]{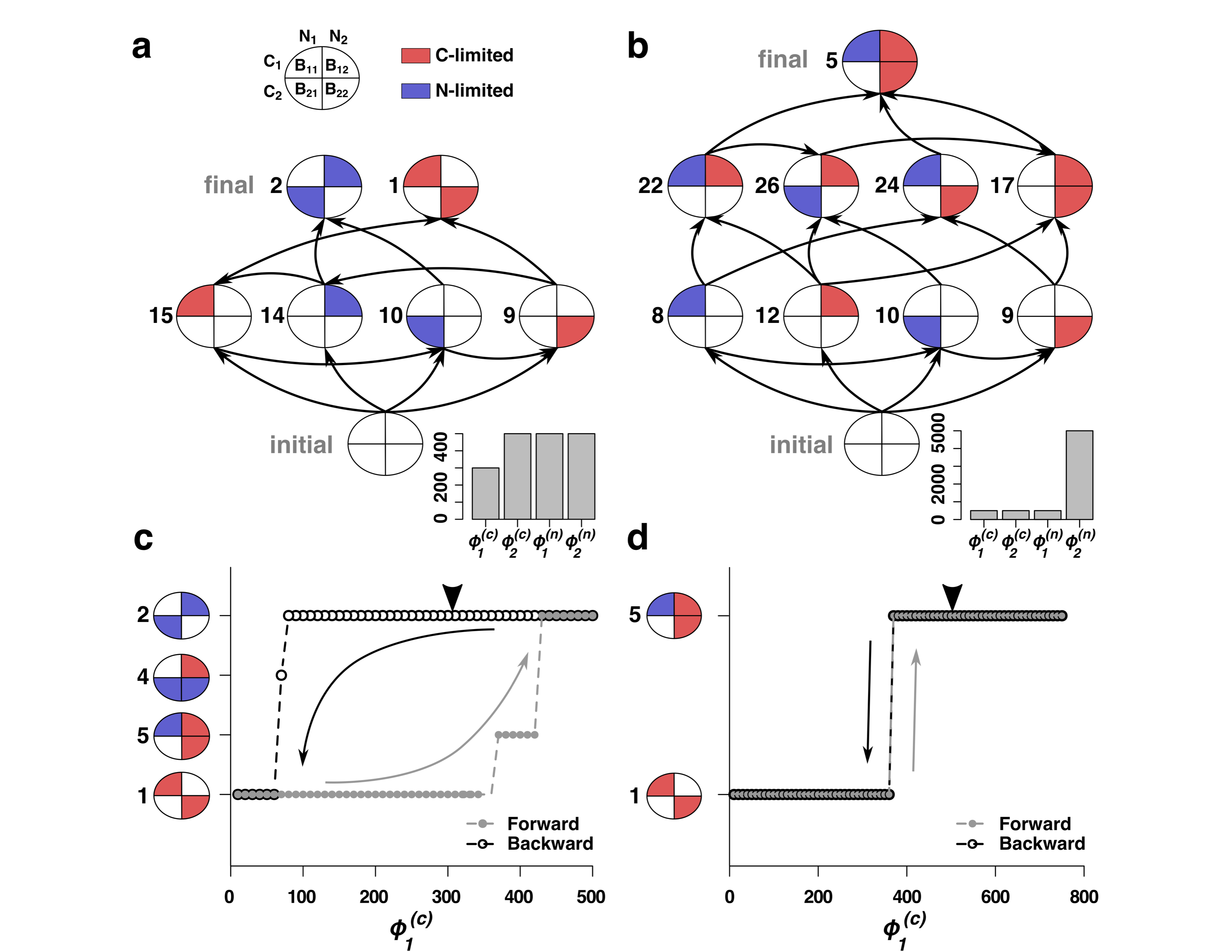}}

\caption{{\bf State transitions triggered by colonization dynamics and changing environment in 2Cx2Nx4S example.} 
(\textbf{a}, \textbf{b}): Transitions triggered by colonization dynamics. We show the graph of all possible state transitions for our 2Cx2Nx4S example (same as in Fig. \ref{fig2n}A) for two different sets of environmental parameters %(i.e., nutrient fluxes) 
shown as barplot in \textbf{a} and \textbf{b} (see Methods Section for details). 
%Dilution rate is same in both cases $\delta=1$. 
Starting from an empty set (i.e., no bacterial species) we randomly select and introduce species from the pool (one at a time) and wait for the system to settle into a dynamical steady state, and this process is repeated until the system reaches an uninvadable state. %This colonization dynamics is repeated for a large number of random order of introduction of species from the pool to obtain all possible transitions (shown as black arrows) between the allowed states at the given nutrient influx (see Methods Section for a detailed description). 
%It can be seen that for environment in panel \textbf{a} we obtained 6 feasible and dynamically stable allowed states connected by a total of 12 transition trajectories permitting 2 alternative uninvadable states (S1, S2) as the terminal end of the colonization dynamics. For the environmental parameters in \textbf{b} we observed 10 feasible stable states connected by 19 transitional trajectories leading to a single uninvadable state (S5).
(\textbf{c}, \textbf{d}): Transitions triggered by changing environment. Here we show the transition between states %for the above 2Cx2Nx4S system 
when one of the parameters $\phi^{(c)}_1$ is modified from its original value (pointed black arrow) in panel \textbf{a} and panel \textbf{b} (shown in panel \textbf{c} and panel \textbf{d} respectively). (\textbf{c}): The environment is set to a low $\phi^{(c)}_1= 10$ units, with other influxes same as in panel \textbf{a}) and on executing colonization dynamics the system settles to the uninvadable state S1. 
%The system is initially prepared at a low $\phi^{(c)}$ (= 10 units, with other influxes same as in panel \textbf{a}); in which case the colonization dynamics drives the system to the uninvadable state S1. 
We next increment $\phi^{(c)}_1$ by a unit of 10 and apply colonization dynamics on the current state of the system and iterate this process until $\phi^{(c)}_1 = 500$.
%$\phi^{(c)}$ was then increased by units of 10 and colonization dynamics was executed on the current state of the system. This process of incrementally increasing $\phi^{(c)}$ and performing the colonization dynamics on the system (obtained at the end of the previous step) was done until $\phi^{(c)} = 500$. 
Grey filled dots indicate the states that the system experiences at the end of each colonization dynamics on its journey from $\phi^{(c)}_1 = 10$ to $\phi^{(c)}_1=500$. One can see that at $\phi^{(c)}_1=370$ the system transitions to the uninvadable state S5 and at $\phi^{(c)}_1 = 430$ it jumps to state S2 and stays there up till $\phi^{(c)}_1=500$. Starting from this state (S2) the above process was repeated but with decreasing $\phi^{(c)}_1$. Black empty circles highlight the states observed by the system until $\phi^{(c)}_1=10$. This completes a full cycle of changing $\phi^{(c)}_1=10$. It can be seen that in the transition from state S2 to state S1 occurs at a much lower value of $\phi^{(c)}_1$ compared to the transition from S1 to S2 when $\phi^{(c)}_1$ was increased, hence displaying the phenomenon of hysteresis. %At $\phi^{(c)}_1=300$ (pointed black arrow) 2 states are feasible, S1 and S2. This is exactly the environment provided in panel \textbf{a} which displayed S1 and S2 as the 2 alternative states. 
(\textbf{d}): %As done in panel \textbf{c}, we performed a similar task of exploring the transitions of states when 
Similarly to panel \textbf{c} we varied $\phi^{(c)}_1$ from a low to high value, keeping all the other nutrient influxes same as in panel \textbf{b}. In contrast to \textbf{c}, we do not observe any hysteretic transitions when $\phi^{(c)}_1$ is changed in the forward and reverse directions. The black arrow points at the environmental conditions from panel \textbf{b}.
}
\label{fig5n}
\end{figure*}
In the course of sequential colonization by species 
selected from our pool (see Methods Section for details), microbial ecosystem 
goes through a series of transitions between several 
feasible allowed states culminating in one of the uninvadable states. 
The set of all colonization trajectories 
can be visualized as a directed graph with 
edges representing transitions caused by 
addition of species from the pool to each of the state. 
Both the set of nodes (selected among all allowed states) 
as well as the set of possible transitions between these 
nodes are determined by the environmental variables 
(nutrient influxes). 

In Fig. \ref{fig5n} we 
show two examples of state transition graphs in our 2Cx2Nx4S case. For one set of environmental parameters 
%%% we already have them in figure caption
% $\phi^{(c)}_1=300, \phi^{(c)}_2=500, \phi^{(n)}_1=500, \phi^{(n)}_2=500$ 
shown in Fig. \ref{fig5n}A, our model 
has 6 feasible and dynamically stable allowed states (plus 1 empty state)
connected by 12 transitions 
triggered by species addition. 
For the same %model
species pool, changing nutrient fluxes 
% to $\phi^{(c)}_1=500, \phi^{(c)}_2=500, \phi^{(n)}_1=500, \phi^{(n)}_2=5000$
(Fig. \ref{fig5n}B) results in a different
set of 10 feasible allowed states 
connected by 19 transitions. 
The uninvadable states are visible as terminal 
ends of directed paths in the top layer of Figs. \ref{fig5n}A,B.
While the fluxes used in  Fig. \ref{fig5n}B allow for 
only one uninvadable state, S5, 
a different set of fluxes used in Fig. \ref{fig5n}A permit 
for two alternative uninvadable states, S1 and S2. 
Which of these two states is realized depends on the 
order in which species were added to the system. 

As shown in Fig. \ref{fig5n}C,D 
transitions between these and other alternative states 
can also be triggered by changing nutrient fluxes. Whenever multistability is present, transitions happen in a hysteretic manner. In Fig. \ref{fig5n}C we show  
a full cycle of changing fluxes, first up (grey line) and then down (black line) 
in, which results in a system going through a series of discontinuous transitions between states and ending up in a different uninvadable state than that it started from. Where as in contrast, in Fig. \ref{fig5n}D there is a unique uninvadable stable state feasible at a certain flux point.
This matches several hallmark properties of alternative 
stable states defined in Ref. \cite{Schroder2005} 
as having ``discontinuity in the response 
to an environmental driving parameter'', 
lack of recovery after a perturbation 
(hysteresis), and ``divergence due to different initial 
conditions''.

%MAYBE HAVE A SECTION WITH SOME LARGE L RESULTS HERE.
%EXAMPLES ARE BELOW.
%
%\textcolor{red}{ Results WC - 3665 (without headlines)}
%
\section*{Discussion}
Ever since Robert May's provocative question ``Will a large complex 
system be stable?'' \cite{may1972will} 
the focus of many theoretical ecology studies has been on dynamical 
stability of steady states in large ecosystems. Unlike the classic 
MacArthur model \cite{butler2018stability}, 
but similar to the original Tilman model \cite{tilman1982resource}, 
our model is characterized by 
a mixture of dynamically stable and unstable states. 
Based on a small sample of examples that we analyzed in detail, 
we found that the stable states in our model generally outnumber the unstable ones. 
For example, in the 2Cx2Nx4S example used above
only one state out of 33 is dynamically unstable, while for the 6Cx6Nx36S example used in 
Fig 2C and Fig 4, we found only $137$ unstable states 
out of $1195$ uninvadable states that we were able to 
classify using our methods. Another interesting observation 
is that for randomly selected carbon and nitrogen yields of individual species 
(defining its C:N stoichiometry) 
with probability around 50\% one ends up with an ecosystem 
%for approximately half of the sets of yields defining 
%stoichiometry of each species in our pool, 
lacking both unstable states as well as alternative stable states.

%... are stable and ... are unstable) %\textcolor{blue}{
%The final count is that the number of stable states is $1022+36=1058$ and the number of unstable ones is $173-36=137$. Feasible volumes of 
% (feasible volumes of $16$ states were too small to 
% categorize them as either stable or unstable using our methods).

In fact, the existence of dynamically unstable states in our 
model always goes hand in hand with multistability.
%, 
%where several alternative stable states are possible for the same 
%environmental parameters. 
%
Indeed, in the simplest case of a bistable ecosystem considered in Ref. 
\cite{may1977thresholds}, a single dynamically unstable steady state always separates 
two stable states of an ecosystem. Depending on perturbation 
this unstable state would collapse to either one of the two alternative stable states
realized for the same environmental parameters. 
Interestingly, in our model we always found $V-1$ unstable states coexisting with 
$V$ alternative stable states for the same environmental parameters. 
While, this result is natural for dynamics maximizing a 
{\it one-dimensional} Lyapunov function 
where $V$ maxima (corresponding stable states) 
are always separated by $V-1$ minima (unstable stable states), 
we currently do not understand why this rule seems to apply to 
our {\it multi-dimensional} system. 
Indeed, in 1D any smooth function bounded from 
above and reaching $-\infty$ at $x=\pm \infty$ always 
has $V$ maxima and $V-1$ minima. In higher dimensions
this property imposes additional constraints 
on indices on critical points other than maxima 
which are dictated by the Morse theory \cite{milnor1963morse}.
We leave the search for this Lyapunov function for future studies.

As we demonstrated above, both unstable states 
and multistability are possible only if different microbes have different 
C:N stoichiometric ratios. The same is true for models in which 
microbes co-utilize multiple substitutable (say carbon) nutrients. 
Indeed, a convex Lyapunov function defined in 
Ref. \cite{macarthur1970species,case1979global,chesson1990macarthur} guarantees that 
for any set of nutrient influxes there 
exists exactly one stable equilibrium.
While the standard MacArthur model has no multistability and, 
as proven in Ref. \cite{butler2018stability}, all of 
its steady states are dynamically stable, its variant in 
which different microbes have different yields on the same nutrient, 
has both these properties \cite{maslov2018unpublished}. %\textcolor{red}{Sergei unpublished - cite this}. 
Different yields of different microbial species 
prevent one from constructing the Lyapunov function used in 
the standard MacArthur model \cite{macarthur1970species}. 
We leave the topic of existence and the functional form of 
the Lyapunov function in our model for future studies. 

% Several methods used in our study, such as e.g. the way to calculate the  
% volumes of feasible nutrient influxes (also known as 
% structural stability \cite{rohr2014structural}) 
% for each of the allowed states, are quite general. In fact such structural 
% stability has been recently calculated for the MacArthur model 
% in Ref. \cite{butler2018stability} and for the Generalized Lotka-Volterra model 
% in Ref. \cite{grilli2017feasibility} (with microbes growth rates playing the 
% role of nutrient influxes in our model).

% Above we presented a detailed study of steady states in a simple model of 
% a microbial ecosystem, growing on a mixture of multiple resources of 
% two essential types, say C and N. Because of several simplifying assumptions, 
% most importantly that of extreme specialization of microbes in our model 
% capable of utilizing just one carbon and one nitrogen source, we were to develop 
% a good mathematical understanding of the model. Most importantly, 
% we were able to formulate two exclusion rules (See Rule 1 and 2 above) 
% fully dictating the repertoire of steady states possible in our model for different 
% environmental conditions (nutrient influxes). 

Using the algorithms of the  
stable matching problem \cite{gale1962college,gusfield1989stable} we 
were able to list all of these states based only on ranked tables of nutrient 
competitive abilities of different microbes. The advantage of this approach 
is that it bypasses the need for precise measurements of the kinetic parameters 
and depends only on the relative microbial preferences toward nutrients. This property is likely 
unique to our model. Indeed, for a popular MacArthur model \cite{macarthur1964competition,macarthur1970species,chesson1990macarthur}
of co-utilization of fully substitutable resources, the relative rankings 
of different microbes for nutrients depend on nutrient concentrations. 
This greatly complicates the task of deducing the ultimate set of allowed 
states, that is to say, all subsets of surviving microbes realized 
for different environmental conditions. 

% %%% moved from intro
% The existence of {\it essential nutrients of two types (C and N)}, 
% each capable of limiting microbial growth (logical {\it AND-gate} 
% on inputs), 
% makes the properties of our model rather different than those 
% of different variants of the MacArthur consumer-resource model \cite{macarthur1964competition,macarthur1970species,chesson1990macarthur}.
% The latter are used to describe microbial 
% ecosystems growing on fully {\it substitutable} nutrients 
% (logical {\it OR-gate} on inputs)
% and are enjoying a recent surge in popularity \cite{goldford2018emergent,
% tikhonov2016communitylevel,tikhonov2017collective,posfai2017metabolic,
% advani2018statistical,butler2018stability}.

To improve mathematical tractability of our model 
we have made a number of simplifying assumptions. 
These can be relaxed in the following variants of our basic model
%to explore some additional interesting aspects of microbial communities, 
some of which are listed below:
(i) A simple generalization of our model is to relax the condition of extreme specialization 
to allow for generalist species, i.e., those with growth rate is given by: 
$$g_{\alpha}=
%\sum_{i,j} 
\min\bigl(\sum_{i \text{ used by }\alpha} \lambda^{(c)}_{\alpha i}c_i, \sum_{j \text{ used by }\alpha} \lambda^{(n)}_{\alpha j}n_j\bigr)$$. 
Here the sum over $i$ (respectively $j$) is carried out over all carbon (respectively nitrogen) sources a given species is capable of using. Here one assumes that multiple substitutable sources are co-utilized as in the MacArthur model. Another possibility is to assume that each species is using its substitutable resources one-at-a-time, as we assumed for multiple carbon sources in Ref. \cite{goyal2018multiple}. Since at any point in time each of the species is using a ``specialist strategy'' growing on a single  carbon and a single carbon source, we expect many of our results to be extendable to this model variant.
%$\lambda(c)_{\alpha i}$ and $\lambda(n)_{\alpha j}$.
Using either of these model variants one can explore environmental conditions (number of resources and their fluctuations in time and space) that would favor specialists or generalists over each other. It will also be interesting to explore how the presence of generalists affects the number of alternative stable states and how the available nutrients are partitioned between the different coexisting species in such a community.
(ii) We worked with a fixed size of species pool for our models. By relaxing this constraint and having a large universe of species to choose from, one can explore interesting aspects of evolution and adaptations in our model setup, e.g., how the presence of multistability affects the process of community assembly.
(iii) One can also introduce cross-feeding between the species, thus generating additional resources in the system and allowing for a larger number of species to coexist, and, hence further increasing the number of alternative stable states.
(iv) Although we have identified that larger variance in C:N stoichiometry of individual species (quantified by $Y^{(c)}_\alpha/Y^{(n)}_\alpha$ in our model) promotes multistability, other factors affecting the likelihood of alternative stable states remain to be identified in future studies.

\begin{acknowledgments}
Part of this work has been carried out at the University 
of Padova, Italy, in August 2018, during a scientific visit 
by one of us (S.M.). 
\end{acknowledgments}

\section*{Authors contributions}
S.M. designed the research; P.P. simulated the computational model; 
Y.F. and S.M. developed the theory for the computational model; 
P.P., S.M., and V.D. analyzed the data; S. M. V. D. and P.P wrote the manuscript; 
and S.M. supervised the study.

\newpage
\section*{methods}
\textbf{Enumeration of all allowed states.}
%\label{Meth:StableMatchingAlgo}

Every allowed state can be converted into a unique bipartite network with the two types nutrients (carbon and nitrogen) as the nodes and links representing the surviving species. The source node of the link represents the rate limiting nutrient of the species and each link is characterized by its $\lambda^{(c)}$ and $\lambda^{(n)}$.
%direction of the links (pointing \textit{from} the rate-limiting nutrient)  
(see Fig. \ref{fig1n}A). a,  
%\textcolor{red}{Each link is characterized by a pair of $\{\lambda^{(c)}_{(i,j)},\lambda^{(n)}_{(i,j)} \}$ corresponding to the nodes of each type.} 
The competitive exclusion rules int the language of networks is stated below:

\begin{itemize}
 \renewcommand{\labelitemi}{\scriptsize$\blacksquare$}
\item \textbf{Rule 1} - All nodes can have at most one outgoing link %\textcolor{red}{(species that are limited by that nutrient)},
\item \textbf{Rule 2} - All incoming links at any node should have a larger $\lambda$ than the $\lambda$ of the outgoing link (if any).
\end{itemize}

%To obtain all such networks we apply the following steps:
To obtain all allowed states 
%enumerate an exhaustive list 
one needs to perform an exhaustive search of all the networks that satisfy the above constraints. We do this in the following way: 
We start with choosing a particular set of ranked values of $\lambda^{(c)}_{(i,j)}$ and $\lambda^{(n)}_{(i,j)}$ for all 
%\textcolor{red}{possible links} 
$L^2$ species in the pool. We then choose any one type of nodes on which the outgoing links will be assigned first (say carbon; the choice carbon as opposed to nitrogen does not affect the final result of this algorithm). We then perform a two-step procedure for links allocation. We first allocate all outgoing links from C-type to N-type nodes by choosing a rank between 1 and $L+1$ for each of the C-type nodes, where rank $L+1$ corresponds to having no link (no species is limited by this nutrient). For this specific combination of outgoing links we generate all allowed sets of incoming links (from N-type to C-type). To implement this step one needs to follow \textit{Rule 2} to filter out prohibited allocations of incoming links. This procedure is guaranteed to find all allowed states for this chosen specific chosen set of outgoing links from C and one of the allowed states will be uninvadable (see \ref{SI-Note3} for details). One repeats this allocation procedure for each combination ranks of outgoing links for the C-type nodes ($(L+1)^L$ possible allocations in total) to get a list of all allowed bipartite networks for the set of $\{\lambda^{(k)}_{(i,j)}\}$.

We used the above procedure to enumerate allowed states for different numers of resources $L$ (see Fig. \ref{fig1n}C). The values of $\{\lambda^{(k)}_{(i,j)}\}$ were chosen randomly between 10 and 100 for these numerical experiments.
%%%% VERONIKA: I think we can drop the detailed example for the initial submission
%%%%%%%%
%To demonstrate the above mentioned process we consider the case for L=2. In Table \ref{tab:Lambdas_L2} we present the values of $\lambda$ of the 4 species on each nutrient. In Fig. \ref{fig:Methods-1} we represent this information in a form that highlights the ranks of the competitive abilities (based on values of $\lambda$) of the species on each nutrient. This representation helps to describe the above mentioned algorithm. To demonstrate an example, in Fig. \ref{fig:Methods-2} we perform step-3 and step-4 that generates two allowed states one of which is an uninvadable state.

%\begin{table}[ht]
%\centering
%\begin{tabular}{|c|c|c}
%  \hline
%  $\lambda^{(c)}_{(1,1)}=41 ,\lambda^{(n)}_{(1,1)}=16 $ & $\lambda^{(c)}_{(1,2)}=35 ,\lambda^{(n)}_{(1,2)}=50 $\\\hline
% $\lambda^{(c)}_{(2,1)}=52 ,\lambda^{(n)}_{(2,1)}=27 $ & $\lambda^{(c)}_{(2,2)}=56 ,\lambda^{(n)}_{(2,2)}=44 $\\\hline
%\end{tabular}
%\caption{\label{tab:Lambdas_L2} $\lambda^{(c)}$, $\lambda^{(n)}$ values of the 4 species for the $K=M=2$ nutrient case. These values were chosen randomly between 10 and 100 and were rounded off to the nearest integer.}
%\end{table}
\bigbreak

\textbf{Feasibility of the allowed states.} \label{Meth:FeasibleVolumeAlgo}

As described in the main text, at the steady state of each allowed state the concentration of the surviving species and %($S_{surv} \leq K+M$) 
and the concentration of the %$K+M-S_{surv}$ 
nutrients not limited by any of the %$S_{surv}$ 
surviving species is completely determined by the $K+M$ mass conservation laws (Eqs. \ref{eqn:conservation_laws}). Hence each allowed state can be uniquely characterized by a set of $K+M$ variables (we define them as $X_p$ for the $p^{th}$ state) consisting of the population of the $S_{surv} \leq K+M$ species and the concentration of the $K+M-S_{surv}$ non-limited nutrients. E.g., for the %$K=M=2$ 
2Cx2Nx4S %nutrient 
case used in the main text %with the particular choice of $\lambda$ as shown in Table \ref{tab:Lambdas_L2}, there are 33 AS out of which 7 are UIS. Fig. \ref{fig:Methods-2} showed the generation of AS-6 (which is also a UIS). For AS-6 
in S5 we have: $X_5 = \{B_{(1,1)}, B_{(1,2)}, B_{(2,2)}, n_2\}$.

Now, given the parameters defining the species (i.e., $\lambda$ and $Y$) and the chemostat dilution constant $\delta$, each state $p$ will have a finite region in the nutrient influx space \\(a $K+M$ dimensional space 
%defined by each of the C influx $\phi_i^{(c)}$ and N influx $\phi_j^{(n)}$
$\{\phi_i^{(c)}, \phi_j^{(n)}\}$)%, i \in 1:K, j \in 1:M\}$})
this state will be feasible, i.e., $X_p > 0$. % for the $p^{th}$ state. 
The volume of this region %in the $K+M$ dimensional influx space 
quantifies the structural stability of the %$p^{th}$ 
state $p$.

%For the %$K=M=2$
%\textcolor{red}{2Cx2Nx4S} case (Fig. \ref{fig:Methods-1}) we obtained the feasible volumes for each of the 33 AS. 
To simplify the process of calculating the feasible volumes we assumed the \textit{high-influx limit}, i.e., $\phi^{(c)}_{i} >> \frac{\delta^2}{\lambda_{\alpha}^{(c)}} $ and $\phi^{(n)}_{j} >> \frac{\delta^2}{\lambda_{\alpha}^{(n)}}$. It means %that the parameters are chosen such 
that if any nutrient is limited by a surviving species in the allowed state, the concentration of that nutrient at the steady state will be negligible compared to what it was before speciation took place. With this reasonably valid \textit{high-influx limit} assumption the mass conservation laws (Eqs. \ref{eqn:conservation_laws}) that are used to obtain the feasible volumes of each of the allowed states can be represented into a compact matrix form. For the allowed state $p$ the matrix equation becomes: 
\begin{equation}
\Phi = \hat{R_p} X_p
\quad , 
\label{eqn:Matrix-form}
\end{equation}
where $\Phi$ is the vector of the $K+M$ nutrient fluxes and $\hat{R_p}$ is a matrix composed of $Y^{-1}$ of surviving species and "1" for each of the non-limiting nutrients in the allowed state $p$. For the S5 in 2Cx2Nx4S example 
%(generated in Fig. \ref{fig:Methods-2}) 
the Eq. \ref{eqn:Matrix-form} expands to:

\begin{gather}
 \begin{bmatrix} \phi_1^{(c)} \\ \phi_2^{(c)} \\ \phi_1^{(n)} \\ \phi_2^{(n)} \end{bmatrix}
 =
  \begin{bmatrix}
   \frac{1}{Y_{(1,1)}^{(c)}} & \frac{1}{Y_{(1,2)}^{(c)}}  & 0 & 0 \\
   0 & 0 & \frac{1}{Y_{(2,2)}^{(c)}} & 0 \\
   \frac{1}{Y_{(1,1)}^{(n)}} & 0 & 0 & 0 \\
   0 & \frac{1}{Y_{(1,2)}^{(n)}} & \frac{1}{Y_{(2,2)}^{(n)}} & 1
   \end{bmatrix}
   \begin{bmatrix} B_{(1,1)} \\ B_{(1,2)} \\ B_{(2,2)} \\ n_2 \end{bmatrix}.
\label{eqn:Matrix-form2}
\end{gather}

\begin{comment}
\begin{gather}
 \begin{bmatrix} \phi_1^{(c)} \\ \phi_2^{(c)} \\ \phi_1^{(n)} \\ \phi_2^{(n)} \end{bmatrix}
 =
  \begin{bmatrix}
   1\Big/Y_{(1,1)}^{(c)} & 0 & 0 & 0 \\
   0 & 1\Big/Y_{(2,1)}^{(c)} & 1\Big/Y_{(2,2)}^{(c)} & 0 \\
   1\Big/Y_{(1,1)}^{(n)} & 1\Big/Y_{(2,1)}^{(n)} & 0 & 0 \\
   0 & 0 & 1\Big/Y_{(2,2)}^{(n)} & 1
   \end{bmatrix}
   \begin{bmatrix} B_{(1,1)} \\ B_{(2,1)} \\ B_{(2,2)} \\ n_2 \end{bmatrix}
\end{gather}
\end{comment}

\bigbreak
\textbf{Monte-Carlo sampling of nutrient influx space.}

Using Eq. \ref{eqn:Matrix-form} it is trivial to compute if an allowed state is feasible at a particular nutrient influx point $\Phi$. To check feasibility of the allowed state $p$ at $\Phi$ we simply multiply the inverse of the matrix $\hat{R_p}$ with the vector $\Phi$. If all the elements of the resulting vector $X_p$ are positive then the %$p^{th}$ 
allowed state $p$is feasible at $\Phi$. If the matrix $\hat{R_p}$ is not invertible i.e., %the determinant of 
$ det(\hat{R_p})=0$, it indicates that this allowed state is not feasible anywhere in the nutrient influx space. 

We imposed a common upper and lower bound on each of the $K+M$ nutrient influxes \big( $\phi_{min} \leq \phi^{(c,n)}_i \leq \phi_{max} $\big) thus restricting the search of volumes of feasible allowed states in a $K+M$ dimensional hypercube in the nutrient influx space. We chose $\phi_{min} = 10, \phi_{max}=1000$. The lower bound ensures that the system is always in the \textit{high-flux limit} as max($\frac{\delta^2}{\lambda_{\alpha}}$) $= 0.1 << \phi_{min}$. This is because $\delta = 1$ and $\lambda_{min}=10$ ($\lambda$ were chosen uniformly between 10 and 100). 
%We obtained the feasible volume of all the 33 AS (for $\lambda$ given in Fig \ref{fig:Methods-1} for the $K=M=2$ \textcolor{red}{2Cx2Nx4S?} nutrient case) using the above defined method with the yields $Y$ given in Table \ref{tab:Yields_L2}. 
%\big[\textit{Reminder: while the number and the type of allowed state depends solely on the values of $\lambda$ of the species, the feasibility volume and the stability of the allowed state depends on the specific values of the $Y$.}\big]
We then randomly spanned a million points in this hypercube and checked feasibility of each allowed state (i.e., for state $p$ we checked that all the elements of $X_p$ are positive). For each state we calculated the total number of points where it was feasible to quantify the volumes of the allowed states.
%%%%%%%%%
%We probably should mention that we also chose some lambda's and yields first.
%%%%%%%%%%%% later
%(\textcolor{red}{For Veronika: describe how the feasible volumes were used to construct figure 2A in the main text. -  We have a sufficient description in results, and PCA is a very  common procedure})

\begin{comment}
%\begin{table}[ht]
%\centering
%\begin{tabular}{c|c|c|c}
% & N$_\textrm{\scriptsize 1}$ & N$_\textrm{\scriptsize 2}$ &  \\\hline
%C$_\textrm{\scriptsize 1}$ & $Y^{C}_{(1,1)}=0.37 ,Y^{N}_{(1,1)}=0.35 $ & $Y^{C}_{(1,2)}=0.64 ,Y^{N}_{(1,2)}=0.50 $\\\hline
%C$_\textrm{\scriptsize 2}$ & $Y^{C}_{(2,1)}=0.47 ,Y^{N}_{(2,1)}=0.40 $ & $Y^{C}_{(2,2)}=0.14 ,Y^{N}_{(2,2)}=0.30 $
%\end{tabular}
%\caption{\label{tab:Yields_L2} Values of $\lambda$ of species on each of the four nutrients. (i,j) represents the microbe growing on $C_i$, $N_j$.}
%\end{table}
\end{comment}

%\begin{table}[ht]
%\centering
%\begin{tabular}{|c|c|c}
%  \hline
%  $Y^{(c)}_{(1,1)}=0.37 ,Y^{(n)}_{(1,1)}=0.35 $ & $Y^{(c)}_{(1,2)}=0.64 ,Y^{(n)}_{(1,2)}=0.50 $\\\hline
%  $Y^{(c)}_{(2,1)}=0.47 ,Y^{(n)}_{(2,1)}=0.40 $ & $Y^{(c)}_{(2,2)}=0.14 ,Y^{(n)}_{(2,2)}=0.30 $\\\hline
%\end{tabular}
%\caption{\label{tab:Yields_L2} Values of carbon and nitrogen Yields of the 4 species for $K=M=2$ nutrient case. The values were chosen randomly between 0.1 and 1.0 and were rounded to two decimal places.}
%\end{table}

\bigbreak
\textbf{Overlap of Volumes}: Two allowed states are said to overlap with each other %they share a common feasibility region in the $K+M$ dimensional hypercube in the nutrient influx space. I.e., two states are said to overlap 
if there exists a set points in the hypercube at which both of them are feasible. We used the data obtained by Monte-Carlo sampling to calculate shared feasibility regions (shared sets of points) between states. We then normalized those numbers for each state by its overall volume (see Fig. \ref{fig3n}A).
%%% VERONIKA: we have the following in the results
%In Fig. \ref{fig3n}B of the main text we show the network of overlaps between the 6 stable uninvadable states (we will discuss the stability of states below) in the $K=M=2$ nutrient case. For each state, the fraction of the feasible volume that overlaps with another state is represented on the edge connecting the two states. 
%In this case the maximum number of coexisting (overlapping) stable UIS at any nutrient influx point never goes above 2, i.e., 2 stable UIS and 1 unstable UIS. 
For 6Cx6Nx36S case we used Gephi 0.9.2 software package to visualize the network and performed modularity analysis to identify densely interconnected clusters in the network\cite{blondel2008fast}. Resolution parameter was set to 1.5 to produce Fig. \ref{fig3n}B.

%\textcolor{blue}{VERONIKA: We have to refer to the methods for this in a proper places in results.}

\bigbreak
\textbf{Dynamical stability of the allowed states.}
%\label{Meth:DynamicStabilityAlgo}

We checked the dynamic stability of the %33 
allowed states %for the 2Cx2Nx4S example 
%considered in the main text. This was done 
in %the following 
two ways:

\begin{enumerate}

\item {\bf Perturbation analysis.} We prepared each allowed state at %some 
one of %their
its feasible influx points and subjected %them 
it to small perturbations of (i) all the $K+M=2$ nutrient concentrations and (ii) the populations of the $S_{surv}$ species in the state. The importance of perturbing the population of only the $S_{surv}$ species should be noted because an invadable state, by definition, will always be dynamically unstable against addition of (at least one) new species from the species pool. And this instability should not render the invadable state as dynamically unstable. Hence we stress that an allowed state will be dynamically unstable if perturbation of any of the nutrient concentrations or the population of any of the $S_{surv}$ species drives the state to a different allowed state.

%%% belongs to the results:
%From the above analysis we found that all the allowed states except one (S7) are dynamically stable. 

%\item {\bf Eigen-values of the Jacobian:} From basic dynamical systems theory we know that for a steady state of a dynamical system to be stable, the real part of all the eigen-values of the Jacobian computed at this steady state should be negative. We used this property to explore the dynamic stability of the 7 uninvadable states. For each of the 7 uninvadable states we \textcolor{red}{used Monte-Carlo sampling} and at each feasible point we computed the Jacobian and all the eigen-values.
%spanned a million points in the $K+M=4$ dimensional nutrient influx space 

%The results of the eigen-value analysis for the 7 uninvadable states corroborates with the stability results obtained in the above perturbation analysis. I.e., except S7 all the other 6 uninvadable states were found to be dynamically stable at the \textcolor{red}{corresponding???} feasible points.

\item {\bf Overlap analysis.} 
%\textcolor{red}{
The dynamic stability of the allowed states can also be inferred from the influx map (a map that gives us the information of all the feasible states possible at each point in the influx space) obtained from our Monte-Carlo simulations.
%based on the information about overlaps between them (as described in the main text). 
We first recognize all states which had a unique presence at at least one influx point
%for which this state is exclusively feasible 
(i.e. no other states are feasible at this flux point). All such states should be dynamically stable by definition. Note that Monte-Carlo samples a finite number of influx points and thus it is possible to miss crucial influx points which could have rendered some of the states as stable and thus leading to assigning some of the stable states as unstable. This false assignment will lead into the violation of the V/V-1 rule (as described in the main text) at some influx points. To correct for this error we go over each unstable state and check if assigning it as stable reduces the list of violated influx points. If it does then we include it in our list of stable states. This method could not infer the stability of 16 uninvadable states which had very small feasible volumes.

%an underestimation of stable states. To correct for this error impose the rule
%discretized our flux space. It can potentially lead to underestimation of stable states. We used V/V-1 rule to verify and correct our estimates of stability.
%}
\end{enumerate}
\bigbreak

\textbf{Yield-variation.}

Since the dynamical stability and the size of feasible volume of the states depends on the choice of values of the yields $Y$, we performed a set of Monte-Carlo sampling experiments for 2Cx2Nx4S and 6Cx6Nx36S examples to explore how the choice of yields governs multistability.
For 2Cx2Nx4S case we performed 4000 Monte-Carlo simulations for a fixed set of $\lambda$ (see Supplementary Table 1) and yields were drawn from uniform distributions with different standard deviations (1000 simulation per standard deviation). We used these simulations to calculate the fraction of influx space where we observed multistability (Fig. \ref{fig4n}A).
For 6Cx6Nx36S example (see $\lambda$ in Supplementary Tables 3-4) we chose 4 different sets of Yields and performed Monte-Carlo simulations for each of them. We further performed overlap analysis for each of these numerical experiments counting the number of uninvadable stable states feasible for each flux point (see Fig. \ref{fig4n}B).
\bigbreak

\textbf{Colonization dynamics.}
%\label{Meth:ColonizationAlgo}

%In this section we will describe the process of sequential colonization of the $K$ carbon, $M$ nitrogen sources by the $S=K$x$M$ species. 
%This generates all the intermediate allowed states the system transitions to in its journey from the \textcolor{red}{empty} %\textit{abiotic state} 
%to the final uninvadable state (shown in Fig. \ref{fig5n} A,B for our 2Cx2Nx4S example). In the empty
%abiotic 
%state the system consists of only the $K+M$ nutrients with no species present. Hence the nutrients are present at their \textit{abiotic concentrations} \big(i.e., C$_i(0)$ $=\phi_i^{(c)}/\delta$ and N$_i(0)$ $=\phi_j^{(n)}/\delta$\big). 
%In other words, the abiotic concentration of a nutrient is the amount of the nutrient present when no species consumes it. 

To study the process of speciation in our model, we implemented the sequential colonization procedure as described below. We first set the system at the abiotic state \big(i.e., n$_i(0)$ $=\phi_i^{(c)}/\delta$ and n$_i(0)$ $=\phi_j^{(n)}/\delta$\big)
%(choose the nutrient influxes $\phi_i^{(c)}$, $\phi_j^{(n)}$ and the the chemostat dilution rate $\delta$). 
We then randomly select one species from our pool of S species and introduce it into the system with a small population density ($10^{-5}$). We then perform a numerical integration of the current system until the system settles into a steady state. If the population density of any of the species at the steady state falls below a predefined threshold ($10^{-7}$) we considered it to be extinct. We keep performing this random selection and introduction of species addition followed by dynamic integration until no new species from the pool can invade, thus giving us an uninvadable state. This colonization dynamics is repeated for a large number of random-order-introduction of species to obtain all possible terminal ends. The set of all steady states obtained in this process are all the allowed states that the system navigates through.
%, at the end of each The sequence of sets of surviving bacteria after each species addition contains all the allowed states the system navigates through.

%\textcolor{red}{
We followed the above procedure for two different sets of environmental parameters in our 2Cx2Nx4S example. First set: $\phi^{(c)}_1=300$, $\phi^{(c)}_2=500$, $\phi^{(n)}_1=500$, $\phi^{(n)}_2=500$, $\delta=1$ (see Fig. \ref{fig5n}A). Second: $\phi^{(c)}_1=500$, $\phi^{(c)}_2=500$, $\phi^{(n)}_1=500$, $\phi^{(n)}_2=5000$, $\delta=1$ (see Fig. \ref{fig5n}B).
This colonization dynamics was repeated for a large number of random order introductions of species from the pool to obtain all possible transitions (shown as black arrows in Fig. \ref{fig5n}A,B) between the allowed states at the given nutrient influx.

To study transitions between uninvadable states in response to environmental perturbations, we started from one of the uninvadable states and varied one of the fluxes in some range ($10 \leq \phi^{(c)}_1 \leq 800$) with some step ($\delta \phi^{(c)}_1$=10) while constantly introducing the random bacterial species from the pool to the system. 
%}

\begin{comment}
Set values to the environmental parameters (the nutrient influxes $\phi_i^{(c)}$, $\phi_j^{(n)}$ and the the chemostat dilution rate $\delta$. 

\begin{itemize}

\item \textbf{Step-1}: Set the system at the abiotic state.

\item \textbf{Step-2}: Randomly select any one species from the pool of $S$ species and introduce it in the system with a small population density ($10^{-5}$ in our case).

\item \textbf{Step-3}: Numerically integrate (the methods of integration is given below) the current system until the system reaches a dynamical steady-state.

\item \textbf{Step-4}: Remove the set of species from the system whose steady-state population density has fallen below a minimum threshold ($10^{-7}$ in the current simulations). The state obtained after the removal of the extinct species is an allowed state that the system is in currently.

\item \textbf{Step-5}: Stop if the system has reached a UIS. If not, then go back to step-2 and and repeat the cycle (step-2 $\Rightarrow$ step-3 $\Rightarrow$ step-4 $\Rightarrow$ step-5) sequentially until the system reaches a UIS.

\end{itemize}
The sequence of states obtained at the end of Step-5 in every cycle are all the allowed states the system navigates through

Also it gives all possible allowed paths from the initial abiotic state to the final UIS.
\end{comment}

%$^*$
The numerical integration for the above process was done in C programming language using the CVODE solver library of the SUNDIALS %(Suite of Nonlinear and Differential/Algebraic Equation Solvers) 
package\cite{hindmarsh2005sundials} downloaded from the website: \sloppy
\url{https://computation.llnl.gov/projects/sundials/sundials-software}.

%%%
\bibliography{c_and_n_citations} 

\newpage
% \clearpage
% \pagenumbering{arabic}
% \setcounter{page}{1}

\renewcommand{\theequation}{S\arabic{equation}}
\setcounter{equation}{0}
\renewcommand{\thefigure}{S\arabic{figure}}
\setcounter{figure}{0}
\cleardoublepage 
\newpage

%%% How about this variant of the table?
\section*{Supplementary tables}
%%% tables for 2x2
\begin{table}[ht]
\centering
\begin{tabular}{|P{0.8cm}|P{0.8cm}|P{0.8cm}||P{0.8cm}|P{0.8cm}|}
  \hline
   & \multicolumn{2}{c||}{$\lambda^{(c)}_{(i,j)}$} & \multicolumn{2}{c|}{$\lambda^{(n)}_{(i,j)}$}\\
  \hline
   & $N_1$ & $N_2$ & $N_1$ & $N_2$\\
  \hline
  $C_1$ & 41 & 35 & 16 & 50 \\
  \hline
  $C_2$ & 52 & 56 & 27 & 44 \\
  \hline
\end{tabular}
\caption{\label{tab:Lambdas_L2} $\lambda^{(c)}_{(i,j)}$, $\lambda^{(n)}_{(i,j)}$ values of the 4 species for the 2Cx2Nx4S model. %These values were chosen randomly between 10 and 100 and were rounded off to the nearest integer.
}
\end{table}

\begin{table}[ht]
\centering
\begin{tabular}{|P{0.8cm}|P{0.8cm}|P{0.8cm}||P{0.8cm}|P{0.8cm}|}
  \hline
   & \multicolumn{2}{c||}{$Y^{(c)}_{(i,j)}$} & \multicolumn{2}{c|}{$Y^{(n)}_{(i,j)}$}\\
   \hline
   & $N_1$ & $N_2$ & $N_1$ & $N_2$\\
  \hline
  $C_1$ & 0.37 & 0.64 & 0.35 & 0.50 \\\hline
  $C_2$ & 0.47 & 0.14 & 0.40 & 0.30 \\\hline
\end{tabular}
\caption{\label{tab:Yields_L2} Values of carbon and nitrogen Yields of the 4 species for the 2Cx2Nx4S model. 
%The values were chosen randomly between 0.1 and 1.0 and were rounded to two decimal places.
}
\end{table}
%%%% tables for 6x6
\begin{table}[ht]
\centering
\begin{tabular}{|P{0.8cm}|P{0.8cm}|P{0.8cm}|P{0.8cm}|P{0.8cm}|P{0.8cm}|P{0.8cm}|}
  \hline
   & $N_1$ & $N_2$ & $N_3$ & $N_4$ & $N_5$ & $N_6$ \\\hline
   $C_1$ & 47.4 & 78.1 & 93.7 & 68.9 & 75.0 & 44.5 \\\hline
   $C_2$ & 89.6 & 68.6 & 33.6 & 77.8 & 16.5 & 90.8 \\\hline
   $C_3$ & 56.5 & 32.2 & 86.2 & 13.1 & 71.1 & 15.5 \\\hline
   $C_4$ & 53.0 & 94.1 & 38.7 & 10.7 & 34.0 & 34.9 \\\hline
   $C_5$ & 25.0 & 49.3 & 76.3 & 18.2 & 54.5 & 51.8 \\\hline
   $C_6$ & 47.1 & 91.9 & 57.7 & 63.0 & 92.2 & 90.0 \\\hline
\end{tabular}
\caption{\label{tab:Lambdas_c_L6} $\lambda^{(c)}_{(i,j)}$ values of the 36 species for the 6Cx6Nx36S model.
%These values were chosen randomly between 10 and 100 and were rounded off to one decimal place.
}
\end{table}
\begin{table}[ht]
\centering
\begin{tabular}{|P{0.8cm}|P{0.8cm}|P{0.8cm}|P{0.8cm}|P{0.8cm}|P{0.8cm}|P{0.8cm}|}
  \hline
   & $N_1$ & $N_2$ & $N_3$ & $N_4$ & $N_5$ & $N_6$ \\\hline
   $C_1$ & 18.3 & 57.7 & 44.5 & 16.0 & 70.4 & 66.8 \\\hline
   $C_2$ & 56.7 & 31.4 & 78.6 & 91.8 & 34.5 & 34.7 \\\hline
   $C_3$ & 42.3 & 53.8 & 84.8 & 99.2 & 79.0 & 44.6 \\\hline
   $C_4$ & 95.3 & 91.4 & 73.1 & 42.9 & 98.8 & 66.7 \\\hline
   $C_5$ & 76.2 & 98.4 & 31.0 & 55.4 & 14.5 & 57.4 \\\hline
   $C_6$ & 37.6 & 79.3 & 58.4 & 71.8 & 26.0 & 84.5 \\\hline
\end{tabular}
\caption{\label{tab:Lambdas_n_L6} $\lambda^{(n)}_{(i,j)}$ values of the 36 species for the 6Cx6Nx36S model. 
%These values were chosen randomly between 10 and 100 and were rounded off to one decimal place.
}
\end{table}
\begin{table}[ht]
\centering
\begin{tabular}{|P{0.8cm}|P{0.8cm}|P{0.8cm}|P{0.8cm}|P{0.8cm}|P{0.8cm}|P{0.8cm}|}
  \hline
   & $N_1$ & $N_2$ & $N_3$ & $N_4$ & $N_5$ & $N_6$ \\\hline
   $C_1$ & 0.72 & 0.59 & 0.15 & 0.11 & 0.13 & 0.75 \\\hline
   $C_2$ & 0.29 & 0.72 & 0.79 & 0.39 & 0.15 & 0.16 \\\hline
   $C_3$ & 0.76 & 0.61 & 0.40 & 0.86 & 0.60 & 0.63 \\\hline
   $C_4$ & 0.87 & 0.68 & 0.64 & 0.27 & 0.80 & 0.51 \\\hline
   $C_5$ & 0.88 & 0.13 & 0.31 & 0.36 & 0.11 & 0.48 \\\hline
   $C_6$ & 0.46 & 0.34 & 0.38 & 0.83 & 0.70 & 0.86 \\\hline
\end{tabular}
\caption{\label{tab:Yields_c_L6} $Y^{(c)}_{(i,j)}$ values of the 36 species for the 6Cx6Nx36S model. 
%These values were chosen randomly between 0 and 1 and were rounded off to two decimal places.
}
\end{table}
\begin{table}[ht]
\centering
\begin{tabular}{|P{0.8cm}|P{0.8cm}|P{0.8cm}|P{0.8cm}|P{0.8cm}|P{0.8cm}|P{0.8cm}|}
  \hline
   & $N_1$ & $N_2$ & $N_3$ & $N_4$ & $N_5$ & $N_6$ \\\hline
   $C_1$ & 0.10 & 0.30 & 0.67 & 0.36 & 0.32 & 0.66 \\\hline
   $C_2$ & 0.83 & 0.30 & 0.47 & 0.30 & 0.40 & 0.58 \\\hline
   $C_3$ & 0.72 & 0.15 & 0.65 & 0.54 & 0.21 & 0.18 \\\hline
   $C_4$ & 0.30 & 0.22 & 0.84 & 0.64 & 0.29 & 0.56 \\\hline
   $C_5$ & 0.55 & 0.16 & 0.77 & 0.42 & 0.22 & 0.25 \\\hline
   $C_6$ & 0.49 & 0.67 & 0.89 & 0.80 & 0.50 & 0.19 \\\hline
\end{tabular}
\caption{\label{tab:Yields_n_L6} $Y^{(n)}_{(i,j)}$ values of the 36 species for the 6Cx6Nx36S model. 
%These values were chosen randomly between 0 and 1 and were rounded off to two decimal places.
}
\end{table}

\section*{Supplementary figures}
\begin{figure}
%\centerline{\includegraphics[width=\linewidth]{fig3n.eps}}
\centerline{\includegraphics[width=\linewidth]{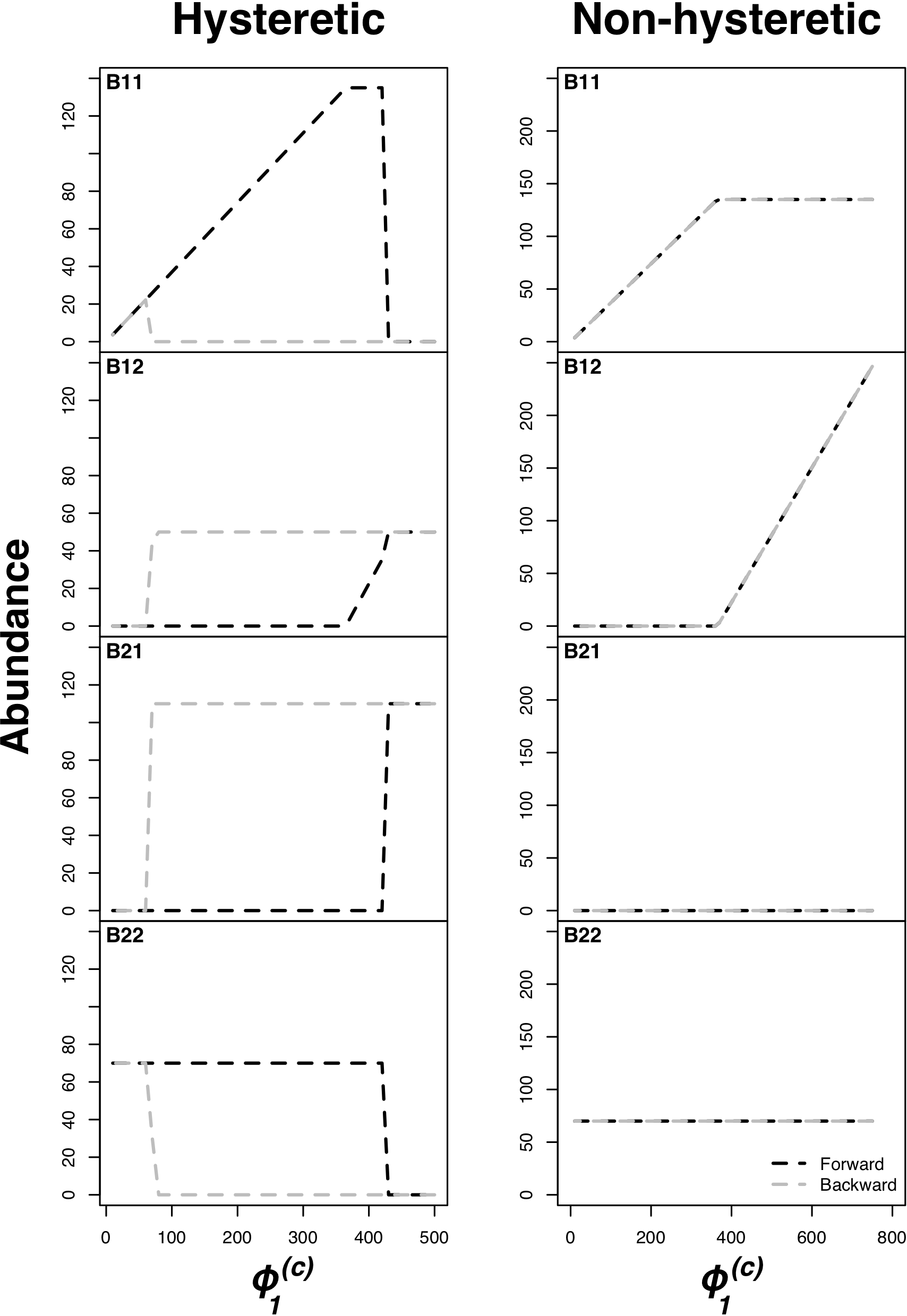}}
\caption{{\bf State transitions triggered by colonization dynamics and changing environment in 2Cx2Nx4S example.} 
Changes in steady state bacterial abundances through flux variation in Fig. \ref{fig5n} (left panel corresponds to hysteretic case in C, right - non-hysteretic transition from D).
}
\label{figs1}
\end{figure}

\newpage

\section*{Supplementary Information}
\renewcommand{\thesubsection}{Supplementary Note \arabic{subsection}}
\subsection{General form of growth laws}\label{SI-Note1}
%\textcolor{red}{correct the spelling mistakes}
It is straightforward to generalize our model to include 
more general functional form for 
growth laws than Liebig's law, $\min(\lambda_{\alpha}^{(c)} c_i,
\lambda_{\alpha}^{(n)} n_j$.  Microbial growth on 
two essential substrates is thought to normally 
follow the Monod's equation for the rate-limiting nutrient:
$g^{(m)}_{\alpha}\min(c_i/(K_{\alpha}^{(c)}+c_i),n_j/(K_{\alpha}^{(n)}+n_j))$ 
(See Ref. \cite{kovrovkovar1998growth} for a discussion of 
limitations of Monod's law).
For low concentrations of the rate-limiting nutrient, say carbon source, 
the Monod's law simplifies to the proportional growth law used throughout this study: $g_{\alpha}=\lambda^{(c)}_{\alpha} c_i$. Microbes' competitive abilities, also known as their 
specific affinities towards each substrate, are related to the parameters of 
Monod's law via 
\begin{equation}
\lambda_{\alpha}^{(c)}= \frac{g^{(m)}_{\alpha}}{K_{\alpha}^{(c)}}; \quad 
\lambda_{\alpha}^{(n)}= \frac{g^{(m)}_{\alpha}}{K_{\alpha}^{(n)}} 
\end{equation}
%$\min(\lambda_{\alpha}^{(c)}c_i,\lambda_{\alpha}^{(n)}n_j)$ would 
%be given by  
In another variant of growth laws, two essential nutrients at low concentrations 
jointly affect the growth rate of the microbe: 
$g^{(m)}_{\alpha} c_i \cdot n_j/[(K_{\alpha}^{(c)}+c_i) \cdot (K_{\alpha}^{(n)}+n_j)]$ 
(see Ref. \cite{bader1978analysis} for a discussion of these and other 
double-substrate growth laws).
For simplicity of mathematical calculation we limit this study to Liebig's law.  However, many of the essential results we obtained (e. g. multistability
phenomenon that can be observed for some ecosystems characterized by specific
sets of parameters) hold for all the growth laws listed above.
In fact, the low concentration version of the previous growth law, where 
$g^{(m)}_{\alpha} c_i \cdot n_j$ has been studied by one of us in context 
of autocatalytic growth of heteropolymers \cite{tkachenko2017onset}. 
The results of this paper are largely consistent with the present study, 
namely, in both cases the system has a large number of steady states
with between $\min(K,M)$ (corresponding to $Z$ in the notation of 
Ref. \cite{tkachenko2017onset}) and $K+M-1$ (corresponding to $2Z-1$).
Regarding why the number of bacterial species can not be larger than the total number of nutrients minus 1, one can prove for any form of the growth laws, that when all yields are the same, states with $K+M$ species have zero feasible volume, that is to say, they are only possible on a lower-dimensional manifold in the $(K+M)$-dimensional space of influxes (this results has been already discussed by Tilman in his special 
case \cite{tilman1982resource}.   
Multistability is also possible in the variant of the MacArthur model
\cite{macarthur1964competition,macarthur1970species,chesson1990macarthur} in which different species can have different yields when growing on the same nutrients. A convex Lyapunov function \cite{macarthur1970species} precluding multistability does not exist in this case. We leave this topic for future studies.

\subsection{Constraints on steady states from microbial and nutrient dynamics}\label{SI-Note2}
A steady state of equations describing the 
microbial dynamics (Eqs. \ref{eqn:dbdt}) is realized
when either $B_{\alpha}=0$ (the species was absent from the system from the start 
or subsequently went extinct) or when its growth rate $g_{\alpha}$ is 
exactly equal to the chemostat dilution rate $\delta$. This imposes constraints 
on steady state nutrient concentrations with the number of constraints 
equal to the number of microbial species present with non-zero concentrations. 
Since, in general, the number of constraints cannot be larger than the 
number of constrained variables, no more than $K+M$ of species could be 
simultaneously present in a steady state of the ecosystem. For Liebig's 
growth law used in this study, each resource can have no more than 
one species for which this resource limits its growth, that is to say, 
which sets the value of the minimum in $\min(\lambda_{\alpha}^{(c)} c_i,
\lambda_{\alpha}^{(n)} n_j$
The steady state concentrations of these resources are 
given by $c^{(*)}_i =\delta/\lambda_{\alpha}^{(c)}$ (if the growth is 
limited by the carbon source) 
and $n^{(*)}_j =\delta/\lambda_{\alpha}^{(n)}$ (if the growth is 
limited by the nitrogen source). Here $\alpha$ is the species 
whose growth is rate-limited by the 
resource in question. In a general case, no more than one species can be 
limited by the same resource (carbon in our example), since the species 
with the largest $\lambda^{(c)}$ would outcompete other 
species with smaller values of $\lambda^{(c)}$ by making the steady state concentration $c^{(*)}_i$ so low that other species can no longer grow on it. Note however, that multiple species $\beta$ could consume the same resource as the rate-limiting 
species $\alpha$, as long as their growth is not limited
by the resource. Each of these species must 
then be limited by 
their other nutrient (a nitrogen source in our example). However, 
their survival requires that carbon concentration set by species $\alpha$ is sufficient for their growth. Thereby, any species growing on a resource in a non-limited fashion must have $\lambda_{\beta}^{(c)} >\lambda_{\alpha}^{(c)}$.

Mathematically, it cane be proven by observing that, since species $\beta$ is limited by its nitrogen resource, one must have
$\lambda_{\beta}^{(c)} c^{(*)}_i >\lambda_{\beta}^{(n)} n^{(*)}_j$. 
At the same time in a steady state, the concentrations of all rate-limiting 
resources are determined by the dilution rate $\delta$ via 
$\lambda_{\beta}^{(n)} n^{(*)}_j=\delta$, and 
$\lambda_{\alpha}^{(c)} c^{(*)}_i=\delta$.  
Combining the above three expressions one gets:
$\lambda_{\beta}^{(c)} c^{(*)}_i >\lambda_{\beta}^{(n)} n^{(*)}_j=
\delta=\lambda_{\alpha}^{(c)} c^{(*)}_i$, or simply
$\lambda_{\beta}^{(c)} >\lambda_{\alpha}^{(c)}$.
The constraints on competitive abilities $\lambda$ 
for species present in a steady state in our model are then:
\begin{itemize}
\item Exclusion Rule 1: Each nutrient (either carbon or nitrogen source) 
can limit the growth of no more than one species $\alpha$.
From this it follows that the number of species co-existing in any given 
steady state cannot be larger than $K+M$, the total number of nutrients. 
\item Exclusion Rule 2: Each nutrient (e.g. specific carbon source) 
can be used by any number of species in a 
non-rate-limiting fashion (that is to say, where it does not constrain 
species growth in Liebig's law). However, any such species $\beta$ 
has to have $\lambda_{\beta}^{(c)} >\lambda_{\alpha}^{(c)}$, 
where $\lambda_{\alpha}^{(c)}$ is the competitive ability of 
the species whose growth is limited by this nutrient. 
In case of a nitrogen nutrient, the constraint becomes 
$\lambda_{\beta}^{(n)} >\lambda_{\alpha}^{(n)}$. 
\end{itemize}

Note that the steady state solutions of equations 
\ref{eqn:dbdt} do not depend on populations 
$B_{\alpha}$ of surviving species. Their steady state populations
$B^{(*)}_{\alpha}$ are instead determined by Eqs. \ref{eqn:dcdt}. 
%However, instead of explicitly solving these 
%equations one could invoke conservation 
%laws tracking each of the resources. 
%In a steady state the supply rate 
Taking into account that, in a steady state, 
the growth rate of each surviving  species is exactly 
equal to the dilution rate $\delta$ of the chemostat, 
after simplifications one gets:  
\begin{eqnarray}
\frac{\phi_j^{(c)}}{\delta} &=&c^{(*)}_i+
\sum_{\text{all }\alpha \text{ using }c_i}
\frac{B^{(*)}_{\alpha}}{Y_{\alpha}^{(c)}}
\nonumber \\
\frac{\phi_j^{(n)}}{\delta}&=&n^{(*)}_j- 
\sum_{\text{all }\alpha \text{ using }n_j}
\frac{B^{(*)}_{\alpha}}{Y_{\alpha}^{(n)}}
\label{eqn:ss_nc}
\end{eqnarray}
As described above, the steady state 
concentration of resources 
are given by $\delta/\lambda^{(c/n)}_{\alpha}$, where 
$\alpha$ are the species rate-limited 
by each resource. In the absence of such species, the 
concentration of a resource is given by anything left 
after it being consumed by surviving species in a non-rate-limiting 
manner. One can show that in this case, 
the resource (e.g. carbon) concentration has to be 
larger than $\delta/\lambda_{\beta}^{(c)}$, where 
$\lambda_{\beta}^{(c)}$ is the smallest affinity 
among microbes utilizing this resource.

One convenient approximation greatly simplifying working with 
equations \ref{eqn:ss_nc} is the "high-flux limit" in 
which $\phi^{(c)}_i \gg \delta^2/\lambda_{\alpha}^{(c)}$ and 
$\phi^{(n)}_j \gg \delta^2/\lambda_{\alpha}^{(n)}$. In this 
approximation one can approximately sets to zero the steady state concentrations of 
all resources that have a species rate-limited by them. The steady state 
concentrations of the remaining resources can take any value as long 
as it is positive. Hence, in this limit the 
equations \ref{eqn:ss_nc} can be viewed as a simple matrix 
test of whether a given set of surviving species limited by a 
given set of resources is possible for a given set of nutrient fluxes. 
Indeed, my multiplying the vector of fluxes with the 
inverse of the matrix $\hat{R}$ composed of inverse yields of surviving
species and 1 for nutrients not limiting the growth of any species one formally
gets the only possible set of steady state species abundances, $B^{(*)}_{\alpha}$, 
and a subset of non-limiting resource concentrations $c^{(*)}_i$ and $n^{(*)}_j$. 
If all of them are strictly positive - the steady state is possible. 
If just one of them enters the negative territory - the steady state 
cannot be realized for these fluxes of nutrients. 

The above rule can be modified to apply even below the 
high-flux limit with the following modifications:
1) Instead of $\phi^{(c)}$ (or $\phi^{(n)}$), one uses 
their ``effective values'' $\tilde{\phi}^{(c)}$ (or $\tilde{\phi}^{(n)}$) introduced in \cite{goyal2018multiple}, determined as 
\begin{eqnarray}
\tilde{\phi}^{(c)}_i&=&\phi^{(c)}_i-\frac{\delta^2}{\lambda^{(c)}_{\alpha(i)}} \nonumber \\
\tilde{\phi}^{(n)}_j&=&\phi^{(n)}_j-\frac{\delta^2}{\lambda^{(n)}_{\alpha(j)}} \quad ,
\label{eq:offset}
\end{eqnarray}
where $\alpha(i)$ is the (unique) species limited by the 
nutrient $i$. If the nutrient is not limiting for any os the  species in the steady state, 
$\alpha(i)$ is the species using the nutrient in a non-limited fashion, which has the {\it smallest} value of $\lambda$. This last rule comes from the observation that in order for a non-limiting resource not to become limiting for a species $\beta$ currently using it in a non-limiting fashion, its concentration cannot fall below 
$\delta/\lambda^{(x)}_{\beta}$. Thus, when checking the feasibility of a given state, the concentration of a non-limiting resource can be written as $\delta/\lambda^{(x)}_{\beta}+$ a positive number, or (more conveniently) the influx of this resource can be offset as described in Eqs. \ref{eq:offset}

% \subsection{The number of allowed and uninvadable states}

% \subsubsection*{Matching problem approach to 
% identifying and counting allowed 
% states}

\subsection{Stable matching approach to 
identifying and counting uninvadable 
states}\label{SI-Note3}

% \subsubsection*{The number of uninvadable states when
% there is exactly one microbe per each pair
% of resources}
First we describe the exact one-to-one mapping between 
all uninvadable steady states (UIS) 
in our model and the complete set of ``stable marriages''
in a variant of a well-known stable marriage or 
stable allocation problem developed by Gale and 
Shapley in the 1960s 
\cite{gale1962college} and awarded the Nobel 
prize in economics in 2012. This mapping provides 
us with constructive algorithms to identify and 
count all uninvadable steady states in our ecosystem.

We start by considering a special case of 
our problem with $L$ carbon and $L$ 
nitrogen sources and a pool of $L^2$ species,  
such that for every pair of sources $c_i$ (carbon) 
and $n_j$ (nitrogen) there is exactly one microbe 
$B_{ij}$ capable of using them.  For the sake of 
simplicity we have switched the notation 
from $B_{\alpha}$ to $B_{ij}$, where $\alpha=(ij)$
is the unique microbe in our pool capable of growing 
on $c_i$ and $n_j$.
Having considered this simpler situation we will return 
to the most general case of unequal numbers 
of carbon ($K$) and nitrogen ($M$) resources and any 
number of microbes from a pool of $S$ species competing 
for a given pair of resources.

In what follows we will refer to a resource as 
{\itshape occupied} if in a given steady state 
there is a microbe for
which this resource is rate-limiting.  In our 
network representation occupied resources have 
an outgoing edge (their out-degree is equal to 1), while 
unoccupied resources have out-degree equal to 0.  
%If such a
% directed edge goes from carbon $c_i$ to nitrogen $n_j$,
% we will say that $c_i$ {\itshape claims} $n_j$ and
% vice versa if the edge goes from nitrogen source to carbon source.
%
% An uninvadable state 
% in the hospitals/residents problem.  The latter is
% a version of the marriage problem, well known
% in the game theory for its multiple applications in
% the fields ranging from computational biology to economics
% to actual practice of resident/hospital or
% student/university assignment.

\subsubsection*{Review of results about stable matchings 
in the hospitals/residents problem}
The hospitals/residents problem \cite{gale1962college} 
is known in various settings.  The one directly relevant 
to our problem
is the following.  There are $L$ applicants for residency positions in 
$L$ hospitals.  A hospital number $i$ has
$L_{i}$ vacancies for residents to fill, $L_{i}$
ranging from zero to $L$.  Each hospital has a list
of preferences in which residency applicants are strictly
ordered by their ranks, from $1$ (the most desirable) to $L$ ,
(the least desirable).  These lists are generally different for different 
hospitals. Each applicant has a ranked list of preferred hospitals 
ranging from $1$ (the most desirable) to $L$ (the least desirable).
Those lists can also vary between applicants.
A {\itshape matching} is an assignment of applicants to hospitals 
such that all applicants got residency and all hospital vacancies are filled. 
A matching is {\itshape unstable}
if there is at least one applicant $a$
and hospital $h$ to which $a$ is not assigned 
such that:
\begin{enumerate}
\item Condition 1. Applicant $a$ prefers hospital $h$ to his/her assigned hospital;
\item Condition 2. Hospital $h$ prefers applicant $a$ to at least one of its assigned applicants.
\end{enumerate}
If such a pair $(a, h)$ exists, it is called ``a blocking
pair'' or ``a pair that blocks the matching''.
A {\itshape stable} matching by definition has
no blocking pairs.  
Gale and Shapley proved that for any set of applicant/hospital 
rankings and hospital vacancies there is at least one 
stable matching \cite{gale1962college}.
Generally the number of stable matching is larger than 
one. For example, for stable marriages and random rankings
the average number of stable matchings is given by 
$L/e \log L$ \cite{gusfield1989stable}. To the best 
of our knowledge, the dependence of this number on 
the distribution of hospital vacancies has not been 
investigated. 
The fact that the actual number of 
uninvadable states is rather close to its lower bound 
(compare black symbols and dashed line in Fig. \ref{fig1n})
indicates that, at least for $L \leq 9$, the number 
of stable matchings averaged over all possible 
in-degree allocations is rather close to 1.

Gale and Shapely not only proved the existence of at least 
one stable matching, but also proposed a constructive algorithm 
on how to find it. Listed below are the main steps in this algorithm
optimized for  for applicants.
each applicants first submits his/her application to the hospital 
ranking $1$ in his/her preference lists. Each hospital considers all 
applications it received so far and accepts all of the applicants 
if their number is less or equal than hospital's announced number of vacancies, 
$L_i$. If the number of applicants exceeds $L_i$, the hospital
gives a conditional admission to the
best-ranking $L_i$ applicants according to hospital's 
own preference list. Each applicant not admitted to their top hospital 
goes a step down on his/her preference
list and applies to the second-best hospital.  The latter admits
this applicant if (1) this hospital has not yet filled all of its vacancies 
or (2) all vacancies are filled, but among the conditionally admitted 
applicants there is at least one who ranks lower (according to hospital's list)
than the new applicant.  Such lower-ranked applicants 
are declined admission and replaced with better ones. They subsequently 
lower their expectations and apply to the next hospital on their list.  
After a number of iterations all applicants are admitted and 
all vacancies are filled so that this process stops.
As Gale and Shapley proved in Ref. \cite{gale1962college}, 
the resulting matching is stable. Furthermore, the theorem 
states that in this matching every applicant gets admitted 
to the best hospital among all stable matchings, while every hospital 
gets the worst set of residents among all stable matchings.
Later research described in Ref. \cite{gusfield1989stable}
describe more complex constructive algorithms allowing one to efficiently 
find all of the stable matchings starting with the applicant-optimal one.

Well developed mathematical apparatus of stable matching problem 
provides an invaluable help in the task of identifying all uninvadable
states in microbial ecosystems. Indeed, without its assistance 
this task would require exponentially longer time.
To connect the problem of finding all uninvadable states to that of
finding all stable matchings between hospitals and residents, 
we start with the following three observations:

1) In any uninvadable steady state, either all carbon
sources or all nitrogen sources (or both) are occupied.  Indeed, if
in a steady state a carbon source $c_i$ and a nitrogen source $n_j$ are not-limiting 
to any microbes, then microbe $B_{ij}$ can always grow and thereby invade 
this state.  Thus uninvadable states can be counted separately:
one first counts the states where all nitrogen sources are occupied, and 
then counts those in which all carbon sources are occupied.  
Double counting happens when both carbon 
and all nitrogen sources are occupied.  We will keep the possibility 
of double counting in mind and return to this problem later.

2) For a pool of species, where for every pair of resources 
there is exactly one microbe using each carbon and each nitrogen.
One can think of each of $L$ carbon (alternatively, nitrogen) 
sources as if it has a list of ``preferences''
ranking all nitrogen (correspondingly carbon) sources. 
Indeed, the ranking of competitive abilities $\lambda^{(c)}_{ik}$ 
of different microbes using the same carbon source $c_i$ but 
different nitrogen sources $n_k$ can be viewed as the ranking of 
nitrogen sources $k$ by the carbon source $i$.
Conversely, the ranking of $\lambda^{(n)}_{mj}$ 
with the same $n_j$ but variable 
$c_m$ can be thought of as ranking of 
carbon sources $c_m$ by the nitrogen source $n_j$.

% assigning number $1$ to the
% largest one, number $2$ to the second-largest, etc,
% down to the number $L$ for the smallest affinity.
% Set the number of $\lambda^{(c)}_{ik}$ thus obtained
% to be the rank of the nitrogen source $n_k$ in the
% list of preferences of the carbon source $c_i$.
% Generate lists of preferences for every source
% (each carbon ranking all the nitrogens, each nitrogen
% ranking all the carbons) in the similar way.  When
% all the sources are given their preference lists
% we make the third observation.  

3) Consider a steady state in which all nitrogen sources
are occupied. In our network representation it corresponds 
to every nitrogen source sending an outgoing link to some 
carbon source. Let $L_i$ be the number of microbes using 
the carbon source $i$ in a non-limiting fashion (the in-degree
of these outgoing links ending on $c_i$). Then, obviously, 
$L = \sum L_i$ (note that some of the terms in this sum 
might be equal to zero). 

One can prove that if the state is uninvadable, then 
the matching given by all edges going from nitrogen sources 
to carbon sources must be stable in the Gale-Shapley sense.
To prove this, let's think of nitrogen sources as ``applicants'' and nitrogen 
sources as ``hospitals'' with their numbers of 
``vacancies'' given by $L_i$. Indeed, any unstable matching has at least one 
blocking pair $(n_j,c_i)$ such that:
\begin{itemize}
\item Condition 1. The nitrogen source (`applicant'') $n_j$ ``prefers'' the carbon source (``hospital'') $c_i$ to its 
currently assigned carbon source (the one used by the current microbe $B_{kj}$ limited 
$n_j$). This means that $\lambda^{(n)}_{ij} > \lambda^{(n)}_{kj}$. 
Thus the microbe $B_{ij}$ can grow on its nitrogen
source (provided that it can also grow on its carbon source).
\item Condition 2. The carbon source (``hospital'') $c_i$ ``prefers'' the nitrogen source (``applicant'') $n_j$ to 
at least one of $L_i$ of its currently assigned carbon sources  (the set of microbes using $c_i$ in a non-rate-limiting fashion). Thereby $\lambda^{(c)}_{ij}$ must be 
larger than the smallest $\lambda^{(c)}$ among these microbes.
According to the Exclusion Rule 2, this smallest $\lambda^{(c)}$
is still larger than $\lambda^{(c)}$ of the microbe limited by $c_i$ 
(if it exists). Thus the microbe $B_{ij}$ can also grow on its carbon 
source.
\end{itemize}
This proves that the microbe $B_{ij}$ corresponding to any blocking pair
can grow on both its carbon and its nitrogen sources, and thereby can 
successfully invade the steady state. This finishes the proof that any 
uninvadable state has to be a stable matching in the Gale-Shapley sense.

However, this does not prove that any stable matching 
corresponds to exactly one uninvadable state.
To prove this we first notice that, up to this point, our candidate 
uninvadable state contained only the 
nitrogen-limited species. We will now supplement it 
with carbon-limited species in such a way that 
1) added species do not violate the exclusion rule 2; 
2) added species render the state completely uninvadable. 
Let is introduce a new notation (applicable to our case in 
which all nitrogen sources are occupied). 
Let $\lambda^{(c)}_{\textrm{min}}(i)$ to 
denote the smallest $\lambda^{(c)}$ among all 
species using $c_i$ in a non-rate-limiting fashion.
The Gale-Shapley theorem only guarantees the protection 
of our state from invasion by a species $(i,j)$ with 
$\lambda^{(c)}_{ij}$  
larger than $\lambda^{(c)}_{\textrm{min}}(i)$
(see the Condition 2 above). To ensure that 
our state is uninvadable by the rest of the 
species, one needs to add some carbon-limited species to 
this state. In order to do this in a systematic way, 
for each $c_i$ we compile the list of all species 
using this carbon source with $\lambda^{(c)}<\lambda^{(c)}_{\textrm{min}}(i)$. 
Each of these species is a potential invader.
Some species could be crossed off from the list 
of potential invaders because they cannot grow on their 
nitrogen source. 
These species have $\lambda^{(n)}$ below 
that of the (unique) species limited by their nitrogen source.
Among the species that remained on the list of 
invaders after this procedure, we select that with the largest 
$\lambda^{(c)}$ and add it to our steady state as 
a $C \to N$ directed edge, that is to say, as a 
carbon-limited species. This will prevent all 
other potential invaders on our list, since they 
have smaller $\lambda_{(c)}$ and thus, following the addition 
of our top carbon-limited species, they would 
no longer be able to grow based on their carbon source.
We will go over all $c_i$ and add such carbon-limited 
species if they are needed. The only scenario when 
such species is not needed if our list of potential invaders 
would turn up to be empty. In this case we will leave 
this carbon source unoccupied. Since for each carbon source 
the above algorithm selects the carbon-limited species (or selects 
to add no such species) in a unique fashion, there is a single  
uninvadable state for every stable matching in the 
Gale-Shapley sense. We are now in a position to predict 
and enumerate all uninvadable states in our model. 

\subsubsection*{Lower bound on the number of uninvadable states}
To count the number of partitions $(L_1, L_2, ..., L_L)$ such 
that $\sum L_i = L$, one can use a well known combinatorial 
method. According to this method, one introduces 
$L-1$ identical ``separators'' (marked with $|$) which are placed between 
$L$ identical objects (marked $\cdot$) separating them into $L$
(possibly empty) partitions. For example, for $L=4$ a partition 
$0,1,0,3$ would be denoted as $|\cdot||\cdot \cdot \cdot$. 
The combinatorial number of all possible arrangements of separators and 
objects is obviously ${{2L-1}\choose{L}}$.
For every such partition the Gale-Shapley theorem guarantees at least
one stable matching (that is, at least one
uninvadable steady state).
The lower bound on the number of uninvadable steady 
states has to be doubled to account for reversal of roles of 
carbons and nitrogens. There is a small possibility 
that we double counted one partition 
$(1, 1, ..., 1)$. Indeed, the unique uninvadable stable state corresponding to 
this partition could in principle be counted both when we start from nitrogen sources 
and when we start from carbon sources. This could happen only 
when the numbers of carbon
and nitrogen sources are equal to each other. More restrictively, 
this partition will be double-counted only if, when we started from C, 
all of the N-sources will send a link back to C, and these links all will end on different 
C-sources. The same has to be true if one starts with N-sources and 
at then sends links back to C. The steady state network in this case will consist of 
one or more loops covering all nutrients.
However, one can prove that, at least for the Gale-Shapley nitrogen-optimal state, 
the last carbon to be picked up would not need to send back a carbon-limited link.
Thus in our task of calculating the lower bound on the number of uninvadable states, 
we don't need to correct for the possibility of double-counting since at least 
one stable matching per partition (namely the Gale-Shapley) would not be double-counted.
Then we have  $N_{UIS} \geq 2{{2L-1}\choose{L}}={{2L}\choose{L}}$. 
The Sterling approximation for this expression is 
$2^{2L}/\sqrt{\pi L}$.  Thus the overall lower bound 
for the number of uninvadable stable states is given by 
\begin{equation}
  \label{NUIS lower bound SI}
  N_{UIS}(L,L) \ge \cdot {{2L}\choose{L}} \simeq \frac{2^{2L}}{\sqrt{\pi L}} \qquad .
\end{equation}

More generally, the number of carbon sources, $K$, is not equal to the number 
of nitrogen sources, $M$. The resource type with a larger number will always 
have at least one resource left without input. Thus here one never 
needs to correct for double counting. Using the same 
reasoning as for $K=M=L$, the lower bound on the number of resources in this case 
is given by ${{K+M-1}\choose{K-1}}+{{K+M-1}\choose{M-1}}={{K+M}\choose{K}}$.
Here, the first term counts the uninvadable steady states in which all nitrogen sources 
are occupied and the partition divides $M$ edges sent by nitrogen sources 
among $K$ carbon sources, which requires $K-1$ ``dividers''. 
The second term counts the number of uninvadable steady states 
in which all carbon sources are occupied. Denoting 
the fraction of carbon resources among all resources as $p=K/(K+M)$ and 
using the Stirling approximation one gets 
\begin{eqnarray}
  \label{NUIS hard case lower bound}
  &&N_{UIS}(K,M) \ge \cdot {{K+M}\choose{K}} \simeq \\ \nonumber 
&\simeq& \frac{\exp\left[(K+M)(-p \log p -(1-p) \log(1-p)\right]}{\sqrt{2\pi (K+M)p(1-p)}} \qquad .
\end{eqnarray}

% In a partition $(L_1, L_2, ..., L_L)$, $\sum L_i = L$,
% there can be from $1$ to $L$ components with the
% values greater than zero.  The number of partitions
% having $H$ nonzero summands is ${{L}\choose{H}}$,
% the ways to construct those $H$ summands (i. e. to
% split the number $L$ into $H$ ordered components)
% number to ${{L - 1}\choose{H - 1}}$.  For each
% such partition Gale and Shapley theorem guarantees at least
% one stable matching (that is, at least one
% uninvadable steady state), so the lower bound
% on the UIS number amounts to $\sum {{L}\choose{H}} \cdot
% {{L - 1}\choose{H - 1}} = {{2L - 1}\choose{L}}$.
% For large $L$ the Stirling's approximation
% gives the number proportional to $2^{2 L - 1} / \sqrt{L}$.
%
% As this is the number of the UIS with all nitrogens
% occupied (leaving at least one carbon free), it should
% be doubled to account for the states where all the
% carbons are occupied.   Therefore we are left with the
% lower bound on the number of the uninvadable
% steady states (NUIS), so that
% \begin{equation}
%   \label{NUIS lower bound}
%   NUIS \ge 2 \cdot {{2L - 1}\choose{L}}
% \end{equation}
% This (sharp) lower bound does not include the states
% in which
% both carbons and nitrogens are occupied as they
% may happen to be nonexistent (e. g. in the cases
% such that each $c_i$ ranks each $n_j$ the same as
% $n_j$ ranks $c_i$).  Such states necessarily contain
% ``loops'' in their graph representation.  Although
% in some special cases there may be quite a lot of UIS
% of this kind, we do not consider them in this
% subsection.

In the case of multiple microbial species using the same
pairs of resources, our version of the
Gale-Shapley resident-oriented algorithm must be further
updated.  Let $M$ be the number of nitrogen sources,
and $K$ --- the number of carbon sources in the ecosystem,
$S$ the number of species in our pool, each requiring a pair of
resources to grow.  As now there may be more than
one microbe that uses a given pair of resources 
$c_i$ and $n_j$, we introduce
the notation $B_{ij}^{(r)}$ for the $r$th microbe
using the same pair of sources $c_i$ and $n_j$.  On average, 
each nitrogen (carbon) source has $S/K$ ($S/M$) 
microbes, which are capable of using it.  
As in the traditional Gale-Shapley algorithm, 
each nitrogen (carbon) source ranks all microbes 
capable of using it by their $\lambda^{(n)}$
($\lambda^{(c)}$).  

The way to identify all uninvadable stable states 
in this case is determined by a variant of the 
stable marriage problem (or rather the hospital/resident 
problem) in which every man (and every woman) 
may have more than one way to propose marriage to 
the same woman (man). In our model this corresponds 
to more than one microbe (a type of marriage) capable 
of growing on the same pair of carbon 
(corresponding to, say, men) 
and nitrogen (corresponding to women) sources. 
You may think of it as if each participant has 
several different ways to propose to the person of the opposite sex
(send flowers, take to a restaurant, etc). 
Each of these proposals is ranked by both parties 
independent of other ways. 
As far as we know, this variant has not been 
considered in the literature yet. However, all of the
results of the usual stable marriage (or hospital-resident)
problem remain unchanged. 

One can easily see that our lower bound (Eq. \ref{NUIS hard case lower bound}) 
on the number of uninvadable states (equal to the number of 
stable marriages in all partitions) remains unchanged. Indeed, it is given by 
the number of partitions and hence depends only on $K$ and $M$ and not 
on $S$. However, for $S \gg K \cdot M$ one expects to have 
many more stable marriages 
%(states with ``loops'') 
for each partition. Thus the lower bound we have established 
is likely to severely underestimate 
the actual number of UIS in the ecosystem.
Indeed, according to the SI section ``The number of uninvadable states in a continuous 
approximation'', the number of uninvadable states 
grows much faster than the lower bound of 
$O(2^{K+M})$.
Future work is needed to connect the stable marriage 
results to those derived in the 
continuous approximation (see Eq. \ref{eq:N_UIS_sp_formula_SI}
below).

% \newpage
% \section{LEAVE THIS FOR FUTURE PAPERS}
\subsection{The number of allowed states and the number of uninvadable states in a continuous 
approximation}\label{SI-Note4}:
We can calculate the number of allowed and, separately, uninvadable states 
in our model in the limit of $K,M \gg 1$ and $S \gg K, M$. 
In this limit, every nutrient has a large average number of microbes 
competing for its utilization.
Throughout this section we assume that each of the resources 
has equal number of microbes capable of using it 
($S/K$ for carbon and $S/M$ for nitrogen). 
Let $r^{(c)}_i$ (respectively $r^{(n)}_j$) be the rank of the (unique if 
present) microbe whose growth is limited by $c_i$ (respectively $n_j$).
The rank is defined as the the number of microbes in 
the pool with value of $\lambda$ larger or equal than $r$. Hence, 
in our pool of species, the most competitive microbe 
for each nutrient has the rank $1$, while the worst one - 
the rank $S/K$ (or $S/M$ for nitrogen resources).
It is convenient to assume that in a special case, where there 
is no microbe limited by the resource, the rank of the resource 
is equal to $S/K+1$ ($S/M+1$ correspondingly). 
In this case all microbes using this 
nutrient would be allowed to grow on it according to our competitive 
exclusion rules. It is also convenient to normalize the ranks as 
$0 \leq x_i=(r^{(c)}_i-1)/(S/K) \leq 1$ and 
$0 \leq y_j=(r^{(n)}_j-1)/(S/M) \leq 1$. 
These normalized 
variables quantify the probability that a randomly selected 
microbe using a given nutrient ($c_i$ for $x_i$ and $n_j$ for $y_j$) 
would be able to grow on it (provided that the second resource 
would also allow for its growth). To calculate the probability $x$ that 
a randomly selected microbe would be able to grow on its carbon source
one has to average $x_i$ over all carbon sources:
$x=\sum_i x_i/K$. Similarly, the probability for a random 
microbe to be able to grow on its nitrogen source 
is given by $y=\sum_j y_j/M$. In what follows we will carry out the 
summation over all possible values of all normalized 
ranks of carbon, $x_i$ ($0:1/(S/K):1$), and nitrogen,
$y_j$ ($0:1/(S/M):1$), sources. In the continuous limit $K/S, M/S \ll 1$ 
these sums can be replaced by integrals over continuous variables 
ranging between $0$ and $1$. For $K \gg 1$, the average rank $x$ 
of all carbon sources has an approximately Gaussian distribution 
with width $1/\sqrt{12K}$, while the average rank $y$ of all 
nitrogen sources has a Gaussian distribution with width $1/\sqrt{12M)}$. Indeed, 
the variance of the uniform distribution between $0$ and $1$ is $1/12$, 
while the variance of the average is reduced by the number of 
variables in the sample. Some of our calculations require 
knowledge of the probability density function outside the region of 
validity of central limit theory. We have also carried out calculations using the exact 
PDF of the sum of $K$ (or $M$ in case of nitrogen) uniformly distributed 
variables known as Bates distribution (see Eq. (2) in 
\begin{verbatim}
 http://mathworld.wolfram.com/
UniformSumDistribution.html
\end{verbatim}
for the exact functional form of the PDF of the Bates distribution).
The results for the Bates distribution were very close to those for 
the Gaussian distribution. 
% In the range $1<K,M \leq 9$ we found little difference between 
% our results using Bates distribution and the Gaussian approximation to it. 
To account for significant difference at $K=M=1$, Fig. \ref{fig1n} 
shows our calculations using the Bates distribution (red and black 
long-dashed lines). Hence, in what follows we will consider only 
the Gaussian case.

\subsubsection*{The number of allowed states in a continuous 
approximation}
Let us first calculate the number of allowed states. 
Consider a state in which $0 \leq K_L \leq K$ of carbon sources and 
$0 \leq M_L \leq M$ of nitrogen sources each have (a unique) microbe 
limited by them (and making them limited). The combinatorial 
number of choices of such microbe-limited resources is given by 
$\binom{K}{K_L} \cdot \binom{M}{M_L}$. 
The number of ways to choose one limiting 
microbe on each of these resources is
given by $S/K$ for carbon resources and $S/M$ for nitrogen resources. 
Indeed, since each species has exactly one carbon (nitrogen) 
source it  could utilize, the number of species per each resource is 
$S/K$ ($S/M$ correspondingly).
The total number of ways to choose 
$K_L+M_L$ microbes in the candidate steady state 
is thus given by $(S/K)^{K_L} \cdot (S/M)^{M_L}$. 
The probability that all these microbes 
would be allowed by their non-limiting resources is given by 
$y^K_L \cdot x^M_L$ (note that the average rank $y$ of 
nitrogen resources is raised to the power of $K_L$ of 
limited carbon sources and vice versa). Indeed, the selection 
of a non-limiting resource is entirely random when one goes over all
possible microbe candidates. The sum over 
all possible values $K_L$ and $M_L$ 
%$0 \leq K_l \leq K$ and 
%$0 \leq M_l \leq M$ 
is simply given by:
\begin{eqnarray}
N_{AS}(x,y)&=&\sum_{K_L=0}^{K} \sum_{M_L=0}^{M} 
\left(\frac{S}{K}\right)^{K_L}\left(\frac{S}{M}\right)^{M_L} \cdot \nonumber \\
&\cdot & \binom{K}{K_L} \binom{M}{M_L} y^K_L \cdot x^M_L \nonumber \\
&=& (1+\frac{S}{K}y)^K \cdot (1+\frac{S}{M}x)^M \qquad .
\end{eqnarray}
Thus the total number of the allowed states (invadable or not) is given 
by the following integral:
\begin{eqnarray}
&&N_{AS}= \nonumber \\
&=&\int_{0}^{1} dx\sqrt{\frac{12K}{2\pi}}
\exp(-6K(x-1/2)^2) (1+\frac{S}{M}x)^M \cdot \nonumber \\ 
&\cdot&\int_{0}^{1}dy\sqrt{\frac{12M}{2\pi}}\exp(-6M(y-1/2)^2)
(1+\frac{S}{K}y)^K \qquad .
\label{eq_n_as_integral}
\end{eqnarray}
This integral can be calculated using the saddle point approximation. 
In the limit $M \sim K \gg 1$ and $M/S \ll 1$, 
the saddle point $x^*$ for the first integral over $x$ is given by 
\begin{equation}
x^*=\frac{1+\sqrt{1+\frac{4M}{3K}}}{4}=\frac{1}{4}(1+\rho_x) \quad .
\end{equation}
Here in order to simplify the notation we introduced a new variable 
$\rho_x=\sqrt{1+\frac{4M}{3K}}$.
The integral over $x$ in the saddle point approximation is then given by 
\begin{equation*}
I_x=\sqrt{\frac{1+\rho_x}{2\rho_x}}\left(\frac{1+\rho_x}{4} \cdot \frac{S}{M}+1\right)^M
\exp \left(-\frac{3K}{8}(\rho_x-1)^2\right) \qquad. 
\end{equation*}
Similarly, the integral over $y$ in the saddle point 
approximation is then given by 
\begin{equation*}
I_y=\sqrt{\frac{1+\rho_y}{2\rho_y}}\left(\frac{1+\rho_y}{4} \cdot \frac{S}{K}+1\right)^K
\exp\left(-\frac{3M}{8}(\rho_y-1)^2\right) \qquad. 
\end{equation*}
Here, $\rho_y=\sqrt{1+\frac{4K}{3M}}$ and is related to $\rho_x$ by 
\begin{equation}
(\rho_x^2-1)\cdot(\rho_y^2-1)=\frac{16}{9} \quad .
\end{equation}
The number of allowed steady states $N_{AS}$ is simply  
the product of $I_x$ and $I_y$. In the symmetric limit of 
the equal number of nutrient sources $M=K=L$ where $\rho_x=\rho_y=
=\rho=\sqrt{7/3} \simeq 1.53$
the formula for the 
number of steady state can be simplified as 
\begin{eqnarray}
&&N_{AS}=I_x \cdot I_y=\nonumber \\
&=&\frac{1+\rho}{2\rho}\left[\left(\frac{1+\rho}{4} \cdot \frac{S}{L}+1\right)
\exp\left(-3(\rho-1)^2/8 \right)\right]^{2L} \simeq \nonumber \\
&\simeq&0.827\left(0.569\frac{S}{L}+0.901\right)^{2L}
\label{eq:nas}
\end{eqnarray}
As one can see the number of allowed states rapidly 
increases with both the number of 
resources of each type, $L$, 
as well as with number of species 
per each resource, $S/L$.
This increase however is much slower than 
that in the number of candidate states
not constrained by the exclusion rule 2.
Indeed, the number of such candidates 
$N_{c}=(1+S/K)^K\cdot (1+S/M)^M$, which 
for $K=M=L$ becomes
$(S/L+1)^{2L}$ (compare this expression to 
Eq. \ref{eq:nas}).

Finally, for $S=L^2$ used in our simulations shown 
in Fig. \ref{fig1n}, the expression for the number of allowed states becomes
$N_{AS}\simeq0.827\left(0.569L+0.901\right)^{2L}$.

\subsubsection*{The number of uninvadable states in a continuous 
approximation}
To calculate the number of uninvadable states, one needs to check if 
each of the allowed states calculated in the Eq. \ref{eq_n_as_integral}
can be invaded by each of the species that are currently not present in the state. 
Fortunately, the notation introduced in the previous section makes this task very easy. 
When calculating the number of allowed states in our model we were going over all 
species present in the state and multiplying 
our formula by the probability that it's {\it non-limiting} resource is 
allowed by our rules of competitive exclusion. This probability is equal to 
$y = \sum_j y_j/M$ for species limited by the concentration of their carbon sources 
and  $x = \sum_i x_i/K$. Here (as before) $x_i$ and $y_j$ are the normalized ranks 
of the species limited by $c_i$ and $n_j$ correspondingly.
$x_i=0$ (or $y_j=0$) corresponds to a situation where there are no species in our pool 
with $\lambda$ larger then the species currently limited by this carbon (or nitrogen) 
source. Conversely,  $x_i=1$ corresponds to a situation where the resource is currently 
not limiting for any of the species in the steady state under consideration. Hence the 
growth of any introduced species on this resource is allowed by the competitive exclusion 
rules. $x_i=1/2$ corresponds to a case where the species with a median value of $\lambda^{c}$ 
is limiting $c_i$ so that exactly half of all species using this resource can grow on it 
(provided that their nitrogen source allows for growth). A species not currently present 
in the steady state can grow in it if and only if {\it both} its carbon and nitrogen sources 
allow for its growth. The probability of this being true for a randomly selected species 
is simply $x \cdot y$, while the probability that it is not allowed to grow by either 
one or both of its nutrients is simply $1-x \cdot y$. The probability that none among $S$ 
species can grow in a given steady state is 
\begin{equation}
\textrm{Prob(State is Uninvadable)}=(1-x y)^S \quad .
\end{equation}
Here in principle we only need to check uninvadability against 
$S-K_L-M_L$ species {\it not present} in the steady state tested for invadability. 
However, in the limit $S \ll K, M$, this difference is small and would be ignored
in our calculations.

\begin{figure}
\centerline{\includegraphics[width=\linewidth]{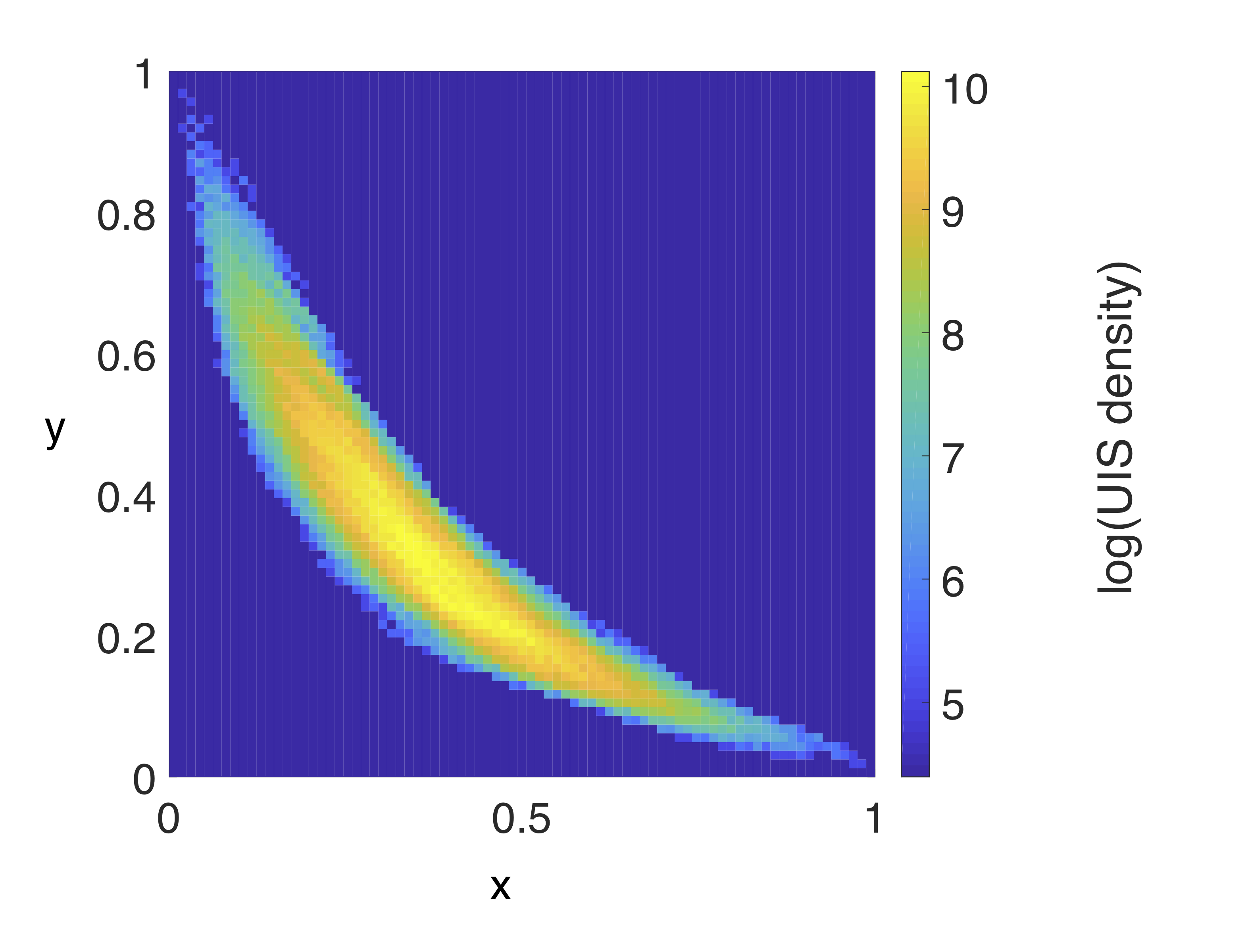}}
\caption{{\bf The density of UIS for $L=9$.} 
The pseudocolor plot shows the natural logarithm of 
the density of uninvadable stable states on the $x-y$ plane, 
where $x$ and $y$ are, respectively, the average normalized ranks of 
carbon- and nitrogen-limited microbes across all resources. }
\label{fig1S_A}
\end{figure}

Thus the number of uninvadable states is given by an integral 
similar to Eq. \ref{eq_n_as_integral}:
\begin{eqnarray}
&&N_{UIS}= \int_{0}^{1} dx \int_{0}^{1}dy \nonumber \\
%\exp(S \cdot \mathcal{L}(x,y))= \nonumber \\
%&=&\int_{0}^{1} dx \int_{0}^{1}dy 
&&(1-xy)^S \nonumber \\
&\cdot& (1+\frac{S}{M}x)^M (1+\frac{S}{K}y)^K \nonumber \\ 
&\cdot&\sqrt{\frac{12K}{2\pi}}\exp(-6K(x-1/2)^2) \nonumber \\
&\cdot& \sqrt{\frac{12M}{2\pi}}\exp(-6M(y-1/2)^2)
\quad .
\label{eq_n_uis_integral}
\end{eqnarray}
The density of stable states on the $x-y$ plane 
described by this equation can be visualized already 
%for relatively small value of 
$L=9$.
%, where our saddle point approximation does not yet work very well. 
For tables of $\lambda^{(c)}$ and  $\lambda^{(n)}$ used in our 
main text we exhaustively identified 81,004 UIS. For each of these UIS we 
calculated $x_i$ and $y_i$ - the average normalized ranks of microbes limited by their 
carbon and nitrogen sources respectively. The natural logarithm, of the 
density of these 81,004 points on the $x-y$ plane is visualized in 
Fig. \ref{fig1S_A}. One can see that most states are localized within 
a ``smile'' stretching from the upper left corner (the top competitors for carbon and 
the weakest competitors for nitrogen) to the lover right 
corner (the top competitors for nitrogen and 
the weakest competitors for carbon)
of the diagram. That means that the average ranks of carbon and 
nitrogen competitive abilities of microbes present in steady
states and limited by these resources are negatively correlated.

The probability of the state being uninvadable couples the integration 
over $x$ and $y$ in the Eq. \ref{eq_n_uis_integral}. This integral, when
written as $N_{UIS}=\int_{0}^{1} dx \int_{0}^{1} dy \exp(S \cdot \mathcal{L})$, 
in the limit of large $S \gg K, M$
can be approximately calculated in the saddle point approximation. 
Here 
\begin{eqnarray}
&&\mathcal{L}(x,y)=\log(1-x\cdot y)+ \nonumber \\
&+&\frac{K}{S}\log(1+\frac{S}{K}y)+ \frac{M}{S}\log(1+\frac{S}{M}x)- \nonumber \\
&-&6\frac{K}{S}(x-1/2)^2-6\frac{M}{S}(y-1/2)^2 +\nonumber \\
&+&\frac{1}{2S}\Big(\log{12M}+\log{12K}-2\log 2\pi \Big)\qquad .
\label{eq:L}
\end{eqnarray}

The integral of Eq. \ref{eq_n_uis_integral} over $x$ is dominated 
by the saddle point $x^*(y)$ obtained by 
solving for x the following equation
\begin{eqnarray}
0=\frac{\partial \mathcal{L}}{\partial x}=-\frac{y}{1-xy}+\frac{1}{1+Sx/M}-12\frac{K}{S}(x-\frac{1}{2}) \qquad.
\end{eqnarray}
We will be interested in solving this equation in the regime 
where $x \simeq 1$ (hence $1/(1+Sx/M) \simeq M/(Sx)$), and $y \ll 1$ (hence $-y/(1-xy) \simeq -y$. 
In this limit one gets a quadratic equation for $x$
\begin{eqnarray}
0=x^2+x\left(\frac{Sy}{12K}-\frac{1}{2}\right)+\frac{M}{12K} \qquad.
\end{eqnarray}
The solution is given by 
\begin{equation}
x^*(y)=\frac{1}{4}\left(1-\frac{Sy}{6K}+\sqrt{\left(1-\frac{Sy}{6K}\right)^2+\frac{4M}{3K}} \right) \qquad .
\label{eq:x*}
\end{equation}
For $M=K$ the maximal value of $x^*(y)$ is reached 
at $y=0$ and is equal to $x^*(0)=(1+\sqrt{7/3})/4=
0.6319$. For large excess of the number of nitrogen sources over that of carbon ones, $M>6K$, $x^*(y)$ can 
reach its maximal value of $1$ even before $y$ hits 0.
In this case, the saddle point disappears and the integral will be dominated by region near $x=1$. 
We will leave for future studies the calculation of the number of uninvadable steady states in this case.
In the limit $M/S \ll 1$, $K/S \ll 1$ the second derivative evaluated at this stable point is 
\begin{equation}
\frac{\partial^2 L}{\partial x^2} \rvert_{x^*(y)}=-12\frac{K}{S}-y^2-\frac{M}{Sx(y)^2} \quad .
\end{equation}

The saddle point integration over $x$ results in 
the following expression for the number of 
uninvadable states
\begin{equation}
N_{UIS}=\int_{0}^{1} dy \quad \exp(S \cdot \mathcal{L^*}(y))
\label{eq:sp_integral_y}
\end{equation}
where 
\begin{eqnarray}
&&\mathcal{L^*}(y)=\log(1-x^*(y)\cdot y)+ \nonumber \\
&+&\frac{K}{S}\log(1+\frac{S}{K}y)+
\frac{M}{S}\log(1+\frac{S}{M}x^*(y))- \nonumber \\
&-&6\frac{K}{S}(x^*(y)-1/2)^2-6\frac{M}{S}(y-1/2)^2+\nonumber \\
&+&\frac{1}{2S}\Big(\log 12M +\log 12K -2\log 2\pi \nonumber \\
%&-&\frac{1}{2S}\left(
&-&\log(12K+Sy^2+M/x^*(y)^2)+\log 2 \pi \Big)
%\right)
\qquad .
\label{eq:L*}
\end{eqnarray}
Here the last term $\Delta \mathcal{L^*}=(1/2S)
(-\log (12K+Sy^2+M/x^*(y)^2)+\log 2 \pi)$ comes from the saddle point integral 
over $x$. In other words, $\mathcal{L^*}(y)=\mathcal{L}(x^*(y), y)+\Delta \mathcal{L^*}(x^*(y), y)$, where 
$\mathcal{L}$ is defined by the Eq. \ref{eq:L}.
% \begin{equation}
% \Delta \mathcal{L^*}=\frac{1}{2S}\left(\log\left(12K+Sy^2+\frac{M}{x^*(y)} \right)
% \end{equation}
The integral over $y$ has two saddle points:
one for small $y \sim K/S$ and hence large $x(y) \sim 1$ and the other for large 
$y \sim 1$ and small $x(y) \sim M/S$. We will calculate only the first saddle point and 
then apply symmetry arguments to extend these results to the second one. Indeed, 
if instead of integrating out $x$ we were to integrate out $y$ first, 
the order of saddle points will change places. Hence, we are 
interested only in the region of small $y \sim K/S$.
The saddle point is determined by $\frac{d\mathcal{L^*}(x^*(y),y)}{dy}=0$. 
Let's first calculate the derivative of the last term (referred to as 
$\Delta \mathcal{L^*}$) in the Eq. \ref{eq:L*}. It is given by 
$d\Delta \mathcal{L^*}/dy=[M(dx^*(y)/dy)/x^*(y)^2-Sy]/[S(12K+M/x^*(y)^2+Sy^2)]$. 
The first term in the enumerator and the first two terms in the denominator 
dominate the expression for $y \sim K/S$ resulting in
\begin{equation}
\frac{d\Delta \mathcal{L^*}}{dy}=\frac{M \frac{dx^*(y)}{dy}/x^*(y)^2}{S(12K+M/x^*(y)^2)} \qquad .
\end{equation}
While $dx^*(y)/dy \sim S/K$ is large, the whole 
expression is still of order of $1/M$ or $1/K$ and hence much smaller than 1. As we will see below, the  
dominant contribution to $d\mathcal{L^*}/dy$ is of order of 1. Hence, $d \Delta \mathcal{L^*}/dy$ 
can be safely ignored. One then has $d\mathcal{L^*}/dy=
d\mathcal{L}(x^*(y),y)/dy=\partial \mathcal{L}(x^*(y),y)/\partial y+\partial \mathcal{L}((x^*(y),y)/\partial x \cdot dx^*/dy$. 
Since the saddle point integration over $x$ 
required that $\partial \mathcal{L}(x^*(y),y)/\partial x=0$, the second term is zero. The first term 
is given by 
$\partial \mathcal{L}(x^*(y),y)/\partial y=-x(y)/(1-x(y)y)+1/(1+(S/K)y)-12(M/S)(y-1/2)$.
In the limit $y \sim K/S$ and $x(y) =O(1)$, the first two terms are of order of 1, while 
the third term can be ignored. Furthermore, the denominator in the first term can be ignored. 
Hence, the saddle point $y^*$ is determined by the following condition
\begin{equation}
\frac{1}{1+(S/K)y^*}= x(y^*) \qquad .
\end{equation}
By introducing the dimensionless variable 
$\tilde{y}^*=(S/K)y^*$ and plugging it into the Eq. 
\ref{eq:x*} one gets 
\begin{equation}
\frac{1}{1+\tilde{y}^*} =\frac{1}{4}\left(1-\frac{\tilde{y}^*}{6}+
\sqrt{\left(1-\frac{\tilde{y}^*}{6}\right)^2+\frac{4M}{3K}} \right).
\label{eq:y*}
\end{equation}
While in general this equation does not have the 
analytic solution, it can be easily solved numerically for any value of $M/K<6$ 
(the saddle point disappears for $M>6K$). To fit our numerical 
simulations of the model with $M=K=L$ and $S=L^2$, we solved 
Eq. \ref{eq:y*} for $M/K=1$:
\begin{eqnarray}
y^*&=&0.71428 \frac{K}{S}\qquad , \label{eq:final_y*}
\\
x^*&=&0.58333 \qquad .
\label{eq:final_x*}
\end{eqnarray}

Note that both $x^*$ and $y^*$ are far away from $1/2$ and hence are 
located where the Gaussian approximation to the 
sum of uniformly-distributed random numbers no longer applies. 
However, the Gaussian was not among the main factors 
deciding the position of the fixed point.
%(its contribution 
%$\sim K$ is much larger than two other 
%terms $\sim S$). 
Thus, the results derived above could still be used.
We confirmed it by carrying the saddle point calculations 
in Eq. \ref{eq:sp_integral_y} using the exact form of the 
Bates distribution describing the sum of $M$ uniformly-distributed random 
numbers. Much as for the number of allowed states, for large $L$ the 
number of uninvadable states calculated using the Bates distribution was very 
close to the same number calculated using the Gaussian distribution.

To further verify the accuracy of our calculations 
of $x^*$ and $\tilde{y}^*$ in Eqs. \ref{eq:final_x*} and 
\ref{eq:final_y*} correspondingly, in Fig. \ref{fig1S_C} 
we plotted $S\mathcal{L}(x,y)$ as a function of $x$ and $y$
for $K=M=L$ and $S=L^2$. The red dot marking the predicted position 
of the saddle point is in excellent agreement with its 
numerically-determined position.
\begin{figure}
\centerline{\includegraphics[width=\linewidth]{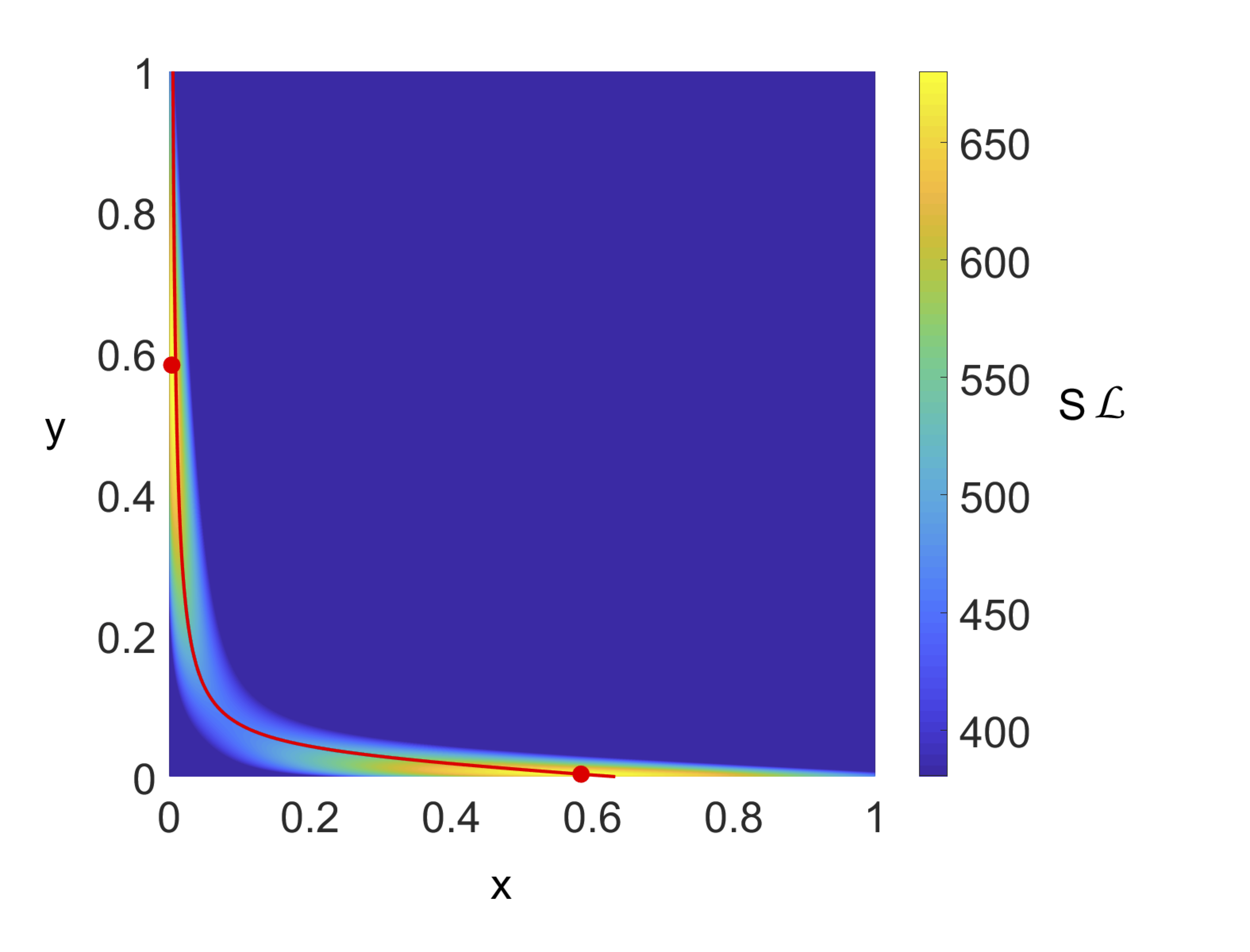}}
\caption{{\bf The logarithm of the density of UIS for $L=200$ 
along with the saddle point line $x^*(y)$ vs $y$.} 
The pseudocolor plot shows $S\mathcal{L}(x,y)$ - the logarithm of the density 
of UIS on the $x-y$ plane calculated for $L=200$ as a 
function of normalized carbon and nitrogen average ranks $x$ and $y$. 
The red line follows $x^*(y)$ vs $y$ described by 
the Eq.\ref{eq:x*}. 
The red dots mark the predicted positions of two saddle points 
according to Eqs. \ref{eq:final_x*} and 
\ref{eq:final_y*} and a symmetric 
one with $x$ and $y$ swapped places.
}
\label{fig1S_B}
\end{figure}
\begin{figure}
\centerline{\includegraphics[width=\linewidth]{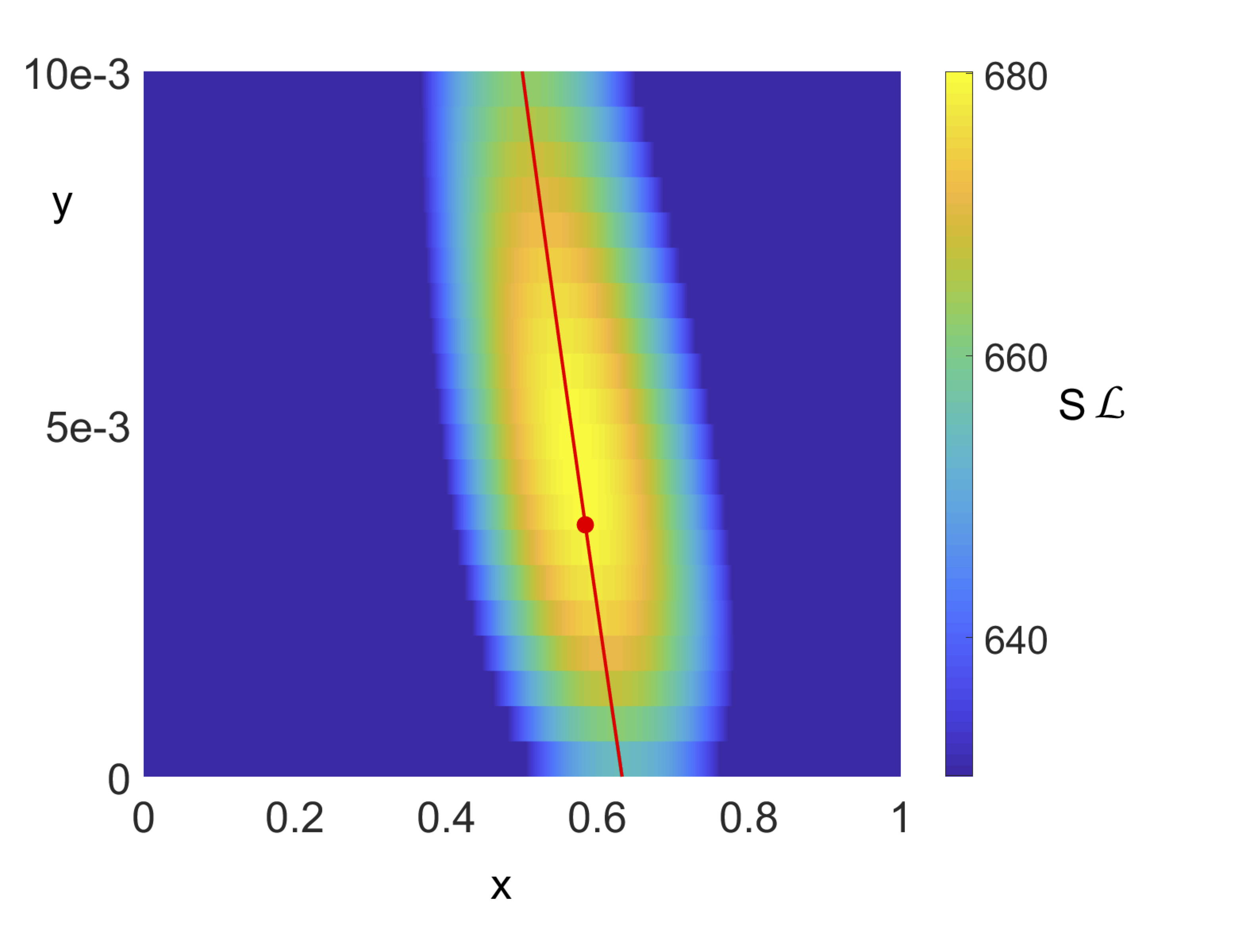}}
\caption{{\bf The logarithm of the density of UIS for $L=200$ 
along with the saddle point position.} 
The pseudocolor plot shows $S\mathcal{L}(x,y)$ - the logarithm of the density 
of UIS on the $x-y$ plane calculated for $L=200$ as a 
function of normalized carbon and nitrogen average ranks $x$ and $y$. 
The red dot marks the predicted position of the saddle point 
according to Eqs. \ref{eq:final_x*} and 
\ref{eq:final_y*}. It is in excellent agreement with its 
numerically-determined position.
}
\label{fig1S_C}
\end{figure}
\begin{figure}
\centerline{\includegraphics[width=\linewidth]{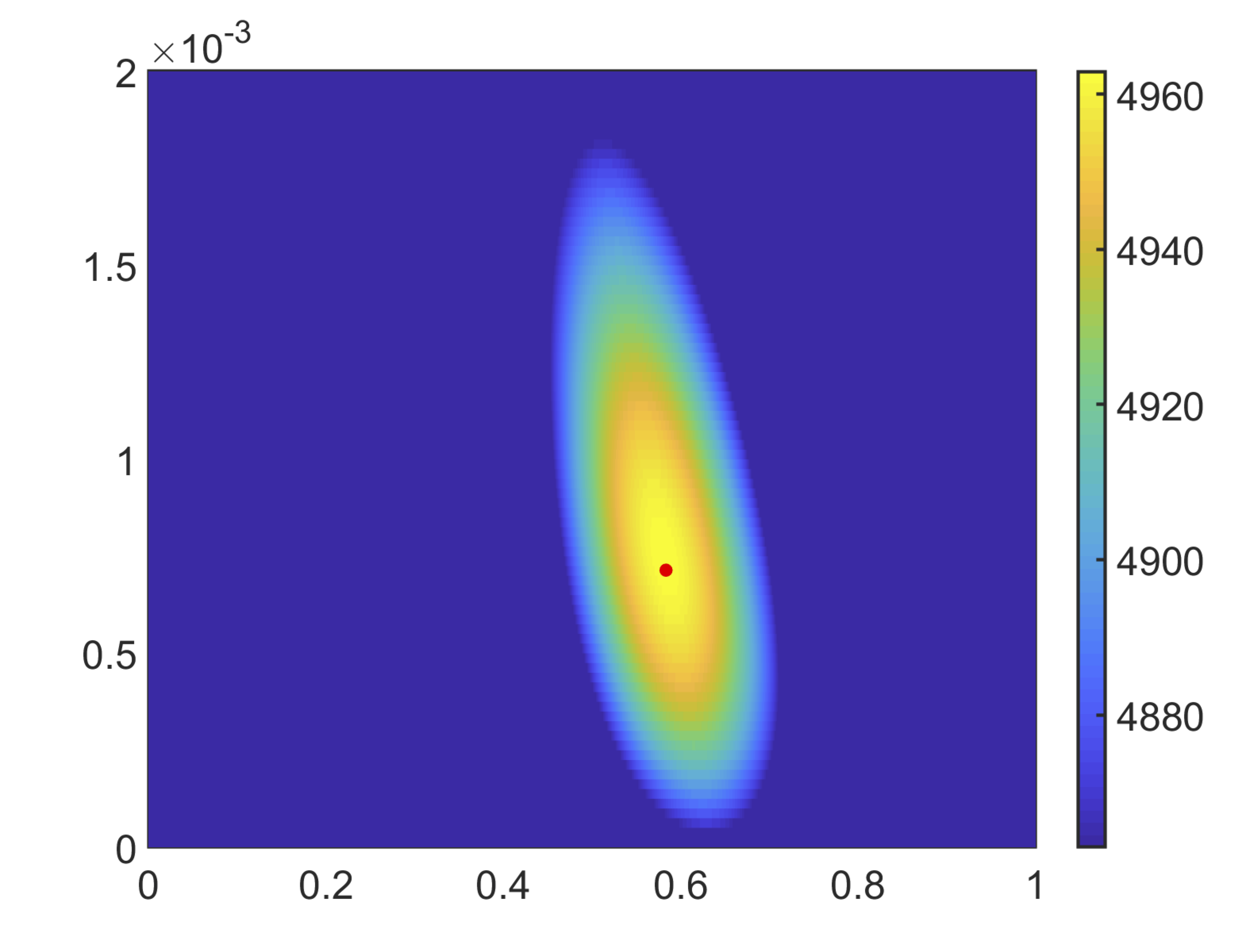}}
\caption{{\bf The density of UIS for $L=1000$ 
along with the saddle point position.} 
The pseudocolor plot shows $S\mathcal{L}(x,y)$ calculated 
for $L=1000$ as a function of $x$ and $y$. 
The red dot marking the predicted position of the saddle point 
according to Eqs. \ref{eq:final_x*} and 
\ref{eq:final_y*} is in excellent agreement with its 
numerically-determined position.
}
\label{fig1S_D}
\end{figure}

Hence, for large $L$ most uninvadable states come from two regions on 
the $x-y$ plane. Let us briefly return to the original notation for which the 
rank of the most competitive microbe using each nutrient is $1$, 
while that of the least competitive one is $S/K$ for carbon sources and $S/M$ for nitrogen sources.
Uninvadable states contributing to the saddle point  
above must have many microbes near the top of the nitrogen ranking table 
with average nitrogen rank being $1+\tilde{y}^*= 1.7143$. This can be realized e.g.
if around 71\% of microbes were the second best competitors 
for their nitrogen resource, while about 29\% were its best competitors In the whole pool. 
There are of course many other solutions all giving the average ranking shown above. On the 
opposite side, the average rank of microbes on their carbon resources is in the middle of the table
$x^*(S/K)=0.5833 L$. The other saddle point corresponds to carbon and 
and nitrogen swapping places with each other.

The second derivative $d^2\mathcal{L^*}/dy^2 \simeq 
d(\partial \mathcal{L^*}/\partial y)/dy=
\partial^2 \mathcal{L^*}/\partial y^2+ \partial^2 \mathcal{L^*}/\partial y \partial x \cdot (dx^*/dy)$. 
Here again we ignored $\Delta \mathcal{L^*}$ 
since its contribution to the derivative is $1/K$ smaller than the above terms.
The first term is given by $\partial^2 \mathcal{L^*}/\partial y^2=-x^*(y)^2/(1-yx^*(y))^2-(S/K)/(1+(S/K)y)^2$. 
Here the second term is much larger. One also has $\partial^2 \mathcal{L^*}/\partial y \partial x=-1/(1-yx^*(y))-yx^*(y)/(1-yx^*(y))^2$. Here the first term is much larger. Given that 
$dx^*(y)/dy \sim S/K$, the dominant 
contributions to $d^2\mathcal{L^*}/dy^2$ from $
\partial^2 \mathcal{L^*}/\partial y^2$ and 
$\partial^2 \mathcal{L^*}/\partial y \partial x \cdot (dx^*/dy)$ are comparable to each other and 
are both of order of $S/K$.
The derivative $dx^*/dy$ is given by
\begin{equation}
\frac{dx^*}{dy}=-\frac{S}{24K}\left(1+\frac{1-\frac{Sy}{6K}}{\sqrt{(1-\frac{Sy}{6K})^2+\frac{4K}{3M}}}\right) .
\end{equation}
Evaluating this expression for $K=M$ at 
$y^*=0.71428K/S$ one gets 
$\frac{dx^*}{dy}|_{y^*}=1.6066$.
Hence, the final expression for 
$d^2\mathcal{L^*}/dy^2$ at the saddle point is
\begin{eqnarray}
\frac{d^2\mathcal{L^*}}{dy^2}|_{y^*}&=&-\frac{S}{K} 
\left( \frac{1}{(1+y^*S/K)^2}+\frac{1}{24}\frac{dx^*}{dy}|_{y^*}\right)= \nonumber \\
&=&-0.2733\frac{S}{K} \qquad .
\label{eq:sp_dydy}
\end{eqnarray}

Plugging Eqs. \ref{eq:final_y*},\ref{eq:final_x*},\ref{eq:sp_dydy} 
into the saddle point estimate of the integral given by 
Eqs. \ref{eq:sp_integral_y},\ref{eq:L*} one gets (term-by-term):
\begin{itemize}
\item
$(1-x^*(y)y^*)^S=(1-x^*\tilde{y}^*\frac{K}{S})^S \simeq 
\exp(-x^*\tilde{y}^*K)=0.6592^K$.
\item
$(1+y^*\frac{S}{K})^K=1.7143^K$.
\item
$(1+x^*\frac{S}{M})^M=(1+0.5833\frac{S}{M})^M$.
\item
$\exp(-6K(x^*-1/2)^2)=0.9592^K$.
\item
$\exp(-6M(y^*-1/2)^2)\simeq \exp(3M/2)=0.2231^M$
\item The saddle point integration over $x$ 
combined with the normalization constant 
of the Gaussian distribution of $x$ generates
\begin{eqnarray}
&&\sqrt{12K}/\sqrt{-S \partial^2 \mathcal{L}/\partial x^2|_{x^*(y^*)}}= \nonumber \\
&&\sqrt{1/(1+M/(12Kx^*)}=0.8963 \nonumber
\qquad .
\end{eqnarray}
\item
The saddle point integration over $y$ combined with the 
normalization constant of the Gaussian distribution of $y$ generates
\begin{eqnarray}
&&\sqrt{12M}/\sqrt{-S \cdot d^2 \mathcal{L^*}/dy^2|_{y^*}}=\nonumber \\
&&\sqrt{12/0.2733}\cdot \sqrt{KM}/S=6.6258/L \quad . \nonumber
\end{eqnarray}
\item Factor $2$ for $K=M$ takes into account that our calculations were done 
for one saddle point in which $y^*=(K/S)\tilde{y}^*\ll1 $ and $x^* =O(1)$. 
The symmetric point located diagonally across this one on the $x-y$ plane would have an identical 
contribution. 
For $K \neq M$ one of these saddle points
would dominate 
the asymptotic formula. 
%We will leave exploring this 
%situation for future studies.
\end{itemize}
Once all these terms are put together one gets the following asymptotic formula 
for the number of uninvadable steady states for $L \gg 1$, 
and $K=M=L$.
\begin{equation}
N_{UIS}(S, L, L)=\frac{11.8769 \, L}{S}\left(0.2419+0.1411\, \frac{S}{L} \right)^L \, .
\label{eq:N_UIS_sp_general_S_SI}
\end{equation}
If in addition, the number of species in the pool is 
given by $S=L^2$ (as used in our numerical simulations for small $L$), one gets
\begin{equation}
N_{UIS}(L^2, L,L)=\frac{11.8769}{L}\left(0.2419+0.1411\, L \right)^L
\label{eq:N_UIS_sp_formula_SI}
\end{equation}

For $K \neq M$ this formula needs to be modified by 
first solving the Eq. \ref{eq:y*} to find the new saddle
points $x^*$ and $y^*$. These values then need to be plugged into 
the bullet list shown above to update the numerical coefficients combined 
in Eq. \ref{eq:N_UIS_sp_general_S_SI}. One fact remains
generally true, however. The leading super-exponential 
contribution would be given by 
\begin{equation}
    N_{UIS}(S,K,M) \sim \left(\frac{S}{M}\right)^M \, ,
\end{equation}
where $M$ corresponds to the resource with the largest number of 
nutrients (nitrogen in this example, where we assumed
that $M>K$).

\begin{figure}
\centerline{\includegraphics[width=\linewidth]{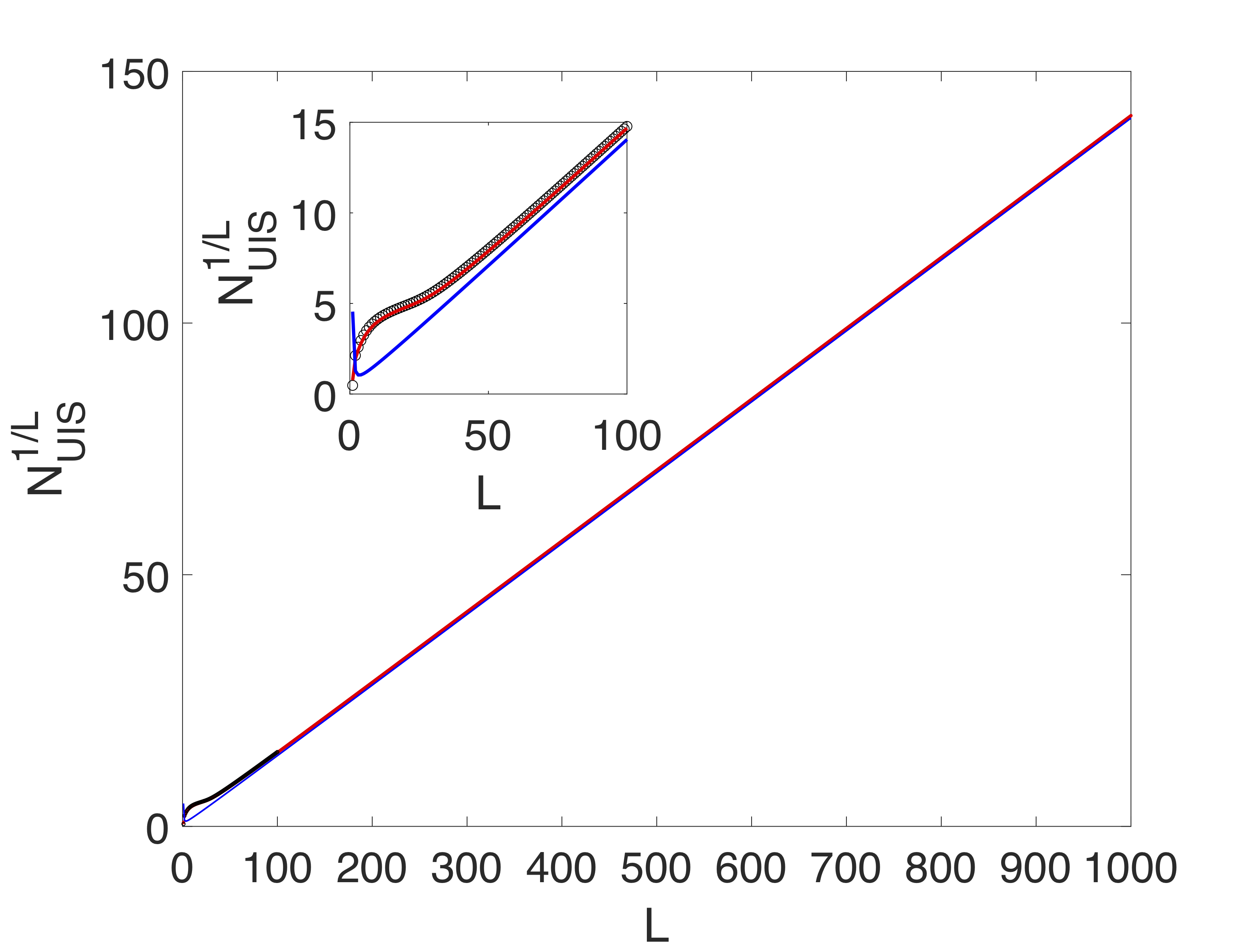}}
\caption{{\bf Saddle point approximation to the number of UIS in the continuous limit.} 
The number of uninvadable states raised to the power of $1/L$ plotted as a function of $L$.
The black symbols correspond to the 2-dimensional numerical integration of Eq. \ref{eq_n_uis_integral}, where 
both $x$ and $y$ range between $0$ and $1$ in steps of $1/L^2$. Because of numerical limitations, 
the integration has been only carried for $L \leq 100$. The red line is the 1-dimensional numerical 
integration of the Eq. \ref{eq:sp_integral_y} over $y$. Now $L$ is extended up to $1000$. The blue 
line is given by the complete saddle point calculation (Eq. \ref{eq:N_UIS_sp_formula_SI}). The 
ratios between either two of these three expressions asymptotically converges to 1.
The inset zooms up on the region $1 \leq L \leq 100$.
}
\label{fig2S}
\end{figure}

To test how well this expression calculated in the limit $L \gg  1$ matches
the numerical integration of Eq. \ref{eq_n_uis_integral}, in Fig. \ref{fig2S}
we compare them for $L$ going up to $1000$ (and $S=L^2$). Fig. \ref{fig2S} plots 
the number of uninvadable states raised to the power of $1/L$ plotted as a function of $L$.
The black symbols correspond to the 2-dimensional numerical integration of Eq. \ref{eq_n_uis_integral}, where 
both $x$ and $y$ range between $0$ and $1$ in steps of $1/L^2$. Because of numerical limitations, 
the integration has been only carried for $L \leq 100$. The red line is the 1-dimensional numerical 
integration of the Eq. \ref{eq:sp_integral_y} over $y$. Now $L$ is extended up to $1000$. The blue 
line is given by the complete saddle point calculation (Eq. \ref{eq:N_UIS_sp_formula_SI}). The 
ratios between either two of these three expressions asymptotically converges to 1.
Note that the saddle point expression in Eq. \ref{eq:N_UIS_sp_formula_SI} is completely 
off for $L \leq 9$ shown in Fig. \ref{fig1n}. 

\subsubsection*{The number of allowed and uninvadable states for 
more than two types of essential resources: continuous representation}
Note that all of the above formulas could be easily generalized to a 
biologically meaningful case of more than two types of essential nutrients.
For example, if one was to add another essential nutrient type (e.g.sources of 
phosphorus), the normalized ranking of $\lambda^{(P)}$ would introduce a new 
variable $0 \leq z <1$. The above formulas would be modified so that the 
probability that a state is not invadable now becomes 
\begin{equation}
(1-x \cdot y \cdot z)^S \qquad , 
\end{equation}
while the combinatorial factor calculating 
the number of allowed states for a 
given value of the average ranks $x$, $y$, and $z$
is given by
\begin{eqnarray}
N_{AS}(x,y,z)&=&(1+\frac{S}{K}\cdot y \cdot z)^K \cdot \nonumber \\
&\cdot& (1+\frac{S}{M}\cdot x \cdot z)^M \cdot \nonumber \\
&\cdot& (1+\frac{S}{P}\cdot x \cdot y)^P 
\end{eqnarray}
Here $P$ is the number of sources of phosphorous in 
the system, and the equation above sums over all 
microbes that are limited by either C, N, or P. 
It takes into account that a microbe 
limited by one resource has to be 
able to grow on the other two 
resources. 

\end{document}